\documentclass{aa}
\usepackage{graphicx}
\usepackage{txfonts}
\usepackage{natbib}
\bibpunct{(}{)}{;}{a}{}{,}
\usepackage{morefloats}
\usepackage{color}
\usepackage{longtable}
\usepackage{epstopdf}
\usepackage{caption}
\usepackage{subfig}
\begin{document}
   \title{ATLASGAL - Kinematic distances and the dense gas mass distribution of the inner Galaxy}


   \author{M. Wienen\inst{1}\thanks{Member of the International Max Planck Research School (IMPRS) for Astronomy and Astrophysics at the Universities of Bonn and Cologne.}, F. Wyrowski\inst{1}, K. M. Menten\inst{1}, J. S. Urquhart\inst{1}, T. Csengeri\inst{1}, C. M. Walmsley\inst{2,}\inst{3}, S. Bontemps\inst{4}, D. Russeil\inst{5}, L. Bronfman\inst{6}, B. S. Koribalski\inst{7} 
       \and F. Schuller\inst{1,}\inst{8}
       }

 \institute{\inst{1}Max-Planck-Institut f\"ur Radioastronomie, Auf dem H\"ugel 69, 53121 Bonn, Germany\\ \email{mwienen@mpifr-bonn.mpg.de}\\
          \inst{2}Osservatorio Astrofisico di Arcetri, Largo E. Fermi, 5, I-50125 Firenze, Italy\\
          \inst{3}Dublin Institute of Advanced Studies, Fitzwilliam Place 31, Dublin 2, Ireland\\
          \inst{4}Universit\'e de Bordeaux, Laboratoire d'Astrophysique de Bordeaux, CNRS/INSU, UMR 5804, BP 89, 33271 Floirac Cedex, France\\
          \inst{5}Aix Marseille Universit\'e, CNRS, LAM (Laboratoire d'Astrophysique de Marseille), UMR 7326, 13388 Marseille, France\\
          \inst{6}Departamento de Astronom\'{\i}a, Universidad de Chile, Casilla 36-D, Santiago, Chile\\
          \inst{7}Australia Telescope National Facility, CSIRO, P.O. Box 76, Epping, NSW 1710, Australia\\
          \inst{8}Alonso de Cordova 3107, Casilla 19001, Santiago 19, Chile}


   \date{Received }

 
  \abstract
  {The formation of high mass stars and clusters occurs in giant molecular clouds. Objects in evolved stages of massive star formation such as protostars, hot molecular cores, and ultracompact HII regions have been studied in more detail than earlier, colder objects. Further progress thus requires the 
  analysis of the time before massive protostellar objects can be probed by their infrared emission. With this in mind, the APEX Telescope Large Area Survey 
of the whole inner Galactic plane at 870 $\mu$m (ATLASGAL) has been carried out to provide a global view of cold dust and star formation at submillimetre wavelengths.}
   {We derive kinematic distances to a large sample of massive cold dust clumps 
    from their measured line velocities. We estimate masses and sizes of ATLASGAL sources, for which the kinematic distance ambiguity is resolved.}
   {The ATLASGAL sample is divided into groups of sources, which are located close together, mostly within a radius of 2 pc, and have velocities in a similar range with a median velocity dispersion of $\sim 1$ km~s$^{-1}$. We use NH$_3$, N$_2$H$^+$, and CS velocities to calculate near and far kinematic distances to those groups.} 
   {We obtain 296 groups of ATLASGAL sources in the first quadrant and 393 groups in the fourth quadrant, which are coherent in space and velocity. We analyse HI self-absorption and HI absorption to resolve the kinematic distance ambiguity to 689 complexes of submm clumps. They are associated with $^{12}$CO emission probing large-scale structure and 
$^{13}$CO (1-0) line as well as the 870 $\mu$m dust continuum on a smaller scale. 
   We obtain a scale height of $\sim 28 \pm 2$ pc and displacement below the Galactic midplane of $\sim -7 \pm 1$ pc. 
Within distances from 2 to 18 kpc ATLASGAL clumps have a broad range of gas masses with a median of 1050 M$_{\odot}$ as well as a wide distribution of radii 
   Their distribution in galactocentric radii is correlated with spiral arms.}
   {Using a statistically significant ATLASGAL sample we derive a power-law exponent of $-2.2 \pm 0.1$ of the clump mass function. This is consistent with the slope derived for clusters and with that of the stellar initial mass function. Examining the power-law index for different galactocentric distances and various source samples shows that it is independent of environment and evolutionary phase. Fitting the mass-size relationship by a power law gives a slope of $1.76 \pm 0.01$ for cold sources such as IRDCs and warm clumps associated with HII regions.
}

\titlerunning{ATLASGAL - Kinematic distances and the dense mass distribution of the inner Galaxy}
\authorrunning{M. Wienen et al.}
   \maketitle
%

\section{Introduction}

\subsection{High mass star formation}
Current theory of star formation distinguishes between the formation of stars with low and high ($>$ 8 M$_{\odot}$) mass. In spite of much effort to understand the process that leads to the formation of high mass stars, very little is known about their earliest phases in contrast to the better understood case of isolated low mass star formation.
Massive stars release a great amount of energy into the interstellar medium through radiation and stellar winds and deposit heavy elements in their late evolutionary phases or via supernova explosions, which leads to a chemical enrichment of the interstellar medium. They thus influence their environment, may trigger the formation of following generations of stars, and also play a dominant role in the evolution of galaxies \citep{2005IAUS..227....3K}. To advance the understanding of high mass star formation, many surveys have been conducted so far. Targeted surveys detected strong radio emission from ultracompact HII regions (UCHIIRs), which are formed when the protostar emits ultraviolet radiation, thus heating and ionizing the remaining molecular cloud \citep{1989ApJS...69..831W,1990ApJ...358..485B,1994ApJS...91..659K,1998MNRAS.301..640W,2012PASP..124..939H}.
Moreover, extensive surveys have been made toward samples that were selected at far infrared wavelengths (IRAS) to identify sites of high mass stars at the earliest evolutionary phase \citep{1997MNRAS.291..261W,2000A&AS..143..269S,2000A&A...355..617M}. In addition, many new massive star forming regions have been revealed in surveys for 6.7 GHz methanol maser emission \citep{2002A&A...392..277S,2002ApJ...566..931S,2011MNRAS.417.1964C,2009MNRAS.392..783G}.

Significant progress in the study of star formation has also been achieved by various large-scale infrared continuum surveys. Dust obscuration of the visible light from stars hampers the detection of the stellar content of the inner Galaxy, which made surveys at longer wavelengths necessary. The Midcourse Space Experiment \citep[MSX;][]{2001AJ....121.2819P} observed the whole Galactic plane within Galactic latitudes of $\pm 5^{\circ}$ between $\sim$ 6 and 25 $\mu$m with a spatial resolution of $\sim 18\arcsec$. It provided the data necessary to investigate interstellar dust, young stellar objects as well as HII regions and Galactic structure. More recently with still increasing resolution and sensitivity the Spitzer Galactic Legacy Infrared Mid-Plane Survey Extraordinaire \citep[GLIMPSE;][]{2003PASP..115..953B} and the MIPS Galactic Plane Survey \citep[MIPSGAL;][]{2009PASP..121...76C} followed. Emission from 3 to 8 $\mu$m with a spatial resolution of $\sim 2\arcsec$ is observed by GLIMPSE over a Galactic latitude range of $\pm 1^{\circ}$ and longitudes between 295$^{\circ}$ and 65$^{\circ}$. As a longer wavelength complement MIPSGAL surveys 278 deg$^2$ of the inner Galactic plane at 24 and 70 $\mu$m with resolutions of 6$\arcsec$ and 18$\arcsec$. Surveys at longer wavelengths are the Bolocam Galactic Plane Survey \citep[BGPS;][]{2011ApJS..192....4A} covering $-10^{\circ} < l < 90^{\circ}$ and $-0.5^{\circ} < b < 0.5^{\circ}$ at 1.1 mm and the Herschel Infrared Galactic Plane Survey \citep[Hi-GAL;][]{2010PASP..122..314M} reaching a Galactic longitude of $\pm 60^{\circ}$ and latitude of $\pm 1^{\circ}$ from 70 to 500 $\mu$m. Another large-scale project is the Red MSX Source \citep[RMS,][]{2013ApJS..208...11L} survey, which searches for massive young stellar objects (MYSOs) in the whole Galaxy. Multi-wavelength observations are conducted to distinguish the MYSOs and UCHIIRs from other red sources, which allows statistical studies of these stages of high mass star formation. 

All these surveys easily find luminous sources in the Galactic plane. However, the disadvantage of these targeted surveys is that they are partial, probing only a particular evolutionary stage.
For example, UCHIIRs detected by their free-free emission probe recent massive star formation. Moreover, many suffer from incomplete statistics. To overcome these drawbacks, the first unbiased submm continuum survey of the whole inner Galactic disk, \textit{the APEX Telescope Large Area Survey of the Galaxy at 870 $\mu$m} \citep[ATLASGAL;][]{2009A&A...504..415S} was conducted. Using the Large APEX Bolometer Camera (LABOCA), observations were made in a Galactic longitude range of $\pm$60$^{\circ}$ and latitude of $\pm$1.5$^{\circ}$ in order to obtain a statistically relevant sample of objects associated with high mass star formation at various stages as well as to study their distribution and to compare physical properties of these clumps. The definition of a clump given by \cite{2000prpl.conf...97W} is adapted in this article. We thus consider clumps to be overdense substructures within molecular clouds that are coherent in Galactic longitude, latitude, and radial velocity in molecular line maps. Clumps exhibit sizes between 0.3 and 3 pc \citep{2007ARA&A..45..339B} and can form whole clusters of stars in high mass star forming regions. Although the ATLASGAL survey is very important to get a global view of star formation at submillimetre wavelengths by identifying all massive clumps forming high mass stars in the inner Galaxy, a main limitation that applies to all continuum surveys is that the distances to the newly found sources are unknown. These are needed to determine important parameters such as masses and luminosities of the clumps as well as to analyse the spiral structure of the Milky Way. To measure distances towards a large sample of high mass star forming clumps located at a vast range of distances, we need an efficient method, which is the kinematic distance estimate \citep{1972A&A....19..354W,2003ApJ...582..756K,2009ApJ...699.1153R,2014MNRAS.437.1791U}.

Ammonia observations of ATLASGAL sources with kinematic distances in the first quadrant of the Galaxy are presented in \cite{2012AA...544A.146W}, where the distance ambiguity is not resolved. We now complement this data with new NH$_3$ measurements towards ATLASGAL sources in the fourth quadrant and add a new distance analysis towards the combined sample. Section \ref{data} describes molecular line and HI data, which were used to derive kinematic distances to ATLASGAL sources. We give details about the rotation curve and the computation of the errors in the distance in Sect. \ref{dist-calc}. Section \ref{kinematic distance} describes the two methods used to resolve the kinematic distance ambiguity, HI self-absorption, and HI absorption. We show our results in Sect. \ref{results} and discuss them in Sect. \ref{discussion}. A summary of our analysis is presented in Sect. \ref{conclusions}. 
\subsection{Kinematic distance}
To obtain an estimate for the kinematic distance requires the measurement of the radial velocity of a source, which can be associated with its galactocentric radius using a model of the Galactic rotation curve \citep[e.g.][]{2009ApJ...700..137R,1993A&A...275...67B}. While the derivation of distances from velocities is straigthforward in the outer Galaxy, it is more challenging for sources with Galactic radii smaller than that of the Sun, thus located in the inner Galaxy. The longitude and velocity of an object determine a unique galactocentric radius, but in general allow for two different values for its distance to the Sun. Furthermore the near and far kinematic distances are equally spaced on the near and far side of the point, where the line of sight is tangent to the orbit of the object and its radial velocity equals the circular velocity. The near and far distances are only at the tangent point the same.

There are many studies, which tried different methods to solve the kinematic distance ambiguity (KDA). In addition to the derivation of distances to high mass star forming regions, pulsar distances are mostly determined using the kinematic distance method. Measurements of HI absorption at 21 cm against bright pulsars leading to lower and upper distance limits to these objects are shown e.g. by \cite{1995ApJ...441..756K} and \cite{1990AJ....100..743F}. A common approach used for HII regions consists of measuring an absorption spectrum against the continuum free-free emission radiated by the HII region. Near and far distances can be distinguished by comparing observed absorption line velocities between the source velocity and the velocity of the tangent point. Previous work analysed H$_2$CO absorption against 6 cm continuum emission to solve the KDA \citep{1972A&A....19..354W,2003ApJ...587..714W,2004ApJS..154..553S}. Since the HI abundance is high towards molecular clouds, KDA solutions using this probe are more reliable than those resulting from H$_2$CO absorption. Distances to small samples of HII regions are determined using HI absorption against radio continuum emission \citep{2003ApJ...582..756K,2003ApJ...587..701F}. To resolve the KDA to compact HII regions from the RMS survey, \cite{2012MNRAS.420.1656U} used the same technique. Recent work by \cite{2014ApJS..212....2G} distinguished between near and far distances to groups of molecular clumps embedded in the same giant molecular cloud with one of them harbouring an HII region using continuum absorption as well. Because this method requires an HII region to be embedded in the molecular cloud, it is only helpful to resolve the KDA for a subsample of ATLASGAL sources. However, we also need a technique appropriate for all molecular clumps independent of them harbouring an HII region.

In addition to the HI absorption, we investigate HI self-absorption (HISA) toward observed NH$_3$ emission of ATLASGAL sources to resolve the KDA. Using this technique we analyse if cold HI in a molecular cloud absorbs warm HI line emission in the interstellar medium. Earlier studies have also used this method, e.g. \cite{2002ApJ...566L..81J} found HISA toward the molecular cloud GRSMC 45.6+0.3, which reveals its near kinematic distance. Recently, \cite{2014MNRAS.437.1791U} examined HI absorption to determine distances to $\sim 800$ RMS sources. In addition, \cite{2009ApJ...699.1153R} analysed HISA together with the HI absorption to resolve the KDA toward 750 molecular clouds observed within the Boston University-Five College Radio Astronomy Observatory Galactic Ring Survey \citep[GRS,][]{2006ApJS..163..145J}. Both techniques are also used by \cite{2006MNRAS.366.1096B} to determine distances to a sample of massive young stellar objects as well as by \cite{2009ApJ...690..706A} for a sample of HII regions. They investigated the confidence of HISA and both find that it agrees with HI absorption in $\sim$ 80\% of their samples. \cite{2009ApJ...690..706A} point out several reasons why the HI absorption is still more certain for the analysis of distances to HII regions. In addition, reliable KDA solutions can be derived from interferometric observations. These avoid contamination from large-scale HI emission \citep{2012MNRAS.420.1656U} and also avoid confusion caused by several regions in the beam, although most ATLASGAL sources are associated with a single HII region and only 23 are associated with multiple HII regions \citep[see Fig. 3 in][]{2013MNRAS.435..400U}. However, interferometric observations of a large ATLASGAL sample, to which we derive kinematic distances, are prohibitive.

This paper uses a consistent method to determine kinematic distances to a large sample of ATLASGAL sources in the first and fourth quadrant. We are aware that our method to derive kinematic distances can result in large distance errors for individual sources. However, our conclusions do not rely on correct individual distances. We focus on the analysis of statistics of parameters derived from the distances, which is not affected by the distance errors, because they average out for our large sample of ATLASGAL sources.

\section{Data sets}
\label{data}
\subsection{Observations}
\label{s:obs}
We mainly use our subsample of ATLASGAL sources, which was observed in the NH$_3$ (1,1), (2,2), and (3,3) inversion lines within a Galactic longitude range from 5$^{\circ}$ to 60$^{\circ}$ to derive kinematic distances. We measured NH$_3$ radial velocities of an ATLASGAL sample in the first quadrant described in \cite{2012AA...544A.146W} and complement it with ammonia observations of a flux-limited sample of ATLASGAL sources in the fourth quadrant down to about 1.2 Jy/beam (Wienen et al. in prep.). For the measurements in the first quadrant we used the Effelsberg 100-m telescope with a beamwidth (FWHM) of 40$\arcsec$ at the frequencies of the NH$_3$ (1,1) to (3,3) lines at $\sim$ 24 GHz and a spectral resolution of about 0.7 km~s$^{-1}$. Sources in the fourth quadrant were observed using the Parkes 64-m telescope with a 13mm receiver as frontend ranging between 16 and 26 GHz and a beamwidth of 60$\arcsec$ at $\sim$ 24 GHz. We had a Digital Filter Bank (DFB3) as spectrometer with a spectral resolution of about 0.4 km~s$^{-1}$. We observed two polarizations of the NH$_3$ (1,1) to (3,3) lines simultaneously in position-switching mode. The source selection as well as data reduction are described in \cite{2012AA...544A.146W} and in Wienen et al. (in prep.). Because we observed only half the number of sources in NH$_3$ in the fourth quadrant than in the first quadrant resulting in 315 radial velocities in the fourth quadrant compared to 752 velocities in the first quadrant, we added other molecular line data for the sources in the fourth quadrant (see Sect. \ref{molecular data}).

\subsection{Molecular line data}
\label{molecular data}
\begin{figure}
\centering
\includegraphics[angle=0,width=9.0cm]{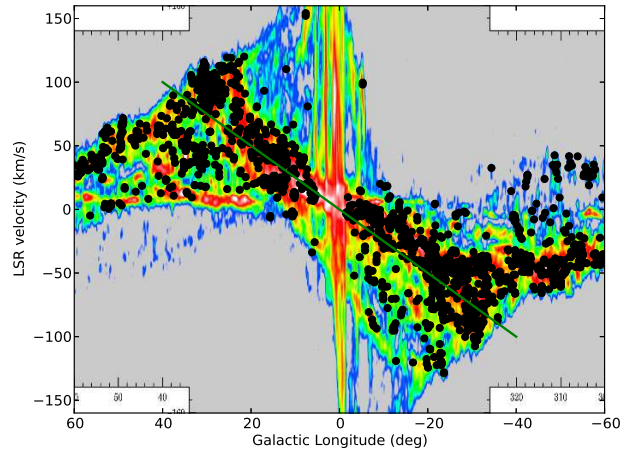}
\caption[comparison of longitude and velocity]{Galactic longitudes of ATLASGAL sources are compared with the radial velocities with CO (1$-$0) emission in the background \citep{2001ApJ...547..792D}. The green line indicates the 5 kpc molecular ring \citep{2006ApJ...653.1325S}.}\label{l-v-ns}
\end{figure}

A subsample of ATLASGAL sources was followed up in molecular lines. From 1067 ATLASGAL sources observed in NH$_3$ HI data were available for 1056 clumps and we thus used radial velocities of 749 ATLASGAL sources in the first quadrant and of 307 clumps in the fourth quadrant from NH$_3$ line measurements to derive kinematic distances. We added N$_2$H$^+$ (1$-$0) observations of ATLASGAL sources in the fourth quadrant (Wyrowski et al. in prep.). In addition to our own data we use velocities from CS (2$-$1) measurements of IRDCs from \cite{2008ApJ...680..349J}, from CS (2$-$1) observations of UCHIIRs from \cite{1996AAS..115...81B}, and from $^{13}$CO measurements of MYSOs from the RMS survey \citep{2007AA...474..891U}. We thus obtain radial velocities of a total of 1065 sources in the fourth quadrant. The different molecular lines are summarized in Table \ref{molecular}, which contains the molecular probe, the number of all sources observed in NH$_3$, and of clumps added from N$_2$H$^+$, CS, and $^{13}$CO measurements, the number of sources, for which HI data are available and which were used to resolve the KDA, their longitude ranges, and a reference. The distribution of the clumps used to determine kinematic distances is overlaid on the CO (1$-$0) emission \citep{2001ApJ...547..792D} in Fig. \ref{l-v-ns}. The straight line illustrates the 5 kpc molecular ring, where star formation is actively going on \citep{2006ApJ...653.1325S}. Most clumps show strong CO emission, which traces the large-scale structure.

\begin{table*}[htbp]
\caption[]{Summary of molecular lines used to obtain radial velocities.}
\label{molecular}
\centering
\begin{tabular}{l c c c l}
\hline\hline
  molecular tracer & number of  & number of & longitude range  & Reference \\ 
                   & all sources & sources with HI data  &  (deg)    &    \\ \hline 
   NH$_3$ (1,1)   & 752 &  749  &  $5 - 60$ & \cite{2012AA...544A.146W}  \\
   NH$_3$ (1,1)   & 315 & 307  &  $300 - 360$ & Wienen et al. (in prep.)  \\
   N$_2$H$^+$ $(1-0)$ & 301  &  301  & $300 - 360$ & Wyrowski et al. (in prep.) \\ 
   CS $(2-1)$   & 146 & 146  & $301 - 358$ & \cite{2008ApJ...680..349J} \\
   CS $(2-1)$   & 111 & 111  & $300 - 355$ & \cite{1996AAS..115...81B} \\
   $^{13}$CO $(1-0)$, $^{13}$CO $(2-1)$ & 200 &  200  & $300 - 350$ & \cite{2007AA...474..891U} \\\hline 
\end{tabular}
\end{table*}

\subsection{Archival data}
\label{archival data}
\subsubsection*{The VLA Galactic Plane Survey}
To look for HI self-absorption and HI absorption of the ATLASGAL sample in the first quadrant we mainly use the VLA Galactic Plane Survey \citep[VGPS,][]{2006AJ....132.1158S}. The HI 21 cm line cubes were created from the survey, which observed the Galactic longitude range from 18$^{\circ}$ to 67$^{\circ}$ and latitude varying from $\pm$1.3$^{\circ}$ to $\pm$2.6$^{\circ}$. The VGPS has an angular resolution of 1$\arcmin$, a spectral resolution of 1.56 km~s$^{-1}$, and an rms noise of 2 K per 0.824 km~s$^{-1}$ channel.

\subsubsection*{The Southern Galactic Plane Survey}
For sources with Galactic longitude $l < 18^{\circ}$ HI data are available from the Southern Galactic Plane Survey \citep[SGPS,][]{2005ApJS..158..178M}. The whole survey was conducted between l=253$^{\circ}$ and 358$^{\circ}$ as well as from l=5$^{\circ}$ to 20$^{\circ}$ with a latitude of $\pm$1.5$^{\circ}$. The SGPS uses the Australia Telescope Compact Array and the Parkes telescope. The angular resolution of the HI data is 2$\arcmin$, the spectral resolution 0.8 km~s$^{-1}$, and the rms sensitivity 1.6 K.

\captionsetup[subfigure]{position=top}
\begin{figure*}[tbp]
\centering

\subfloat[\hspace*{10cm}]{\includegraphics[angle=0,width=12.0cm]{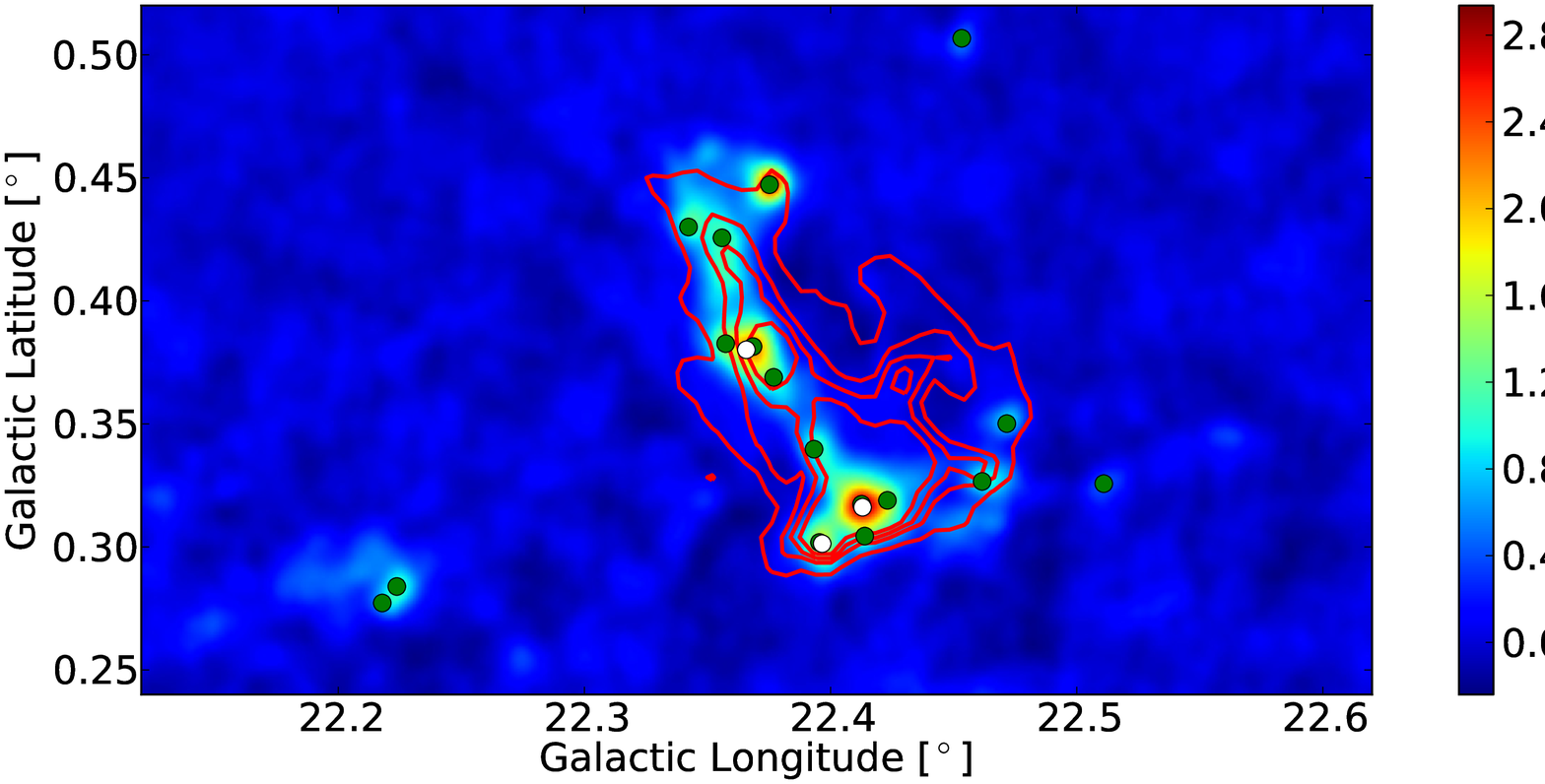}}\vspace*{0.5cm}

\subfloat[\hspace*{15cm}]{\includegraphics[angle=0,width=8.0cm]{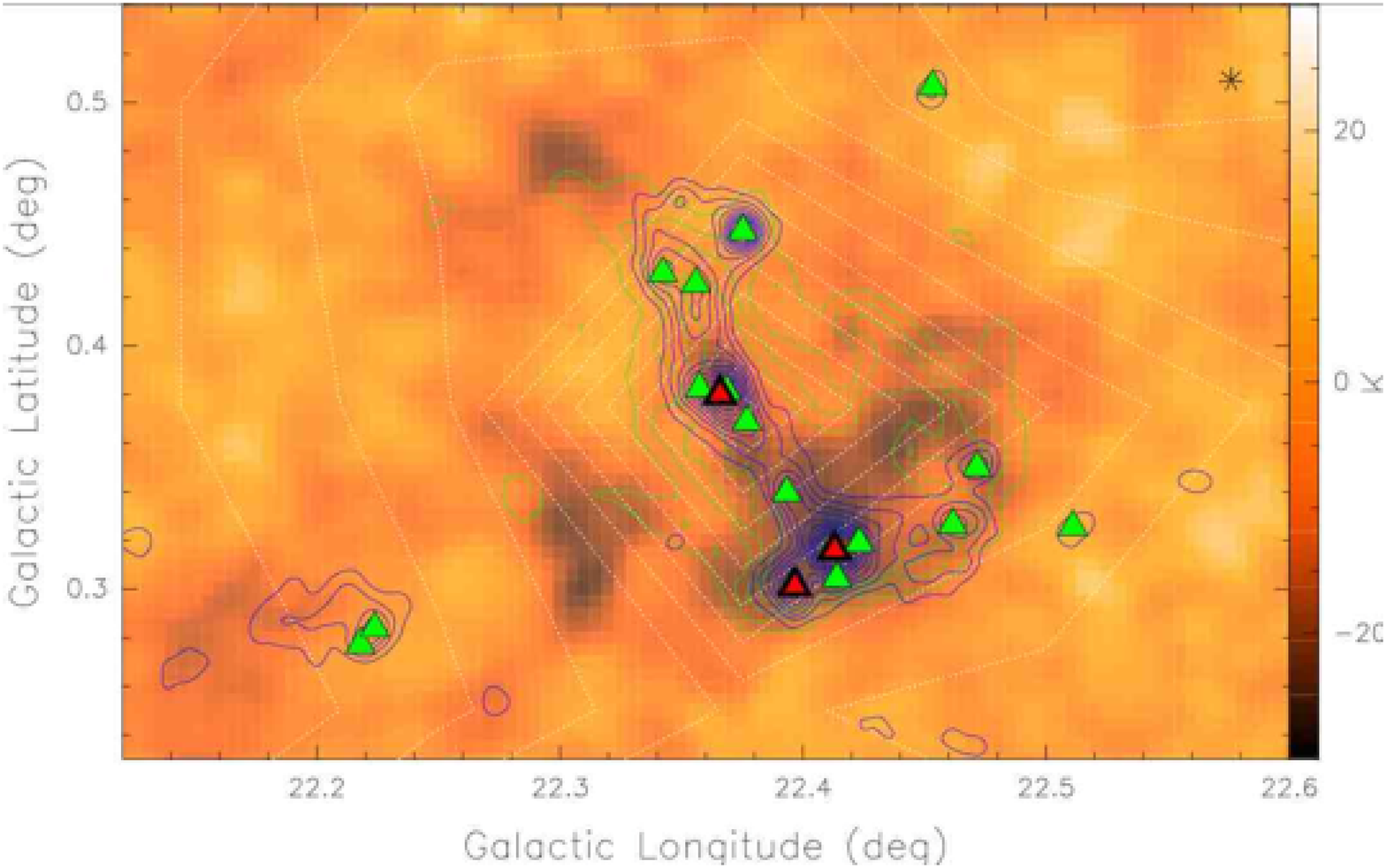}

\includegraphics[angle=0,width=8.0cm]{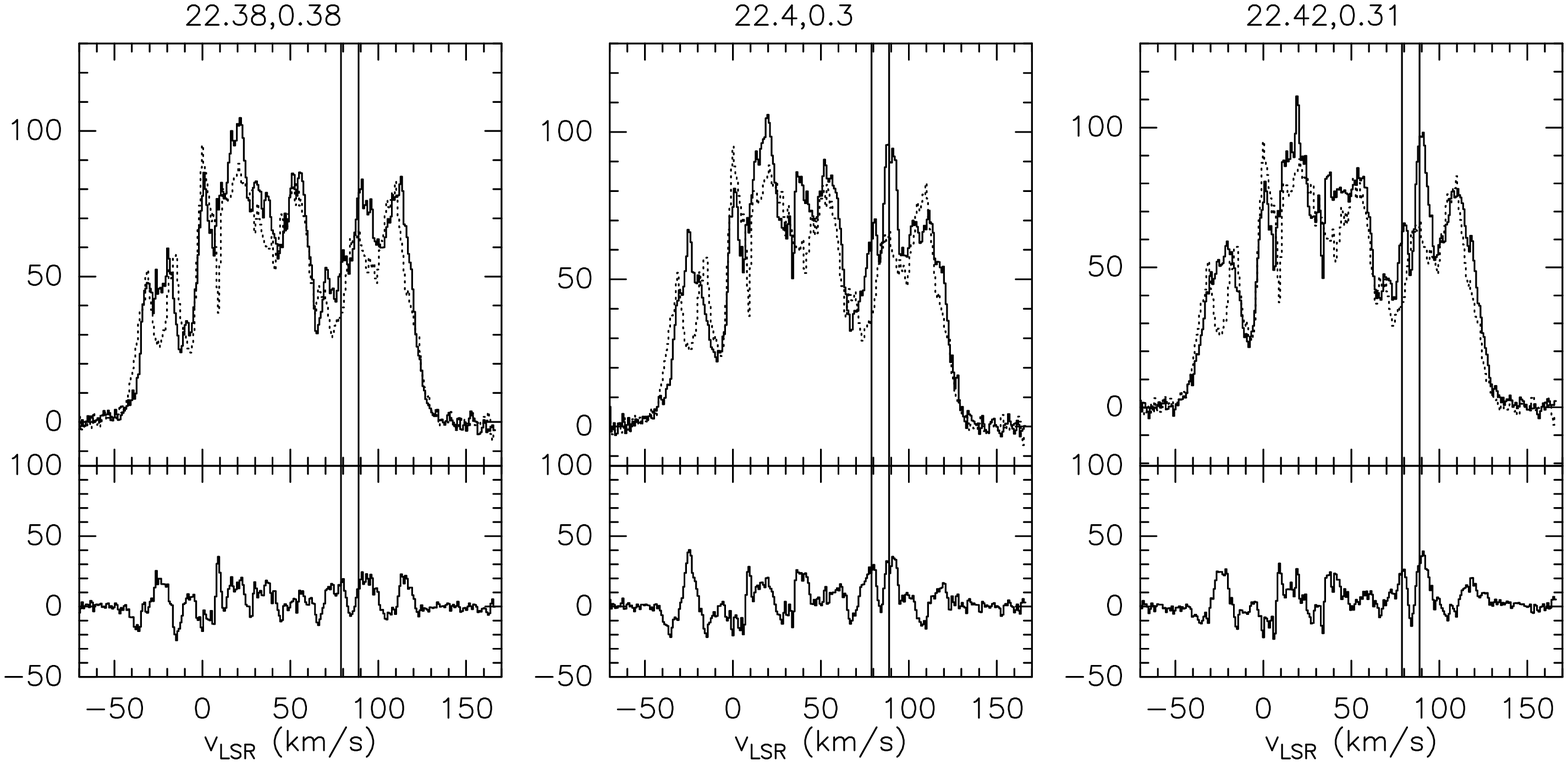}\vspace*{0.5cm}
}

\subfloat[\hspace*{10cm}]{\includegraphics[angle=0,width=7.0cm]{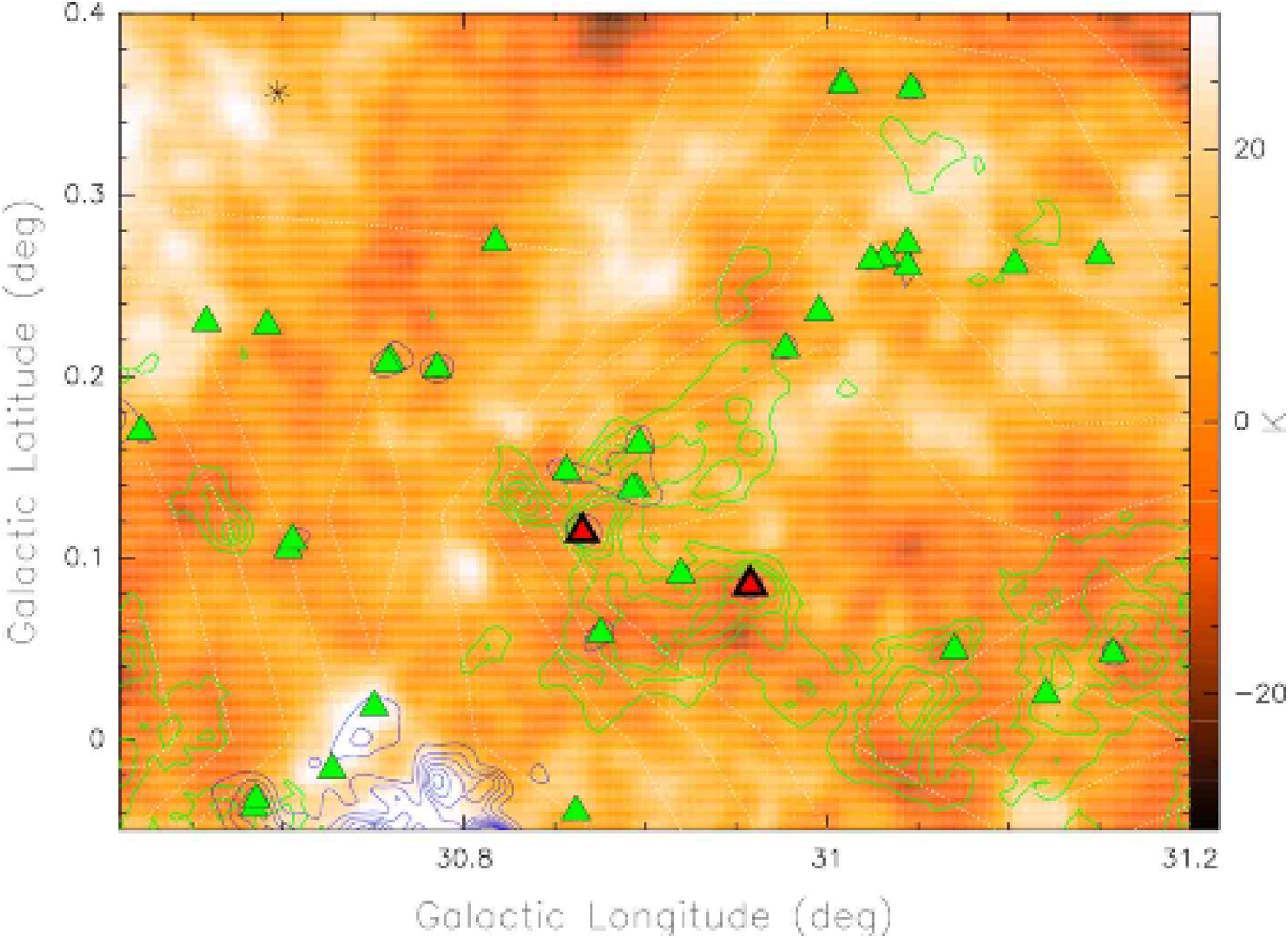}}

\caption[HISA of a complex]{\textbf{a)} Example of a complex in the first quadrant: The 870 $\mu$m dust continuum is shown in the background probing the small-scale molecular cloud structure. We illustrate ATLASGAL sources with observed velocities of $\sim$ 84 km~s$^{-1}$ as white dots and ATLASGAL sources without known velocities as green dots. The $^{13}$CO emission at the velocity of the complex is indicated as red contours. \textbf{b)} The panel on the left displays the HI emission map of the region shown in \textbf{a}: The HI intensity is integrated in the velocity interval around the group velocity and in intervals at smaller and larger velocities. The subtraction of these maps (see Sect. \ref{HISA analysis}) reveals HI self-absorption plotted in the background with the 870 $\mu$m dust continuum overlaid as blue contours, the $^{13}$CO (1$-$0) emission as green contours, and the $^{12}$CO intensity as white contours. We show ATLASGAL sources with measured velocities as red triangles and ATLASGAL sources without velocities as green triangles. We extracted HI lines at the location of observed source and at one offset position, which is indicated by the star on the map. The HI ''on-source'' spectra are illustrated as solid lines and the ''off-source'' spectrum as dashed line in the panel on the right. The labels on the top display the galactic coordinates of observed source positions. Vertical lines in the spectra indicate the variation of the source velocity resulting from the velocity dispersion of clouds or streaming motions. The difference of the on-source and off-source spectra is shown below the HI spectra. \textbf{c)} Example of another region, which illustrates HI emission. The HI map in the background is produced and overlaid with the same contours and symbols as described for panel \textbf{b}.}\label{HISA}
\end{figure*}

\section{Grouping of ATLASGAL sources with known velocities}
\label{grouping}
We describe in this section our method to group sources that cluster in position-velocity space according to the friends-of-friends algorithm \citep{1982ApJ...257..423H,1993MNRAS.261..827M,2006ApJS..167....1B}. As a first step we search around each ATLASGAL source within a maximum distance and velocity interval for associated sources. We accumulate ATLASGAL sources to the same group until no new source that obeys the two criteria to the neighbouring sources is found. All distances between at least two individual sources within one group must therefore be smaller than the maximum distance and the velocities of these must be within a given velocity interval. As a next step we vary the maximum distance and the velocity interval to analyse the effect of these input parameters on the grouping, which leads to different divisions of the ATLASGAL sources into groups. More details about that are given in Appendix \ref{grouping details}. These divisions are investigated to reach a better mapping between our grouping and properties of known molecular cloud complexes (see Appendix \ref{gmc}). We refer to sources, which are gathered together according to their location and kinematics, as a group and denote its spatial extent as a complex, when we compare its structure with the ATLASGAL dust continuum emission. An example of such a complex within $22.1^{\circ} < l < 22.6^{\circ}$ and $0.25^{\circ} < b < 0.5^{\circ}$ is given in the top panel of Fig. \ref{HISA}. The submm dust continuum emission is plotted in the background and traces the filamentary structure within this molecular cloud complex. It contains three ATLASGAL sources with velocities around 84 km~s$^{-1}$ known from NH$_3$, N$_2$H$^+$ or CS observations overlaid as white dots and ATLASGAL sources without known velocities illustrated as green dots. The red contours show the $^{13}$CO emission from the GRS survey \citep{2006ApJS..163..145J} at the velocity of the complex illustrating that the ATLASGAL clumps belong to one complex, which is coherent in space and velocity.

Using our method we obtain 296 groups in the first quadrant and 393 groups in the fourth quadrant. The group number, the mean position, number of sources per group, the mean velocity, the velocity dispersion, and the size of the groups are given in Table \ref{complex-vmean}: No velocity dispersion and size are calculated for 
sources, which are not associated with another ATLASGAL source within 0.3$^{\circ}$ and 10 km~s$^{-1}$.

\begin{table*}
\begin{minipage}{\textwidth}
\caption{Properties of complexes (see Sect. \ref{grouping}). The full table is available at CDS.}              
\label{complex-vmean}      
\centering                                      
\begin{tabular}{l l l c l l l l c}          
\hline\hline                        
Complex & mean longitude  & mean latitude  & number of sources & mean velocity  & mean velocity dispersion  & size  \\ 
     &  (deg) & (deg) & per complex & (km~s$^{-1}$) & (km~s$^{-1}$) & (pc)   \\    
\hline                                  
 1 & 5.2732 & 0.0662 & 9 & 9.9 & 1.4 & 22 \\
2 & 5.6161 & -0.0872 & 2 & -25.7 & - & - \\
3 & 5.8834 & -0.4295 & 22 & 11.8 & 3.9 & 28 \\
4 & 5.9901 & -1.2114 & 13 & 12.3 & 2 & 104 \\
5 & 6.1909 & -0.3626 & 2 & -33.8 & - & - \\
6 & 6.2288 & -0.079 & 8 & 14.8 & 3.3 & 39 \\
7 & 6.6718 & -0.203 & 16 & 16.7 & 5.3 & 22 \\
8 & 7.1655 & 0.132 & 1 & 80.9 & - & - \\
9 & 7.3017 & -0.1621 & 4 & 22 & 3.5 & 47 \\
10 & 7.2911 & -0.5397 & 4 & 19.2 & 1.5 & 3 \\
11 & 7.3395 & -0.01 & 2 & 31.5 & - & 3 \\
12 & 7.5809 & -0.1494 & 5 & 153.2 & 1.1 & 23 \\
13 & 8.0089 & -0.2355 & 5 & 40 & 0.8 & 26 \\
14 & 8.1533 & 0.2662 & 9 & 18.7 & 0.1 & 13 \\
15 & 8.5578 & -0.3362 & 17 & 37.4 & 1.1 & 57 \\
16 & 8.4959 & -0.9814 & 2 & 16 & - & 0.4 \\
17 & 8.9549 & -0.5346 & 1 & 20.2 & - & - \\
18 & 9.0374 & -0.521 & 1 & 37.1 & - & - \\
19 & 9.2132 & -0.1991 & 1 & 42.1 & - & - \\
20 & 9.8551 & -0.732 & 4 & 27.6 & - & 5 \\

\hline                                             
\end{tabular}
\end{minipage}
\end{table*}

\section{Kinematic distance estimation}
\label{dist-calc}
{We calculated kinematic distances to ATLASGAL sources, for which we had determined the LSR velocity, using the \cite{1993A&A...275...67B} rotation curve, which assumes 8.5 kpc for the distance of the Sun to the Galactic centre, R$_0$, and 220 km~s$^{-1}$ for the circular rotation velocity at the position of the Sun, $\Theta_0$. A more recent model using the revised rotation parameters of the Milky Way has been presented by \cite{2009ApJ...700..137R}. They fit their trigonometric parallax measurements of masers in high mass star forming regions to a model of the Galaxy and give their results for flat and more complex rotation curves in Table 3 of \cite{2009ApJ...700..137R}. A comparison of measured HI terminal velocities in the fourth quadrant of the Milky Way observed by \cite{2007ApJ...671..427M} with tangent point velocities calculated from the different rotation curves revealed that the measurements are best fitted by the model from \cite{1993A&A...275...67B}. Moreover, \cite{2009ApJ...700..137R} fit rotation curves to measured parallaxes of masers in 18 high mass star forming regions, which are observed as part of the BeSSeL survey only in the first and second quadrants of the Galaxy so far. \cite{2009ApJ...693..397R} aim at extending their measurements to obtain distances in the first and fourth quadrant. Of course by now more sources have been observed\footnote{http://www3.mpifr-bonn.mpg.de/staff/abrunthaler/BeSSeL/index.shtml}. The ATLASGAL sources are better represented by the \cite{1993A&A...275...67B} rotation curve, before trigonometric parallax measurements will improve the rotation curves in the inner Galaxy. To be consistent for all ATLASGAL clumps we use one rotation curve for the whole Galaxy.

We get two solutions from the rotation curve for every radial velocity of sources inside the solar circle, corresponding to a near and far distance. For objects outside the solar circle only one distance is physically possible. For sources located at the tangent point the near and far distance are similar. We give the source name, LSR velocity, the KDA solution (see Sect. \ref{kinematic distance}), the kinematic distance calculated from the LSR velocity of each source, the distance of the group, in which the clump is located, its errors, the gas mass, radius (see Sect. \ref{mass size}), and the group number in Table \ref{v11-dist}. The distance of the group is computed from the mean of all source coordinates and velocities inside the group. To avoid any influence of peculiar motion we recommend the distance of the group, which is used to determine gas masses and sizes. Two of the sources ($l,b = 5.62, -0.08; l,b = 6.19, -0.36$) have velocities inconsistent with the rotation model and we do not assign a distance to them.

\begin{table*}
\begin{minipage}{\textwidth}
\caption{Kinematic distances to ATLASGAL sources determined using HISA and HI absorption, gas masses, and radii (see Sect. \ref{dist-calc}). The full table is available at CDS.}              
\label{v11-dist}      
\centering 
\renewcommand{\arraystretch}{1.4}
\begin{tabular}{l l c l l l l c}          
\hline\hline                        
Name & $\rm v$(1,1) & KDA & Distance & Distance$_{\mbox {\tiny complex}}$ & M$_{\mbox{\tiny gas}}$ &  radius  & Complex \\ 
     &  (km~s$^{-1}$) & solution & (kpc) & (kpc) & (M$_{\odot}$) & (pc) &  \\    
\hline                                  
 G5.35+0.10 & 11.21 & n & 3.04 & $2.75^{+1.07}_{-1.67}$ & 2.72 & 0.31 & 1 \\
G5.39+0.19 & 10.58 & n & 2.91 & $2.75^{+1.07}_{-1.67}$ & 2.35 & 0.19 & 1 \\
G5.64+0.24 & 7.94 & n & 2.28 & $2.75^{+1.07}_{-1.67}$ & 3.01 & 0.81 & 1 \\
G5.62-0.08 & -25.74 & f & - & - & - & - & 2 \\
G5.83-0.51 & 16.22 & n & 3.64 & $2.94^{+0.96}_{-1.42}$ & 3.24 & 1 & 3 \\
G5.83-0.40 & 7.64 & n & 2.16 & $2.94^{+0.96}_{-1.42}$ & 2.80 & 0.78 & 3 \\
G5.89-0.39 & 9.18 & n & 2.47 & $2.94^{+0.96}_{-1.42}$ & 3.14 & 0.64 & 3 \\
G5.89-0.29 & 10.11 & n & 2.65 & $2.94^{+0.96}_{-1.42}$ & 2.74 & 0.56 & 3 \\
G5.89-0.32 & 10.11 & n & 2.65 & $2.94^{+0.96}_{-1.42}$ & 2.65 & 0.58 & 3 \\
G5.90-0.44 & 9.1 & n & 2.45 & $2.94^{+0.96}_{-1.42}$ & 2.99 & 0.71 & 3 \\
G5.90-0.43 & 6.64 & n & 1.91 & $2.94^{+0.96}_{-1.42}$ & 3.11 & 0.66 & 3 \\
G5.91-0.54 & 14.79 & n & 3.41 & $2.94^{+0.96}_{-1.42}$ & 2.46 & 0.36 & 3 \\
G6.13-0.63 & 15.85 & n & 3.49 & $2.94^{+0.96}_{-1.42}$ & 2.94 & 1 & 3 \\
G6.21-0.59 & 18.38 & n & 3.78 & $2.94^{+0.96}_{-1.42}$ & - & - & 3 \\

\hline                                             
\end{tabular}
\end{minipage}
\end{table*}

\subsection{Calculation of errors in kinematic distances}
\label{dist error}
Errors in the distance computation originate from the deviation of the source velocity from the values one calculates assuming circular rotation. This difference is produced by the random velocity dispersion within molecular clouds and in addition systematic streaming motions, when gas flows through spiral arms. We adopt 5 km~s$^{-1}$ given by \cite{1993A&A...275...67B} as average for the random part found in literature. \cite{2007ApJ...671..427M} determined 6.6 km~s$^{-1}$ for the standard deviation of velocity residuals between SGPS data and the \cite{1993A&A...275...67B} rotation curve to estimate the amount of streaming motions in the inner Galaxy. Using the random and systematic contributions to non-circular motions we calculate the error in the radial velocity $\delta$v = $\sqrt{5^2 + 6.6^2}$ km~s$^{-1}$ = 8.28 km~s$^{-1}$. We vary the LSR velocity of each clump by $\delta$v, the distance uncertainty is then the difference between the distance caculated with $\rm v_{\mbox{\tiny LSR}}$ and the distance obtained with $\rm v_{\mbox{\tiny LSR}}\pm 8.28$ km~s$^{-1}$. To determine errors in distances to the groups we take into account that the random velocity dispersion decreases for groups with many sources and vary the mean radial velocity of a group, $<\rm v_{\mbox{\tiny LSR}}>$, by 
\begin{eqnarray}\label{delta v}
\delta \rm v_{\mbox{\tiny group}} = \sqrt{(6.6 \ \rm km~s^{-1})^2+ \left(\sigma_{\mbox{\tiny v}}/\sqrt{N}\right)^2}
\end{eqnarray}
with the velocity dispersion of sources in a group, $\sigma_{\rm v}$, and the number of clumps in a group, $N$. For the calculation of the error we distinguish between ATLASGAL sources, which are located close to the tangent point, and those that are not. In the first quadrant $\sim$ 10\% of the groups ($\sim$ 5\% in the fourth quadrant) have a mean radial velocity slightly smaller than the tangent point velocity, but the velocity used for the error calculation ($<\rm v_{\mbox{\tiny LSR}}>$+$\delta$v$_{\mbox{\tiny group}}$) is larger than the tangent point velocity. Approximately 5\% of the groups in the first and fourth quadrant have mean radial velocities slightly larger than the tangent point velocity, which likely results from streaming motions, while the velocity used to estimate the distance uncertainty ($<\rm v_{\mbox{\tiny LSR}}>-\delta$v$_{\mbox{\tiny group}}$) is still smaller than the tangent point velocity. Approximately 1\% of the groups in the first and fourth quadrant have mean radial velocities and $<\rm v_{\mbox{\tiny LSR}}>-\delta$v$_{\mbox{\tiny group}}$ greater than the tangent point velocity. These ATLASGAL sources, which lie close to the tangent point, have a near and far distance similar to the distance computed with the tangent point velocity. 

Most sources, 85\% in the first quadrant and 89\% in the fourth quadrant, are not located close to the tangent point. The near distance can vary by a large amount for clumps with radial velocities close to 0 km~s$^{-1}$ as a result of streaming motions. Hence, the near distance is unknown for the subsample with velocities between $\sim -6$ km~s$^{-1}$ and $\sim$ 8 km~s$^{-1}$ in the first quadrant and from $\sim -9$ km~s$^{-1}$ to $\sim$ 6 km~s$^{-1}$ in the fourth quadrant.
There are a few sources with extreme velocities: Some clumps in the fourth quadrant with longitude at $\sim 355^{\circ}$ have radial velocities around 100 km~s$^{-1}$. These are located in the surrounding of Bania's Clump 1 \citep{1977ApJ...216..381B}, which have mostly a non-circular velocity and for which we do not determine a distance. \cite{2010MNRAS.404.1029C} consider that they might be in the Galactic bar because of their high radial velocities.

Further estimates of distances to HII regions in the extreme inner Galaxy were made by \cite{2013ApJ...774..117J}. They use HI absorption to probe the longitude-velocity distribution of features within a Galactic longitude range up to 10$^{\circ}$ and give lower limits to the line-of-sight distance by comparing the HI absorption velocity with the radio recombination line velocity of HII regions.

\section{Resolving the Kinematic Distance Ambiguity}
\label{kinematic distance}
Two methods, one using HI self-absorption and the other HI absorption, gives us an indication whether to chose the near or the far distance of ATLASGAL sources. While HISA can be used for all molecular clumps, the HI absorption needs a background HII region to be associated with the ATLASGAL source. We resolve the kinematic distance ambiguity by combining our results from the two techniques, which are described in the next sections. Similar studies have already analysed the two methods to estimate distances to smaller samples or mostly limited to the first quadrant of the Galaxy \citep{2009ApJ...690..706A,2009ApJ...699.1153R}. Our KDA resolutions to a large sample of ATLASGAL sources in the first and fourth quadrant give an unbiased 3D view of massive star forming clumps in the Galaxy. The high number of kinematic distances allows us to derive important parameters such as masses and sizes of molecular clouds in the first and fourth quadrant and to study their statistics.

Based on the groups found in Sect. \ref{grouping} we define complexes on their association with extended $^{12}$CO and $^{13}$CO emission and association with ATLASGAL sources without known velocities.

\subsection{HI self-absorption}
\label{HI self-absorption}
The HI gas within dense molecular clouds is cold with a temperature of about 10 K because it is shielded from the external interstellar radiation field. In contrast, warm atomic hydrogen is distributed throughout the ISM and has a temperature of $\sim$ 100 K \citep{1962ApJ...135..151C}. When emission from the warm HI background at the far distance is absorbed by cold foreground HI at the same velocity, HI self-absorption occurs. This results in an absorption line in the HI 21 cm spectrum at the velocity of the molecular tracers observed towards the ATLASGAL clumps such as NH$_3$, N$_2$H$^+$ or CS. A strong absorption therefore hints at the near kinematic distance. Because the diffuse warm HI gas is abundant in the whole Galactic plane, it is also located in the background of a cloud at the far distance. However, the radial velocity increases with the distance only up to the tangent point and decreases afterwards. The warm HI background is therefore at lower velocities than the clump, which leads to no HI self-absorption toward this source. In addition, foreground radiation is emitted by warm HI gas at the same velocity as the observed clump. If HISA is not seen, the probability of the far kinematic distance is therefore increased.

However, this method is affected by a few uncertainties. It relies on the approximation of a uniform distribution of HI gas, although an absorption might also arise from a lack of HI emission in some specific location. Previous HISA studies \citep[e.g.][]{1979A&AS...35..129B} found criteria, which increase the confidence of the technique, such as the limitation of the Full Width at Half Maximum of the absorption to a few km~s$^{-1}$ ($<$ 5 km~s$^{-1}$) or absorption lines should have an intensity of at least 10 K to separate HISA from background variations. \cite{2009ApJ...690..706A} translated this temperature requirement into a column density of 6 $\times$ 10$^{21}$ H$_2$ molecules needed to have the minimum HI column density (cm$^{-2}$) for HISA. The ATLASGAL clumps fulfill this criterion because for the sources that are observed in NH$_3$, \cite{2012AA...544A.146W} obtain values of the H$_2$ column density between $2.6 \times 10^{21}$ and $1.8 \times 10^{23}$ cm$^{-2}$ with an average of $1.7 \times 10^{22}$ cm$^{-2}$, which lies above this column density limit.

\subsection{HI Self-absorption analysis}
\label{HISA analysis}
To examine if HISA exists toward a complex of ATLASGAL sources, we create maps of average HI intensity at the velocity of the complex and look for HI absorption features toward source positions with known velocities. If absorption is found, this indicates the near kinematic distance of sources within the complex. The absence of any absorption favours the far kinematic distance solution. To increase the HI intensity contrast in case of an absorption, we first average the HI emission in the velocity range of sources in a complex (the ''on'' map), from which we subtract an HI map containing the HI intensity in a velocity interval different from that of the complex (the ''off'' map). To create the HI on map, we average the HI 21 cm line cubes over the velocity range $\rm v \pm \Delta \rm v$ with the mean radial velocity of sources in the complex, $\rm v$, and the mean of their linewidths, $\Delta \rm v$. Two off maps are built by averaging the HI intensity over two times the mean linewidth at a sligthly smaller and larger velocity than the mean radial velocity of the complex, over the velocity intervals $\rm v - \Delta \rm v - 4 \pm \Delta \rm v$ km~s$^{-1}$ and $\rm v + \Delta \rm v + 4 \pm \Delta \rm v$ km~s$^{-1}$. The two maps at the offset velocities are averaged and then subtracted from the on map.

In addition, we extract the HI 21 cm spectrum for source positions in each complex that show a NH$_3$, N$_2$H$^+$ or CS emission peak (on spectrum), and compare it to the HI 21 cm line at an offset position in the same complex, free from emission of the mentioned molecular tracers at the velocity of the complex. One would reveal HISA as an absorption line in the HI on spectrum in contrast to the spectrum at the offset location. However, an absorption might also result from a fluctuation of the HI background at the source velocity, which adds some uncertainty. We therefore use this spectral analysis of HISA as additional information to the HI average intensity map, the HI spectrum is advantageous if it is not clear from the HI map whether it shows an absorption or not. If the KDA resolutions of both disagree, we decide for the distance assigned by the analysis of the HI map.

Panel \textbf{b)} of Fig. \ref{HISA} shows the HI map of the complex plotted in panel \textbf{a)}, it results from the subtraction of the maps at offset velocities from the map at the mean radial velocity of the complex. The clumps with observed velocities around 84 km~s$^{-1}$ are indicated as red triangles and sources without known velocities as green triangles. To define the boundary of a complex we overlaid $^{12}$CO $(1-0)$ emission from \cite{2001ApJ...547..792D} tracing the large-scale structure on the HI map, illustrated as white contour in Fig. \ref{HISA}. The $^{13}$CO (1$-$0) line and the 870 $\mu$m dust continuum are shown as green and blue contours to probe molecular clumps, which are identified by ATLASGAL sources, on a smaller scale. We assign ATLASGAL sources without known velocities, which are traced by contours of $^{12}$CO emission, to the same complex. We give the mean of the observed velocities of sources within a group as determined in Sect. \ref{grouping} to the sources, which have originally no measured velocity. We plot HI spectra of the region shown in panel \textbf{b)} on the right as solid lines, which are extracted at the position of three observed sources. The HI spectrum at one offset position, indicated by the star on the map, is added as dashed line to the HI spectra. The coordinates of the extracted HI lines are given above the spectra. The vertical lines around the source velocity of the HI spectra indicate its variation due to velocity dispersion of clouds and streaming motions. The difference of the on-source and off-source spectra is illustrated below the HI lines. Because absorption is revealed in the map as well as in the spectra at the source velocity, the complex in panel \textbf{b)} is assigned to the near kinematic distance.

In addition, another HI map is plotted for a complex within $30.5^{\circ} < l < 31.3^{\circ}$ and $-0.05^{\circ} < b < 0.45^{\circ}$ in panel \textbf{c)} of Fig. \ref{HISA}. It contains observed ATLASGAL sources with velocities around 40 km~s$^{-1}$ illustrated as red triangles and a large number of ATLASGAL sources without known velocities as green triangles. We assign the mean velocity of the complex of 39.81 km~s$^{-1}$ (see Table \ref{complex-vmean}) to the sources with no measured velocity. The HI map of this complex reveals HI emission and no absorption is visible in the HI spectra as well, which hints at the far kinematic distance.

\subsection{HI absorption}
\label{continuum absorption}
Another technique to resolve the kinematic distance ambiguity is the analysis of absorption toward sources with strong radio continuum emission, e.g. HII regions. Their thermal free-free continuum emission at 21 cm, which has a higher brightness temperature than that of neutral HI gas in clouds, is absorbed by HI in molecular clouds located between the HII region and the observer. 
The absolute value of the radial velocity increases with the distance from the Sun up to the tangent point velocity and decreases at larger distances. Foreground clouds, which absorb the 21 cm continuum of an HII region at the near kinematic distance, thus have HI absorption lines at velocities up to that of the HII region, which is known from observations of NH$_3$, N$_2$H$^+$ or CS used in this article (see Sect. \ref{molecular data}) in the same molecular cloud. For a 21 cm continuum source located at the far kinematic distance, HI absorption at velocities up to the tangent point velocity is revealed.

\cite{1972A&A....19..354W} analysed first H$_2$CO absorption against 21 cm continuum emission as a tool to resolve the KDA for 28 HII regions. However, H$_2$CO is not as broadly distributed as HI and probes rather molecular than atomic gas. Since the inner Galaxy is filled by a much larger amount with atomic than molecular gas, there might be not enough H$_2$CO in foreground molecular clouds leading to its non-detection near the tangent point. The absorption spectrum toward an HII region may consequently not exhibit high velocities and the near distance might be assumed, although the object is located at the far distance. Since the HI abundance is high in molecular clouds, KDA solutions using this probe are more reliable than the HISA method \citep{2009ApJ...690..706A}.

\captionsetup[subfigure]{position=top}
\begin{figure*}[tbp]
\centering

\subfloat[\hspace*{15cm}]{\includegraphics[angle=-90,width=8.0cm]{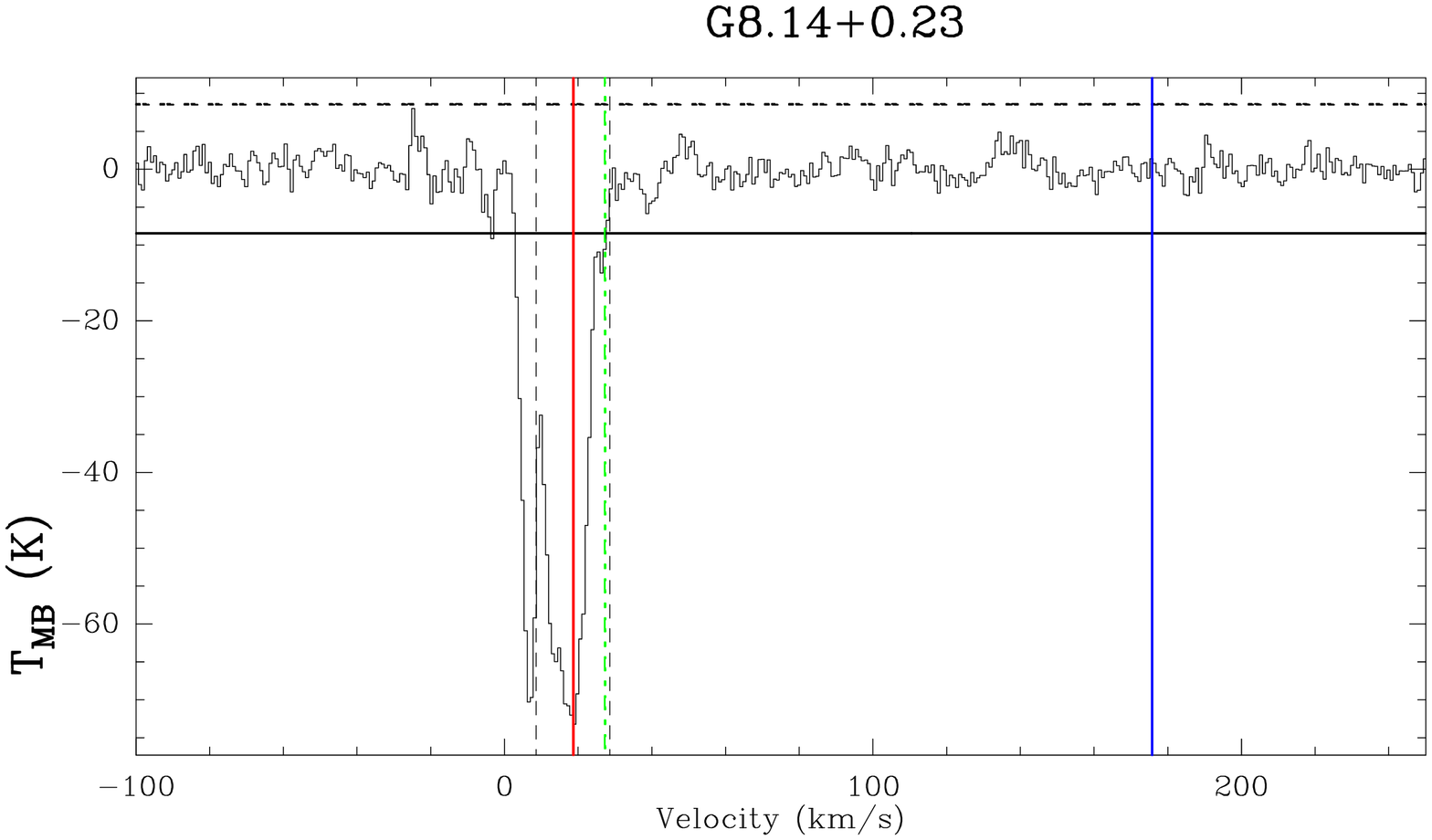}\hspace*{0.5cm}

\includegraphics[angle=-90,width=8.0cm]{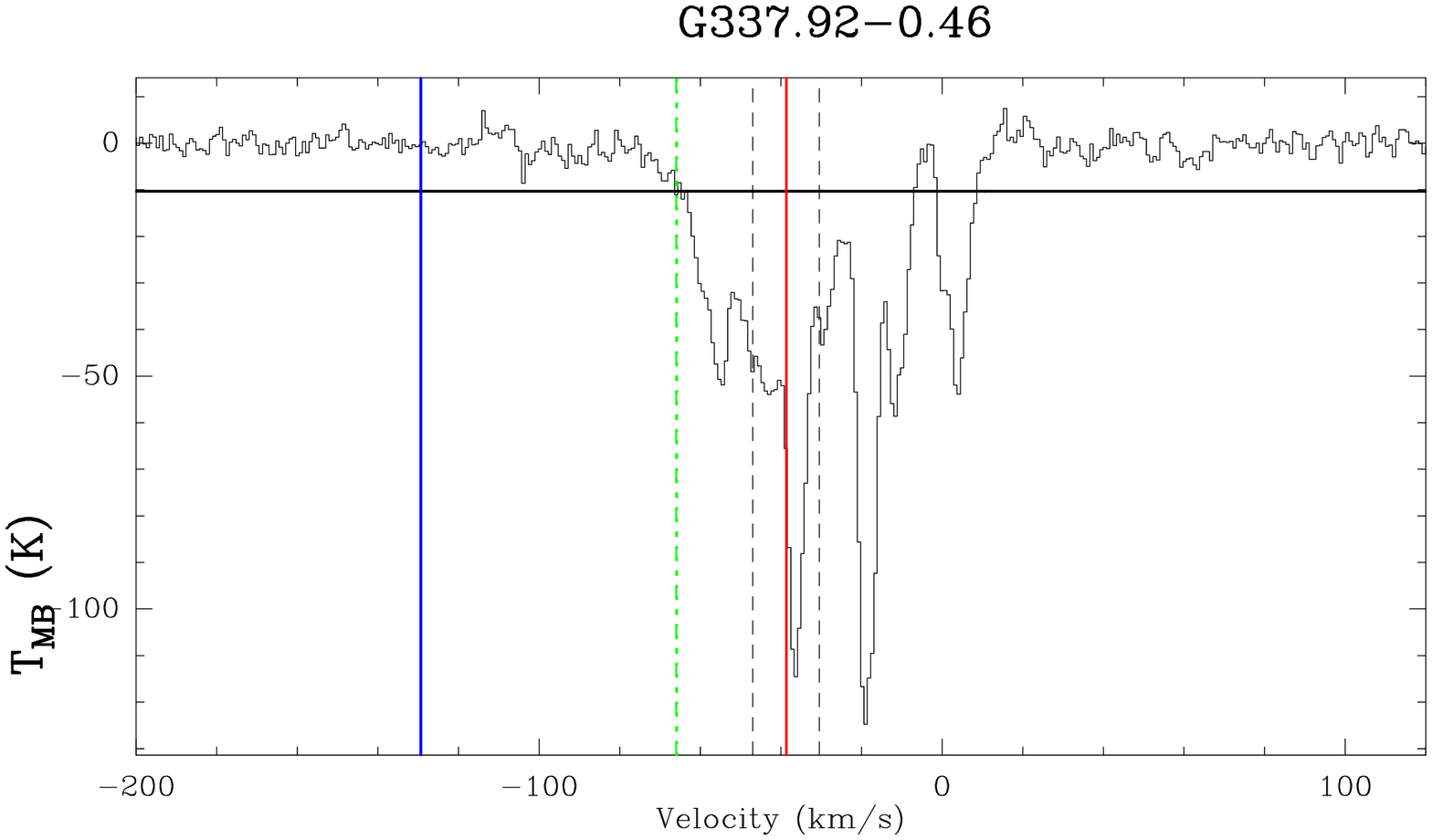}
}

\subfloat[\hspace*{15cm}]{\includegraphics[angle=-90,width=8.0cm]{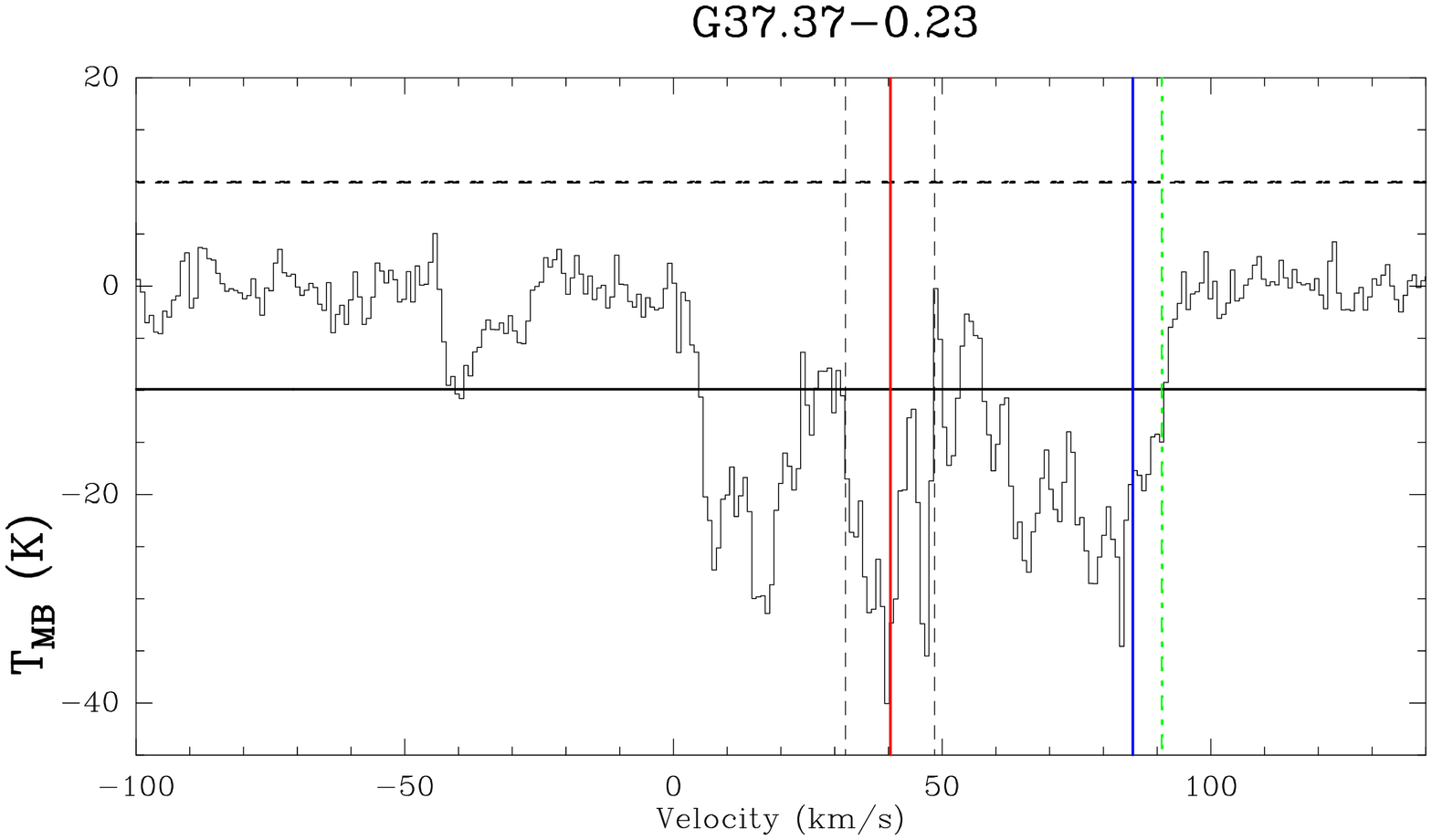}\hspace*{0.5cm}

\includegraphics[angle=-90,width=8.0cm]{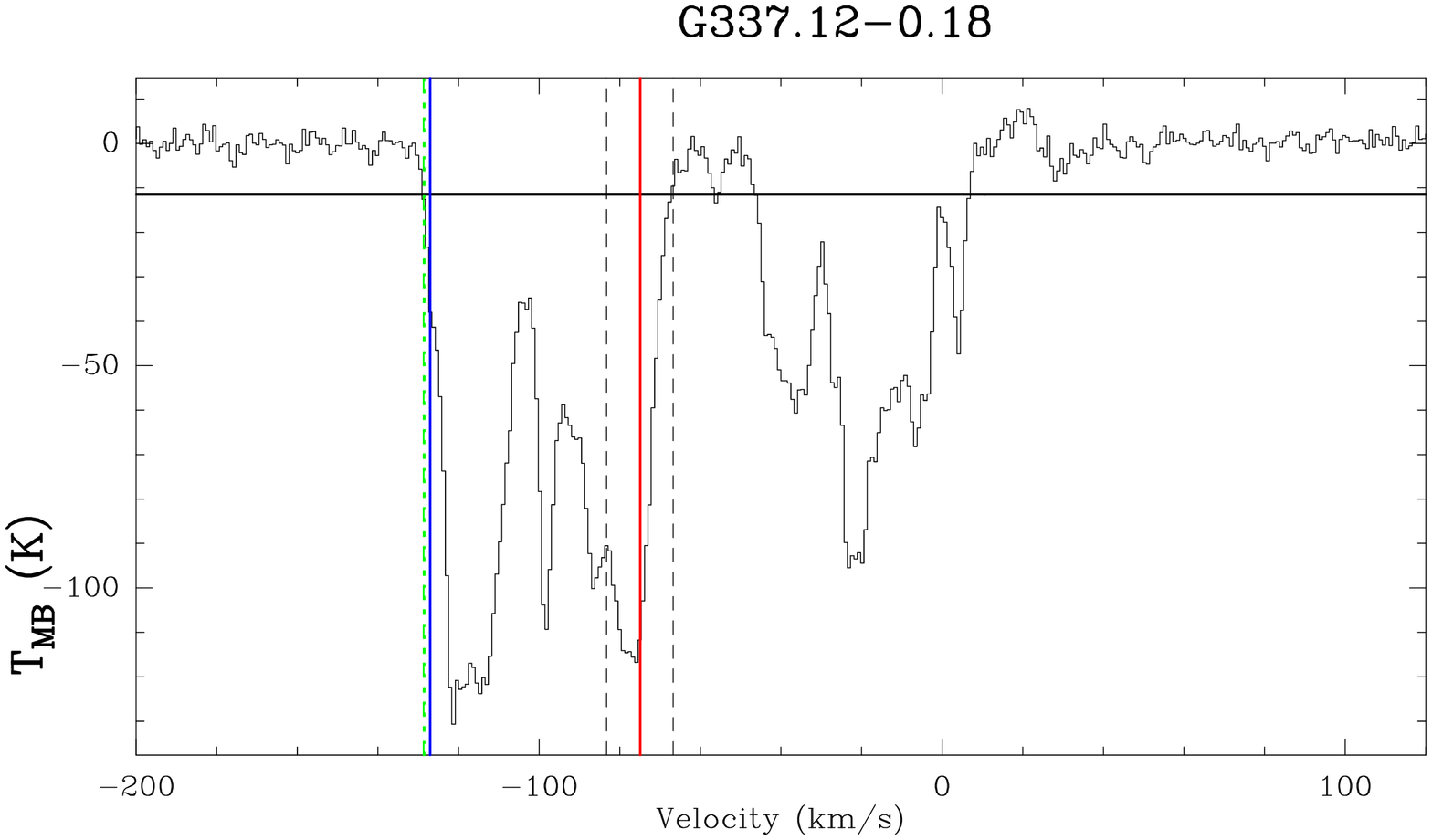}
}

\subfloat[\hspace*{15cm}]{\includegraphics[angle=-90,width=8.0cm]{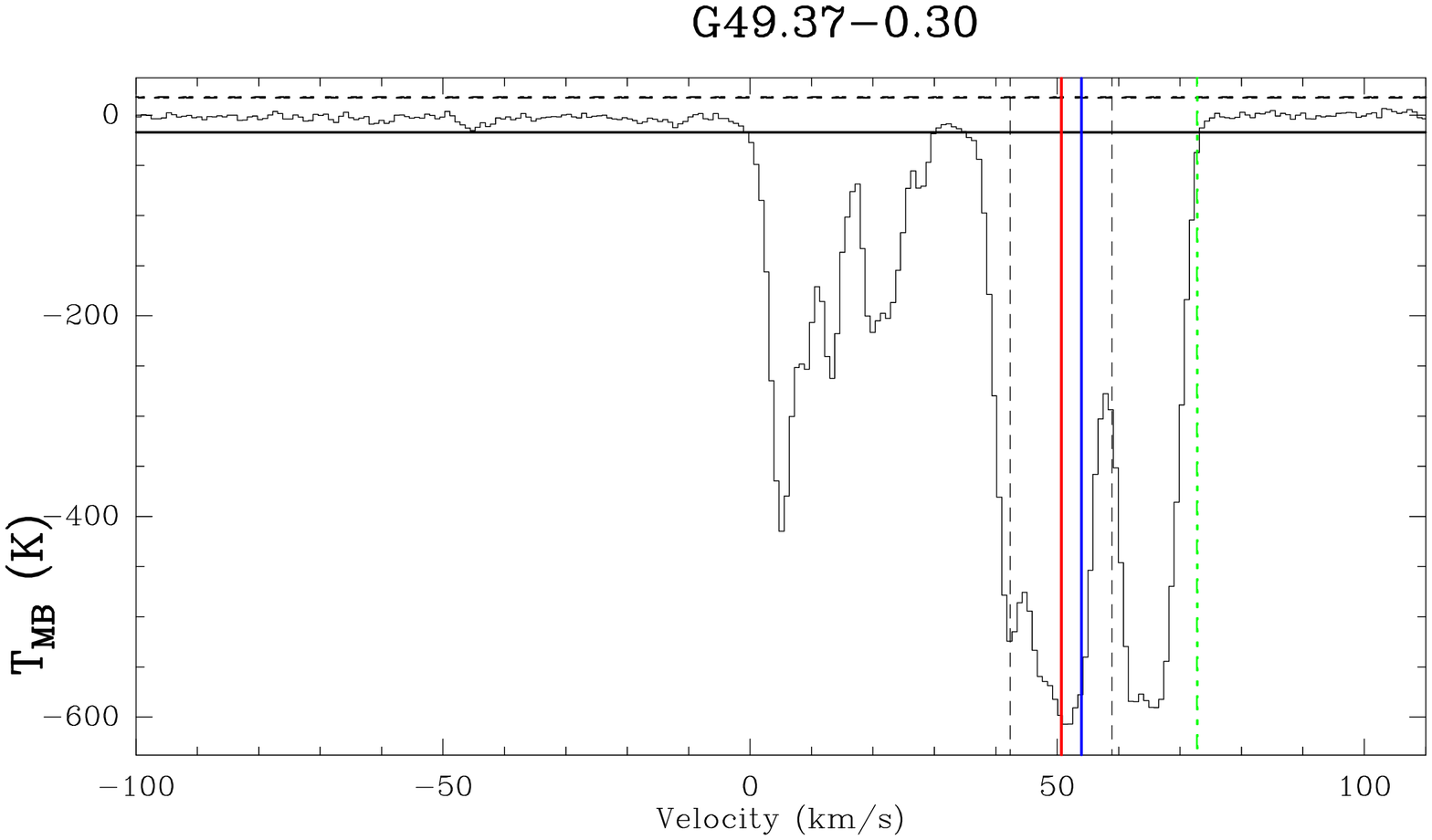}\hspace*{0.5cm}

\includegraphics[angle=-90,width=8.0cm]{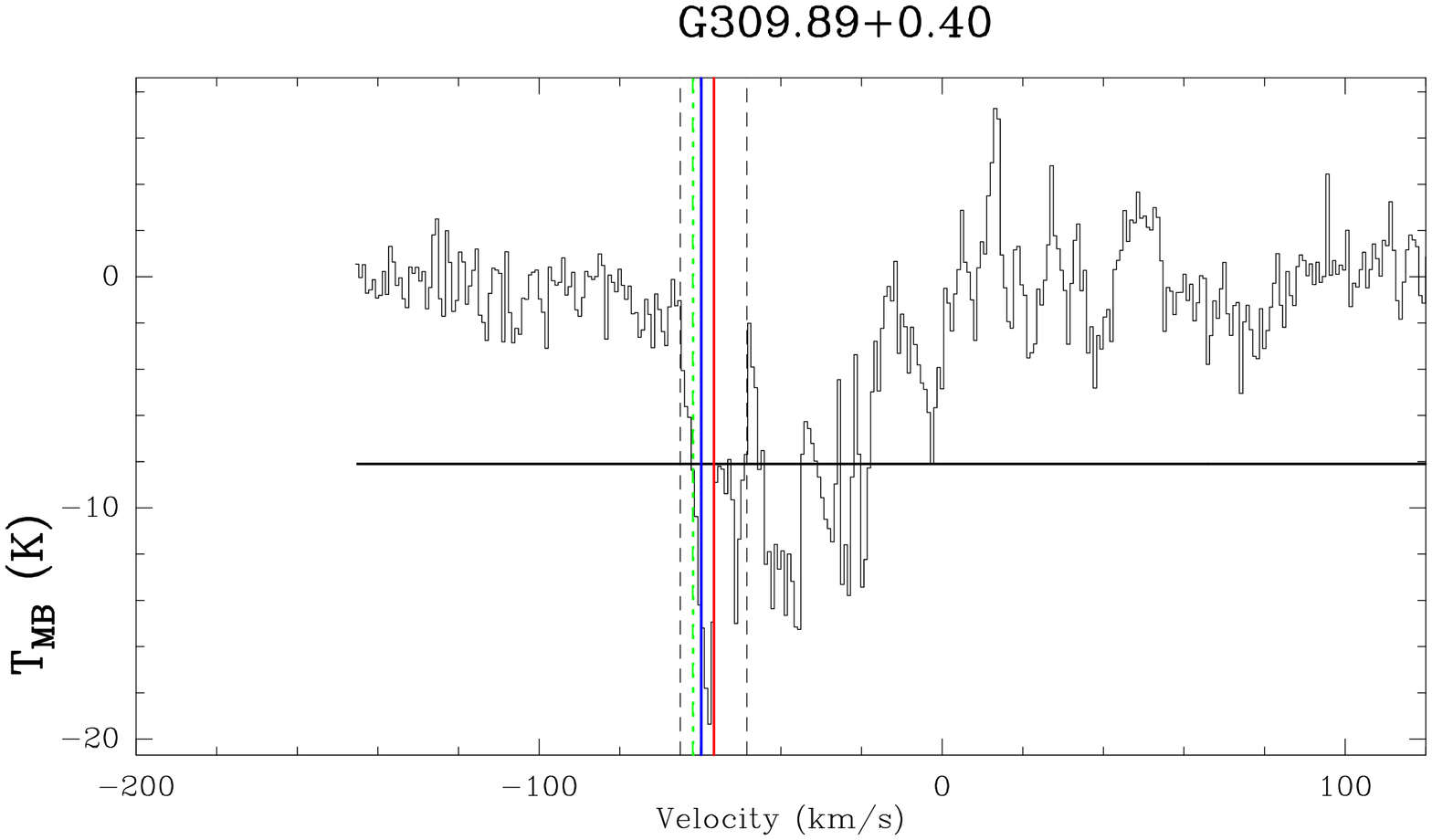}
}

\caption[HI absorption]{Example of HI spectra towards HII regions in the first and fourth quadrant extracted from 21 cm line cubes from the VGPS and SGPS. The source velocity is illustrated as solid red vertical line, the tangent point velocity as solid blue vertical line, and the velocity, at which the absorption is below 5$\sigma$, as dashed-dotted green vertical line. This limit of $-5 \sigma$ is indicated by the solid black horizontal line. In addition, the maximum positive intensity of HII regions in the first quadrant must be lower than 5$\sigma$, which is shown as dashed black horizontal line. The dashed black vertical lines $\pm$8.28 km~s$^{-1}$ around the source velocity display its variation because of the velocity dispersion of clouds and streaming motions (see Sect. \ref{continuum absorption method}). Examples of sources in the first quadrant are shown on the left and of clumps in the fourth quadrant on the right. \textbf{a)} There is a large difference between the tangent point velocity and the velocity, up to which HI absorption is seen and which is close to the source velocity for G8.14$+$0.23 and G337.92$-$0.46, which are thus located at the near distance. They are located in the diagonal right region in Fig. \ref{kolpac-ns} (red points) confined by the amount of streaming motions. \textbf{b)} Our analysis reveals HI absorption at velocities close to the tangent point velocity for G37.37$-$0.23 and G337.12$-$0.18, which are assigned to the far distance and indicated as blue points in the horizontal right part in Fig. \ref{kolpac-ns}. \textbf{c)} We locate G49.37$-$0.30 and G309.89$+$0.40 at the tangent point because the source velocity is within 8.28 km~s$^{-1}$ around the tangent point velocity. They are displayed as purple points in Fig. \ref{kolpac-ns}.}\label{21cm continuum}
\end{figure*}

\subsection{HI absorption method}
\label{continuum absorption method}
To utilize this technique we use HI 21 cm line cubes from the VGPS and SGPS data sets. To recognize if an absorption dip in the HI spectrum is produced by HI absorption toward an HII region we calculate the difference between the HI spectrum extracted at the position of the source and in a region around it. Assuming that the HI emission is relatively smoothly distributed and extended we subtract the HI line emission and the resulting spectrum reveals absorption lines in case of an HII region. To derive the HI spectrum in an envelope around the source we average the emission within 120$\arcsec$. We also tried out a larger size of 420$\arcsec$ to extract more HI emission inside a larger region around the source. However, this has a slightly different shape than the emission coming from the source itself producing HI intensity fluctuations. These might be considered as absorption dips with velocities exceeding the source velocity and thus indicating the far distance instead of the near distance. To avoid those HI intensity fluctuations resulting from the emission in the larger region around the source, we decided to use the small envelope. Examination of the HI spectra reveals that we can only use them to resolve the KDA if they satisfy the following criteria:
\begin{enumerate}
 \item We estimate the noise level $\sigma$ from part of the spectrum free from absorption and use only spectra with absorption features over 5$\sigma$. Moreover, different shapes of HI lines at the source position and in the surrounding result in intensity fluctuations around 0, when the two spectra are subtracted. A dip in the HI spectrum must fall below $-$1.5 times the maximum positive intensity to be recognized as absorption line. The fluctuations should not exceed 5$\sigma$ for sources in the first quadrant. This is not set in the fourth quadrant because inspection of HI spectra for a sample of known HII regions shows that too many sources would not be considered as HII regions.
 \item To assure that the source is indeed associated with a strong HII region the continuum flux is extracted from 21 cm continuum emission maps. A source is identified as HII region if the mean continuum level is higher than 3 to 5$\sigma$, where $\sigma$ is the continuum noise ($\sim 100$ mJy/beam for the sample in the first quadrant and $\sim 600$ mJy/beam for the sources in the fourth quadrant). The different flux limits result from different sensitivities and beam sizes of the VGPS and SGPS described in Sect. \ref{archival data}. 
 \item There must be an HI absorption line at the velocity of the molecular cloud, which harbours the ATLASGAL clump consisting of cold gas observed in NH$_3$, N$_2$H$^+$, or CS. This assures that the HII region is also embedded in the molecular cloud.
\end{enumerate}
We obtain 65 HII regions in the first quadrant and 146 HII regions in the fourth quadrant, for which we can distinguish between near and far distances using the method from \cite{2003ApJ...582..756K}. They determined the source velocity ($\rm v_{\rm s}$), the velocity of the tangent point ($\rm v_{\rm t}$), and the maximum velocity of absorption lines in the 21 cm continuum ($\rm v_{\rm a}$) of a sample of HII regions located in the first quadrant. Their comparison of $\rm v_{\rm t} - \rm v_{\rm s}$ and $\rm v_{\rm t} - \rm v_{\rm a}$ showed that the sources are mostly located in two regions, where $\rm v_{\rm t} - \rm v_{\rm a}$ is around 0 and where $\rm v_{\rm t} - \rm v_{\rm a}$ is increasing with rising $\rm v_{\rm t} - \rm v_{\rm s}$. To assign kinematic distances to those regions, \cite{2003ApJ...582..756K} ran simulations with 10000 HII regions to analyse the velocity differences, which yields that sources with $\rm v_{\rm t} - \rm v_{\rm a}$ close to 0 are located at the far kinematic distance, while those with growing $\rm v_{\rm t} - \rm v_{\rm a}$ and $\rm v_{\rm t} - \rm v_{\rm s}$ are at the near distance. 

We obtain the source velocity of the HII regions from NH$_3$, N$_2$H$^+$, and CS emission associated with the dust clumps (see Sect. \ref{molecular data}). The tangent point velocity is calculated using the rotation curve by \cite{1993A&A...275...67B} for the longitude range from $-20^{\circ}$ to $60^{\circ}$, while it is computed in the fourth quadrant with the linear function by \cite{2007ApJ...671..427M}, who fitted HI terminal velocities as a function of Galactic longitude from 300$^{\circ}$ to $\sim 339^{\circ}$. Figure \ref{21cm continuum} shows HI absorption line spectra of ATLASGAL sources in the first and fourth quadrant placed at the near and far kinematic distance and at the tangent point. The tangent point velocity is indicated as solid blue vertical line and the source velocity as solid red vertical line. We determine the noise level of the spectra, $\sigma$, and give the limit of 5$\sigma$, which should not be exceeded by the maximum positive intensity of the sources in the first quadrant, as dashed black horizontal line. We measure the maximum velocity in the first quadrant at which the absorption is below 5$\sigma$, which is illustrated by the solid black horizontal line, and lower than $-1.5$ times the maximum positive intensity. The same is considered in the fourth quadrant, where the absorption velocity is the minimum velocity, which is indicated by the dashed-dotted green vertical line in Fig. \ref{21cm continuum}. The velocity dispersion of clouds and streaming motions, which result in differences of the source velocity from circular rotation, are shown as dashed black vertical lines $\pm$8.28 km~s$^{-1}$ around the source velocity \citep{1971A&A....10...76B,2007ApJ...671..427M}. Seventeen sources of our sample in the first quadrant and seven clumps in the fourth quadrant have a source velocity within 8.28 km~s$^{-1}$ of the tangent point velocity or lie beyond the tangent point, for which we assign the distance of the tangent.

To distinguish between near and far distance we plot the velocity differences $\rm v_{\rm t} - \rm v_{\rm a}$ against $\rm v_{\rm t} - \rm v_{\rm s}$ for our subsamples of HII regions in the first and fourth quadrant in Fig. \ref{kolpac-ns} as shown in \cite{2003ApJ...582..756K} for simulated HII regions. They studied the quality of their KDA resolutions, which shows that 90\% of sources in the diagonal right part bounded by solid red lines in Fig. \ref{kolpac-ns} are at the near distance and 90\% of HII regions in the horizontal right region, confined by solid red lines as well, are at the far distance. Those sources are assigned to distances with a high degree of confidence. The dashed red lines around them indicate additional space taking into account streaming motions of 8.28 km~s$^{-1}$. According to \cite{2003ApJ...582..756K} the triangle at the lower left part of the plot contains sources, for which the resolution of the kinematic distance ambiguity is more uncertain. The dotted line in the triangle separates 21 cm continuum sources at the near distance in the upper region from those at the far distance below the dotted line with a lower degree of confidence. The velocity differences of identified HII regions in the first quadrant are shown as black points in Fig. \ref{kolpac-ns} and those of HII regions in the fourth quadrant in green, for which we show the inverted velocity differences to be able to present the two samples in one plot. Six sources in the first quadrant and 34 sources in the fourth quadrant lie between the horizontal and diagonal regions. In the sample from \cite{2012MNRAS.420.1656U} 10 HII regions are also in this location in the same plot of velocity differences. They took their distance estimate from the literature, mostly from spectrophotometric measurements and H$_2$CO absorption, which suggests the near distance for these sources. As a consequence we adopt their KDA resolution and place the sources from our sample between the horizontal and diagonal regions in Fig. \ref{kolpac-ns} at the near distance.

\begin{figure}
\centering
\includegraphics[angle=0,width=9.0cm]{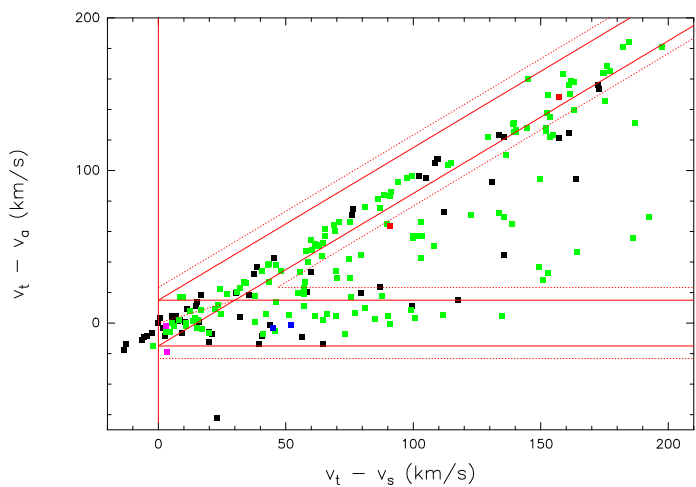}
\caption[HI absorption]{Comparison of the differences between the tangent point velocity ($\rm v_{\rm t}$), the maximum velocity of absorption lines in the 21 cm continuum ($\rm v_{\rm a}$), and the source velocity ($\rm v_{\rm s}$) of HII regions. The subsample in the first quadrant is shown in black and sources in the fourth quadrant in green. To be able to display the two in one plot, we invert the velocity differences of HII regions in the fourth quadrant. The solid and dotted lines confine regions, where 21 cm continuum sources are assigned near and far distances using the method from \cite{2003ApJ...582..756K}. The examples of ATLASGAL sources shown in Fig. \ref{21cm continuum} are indicated: Clumps at the near distance are shown in red, sources at the far distance in blue, and clumps at the tangent point in purple.}\label{kolpac-ns}
\end{figure}

\subsection{Methods to resolve the KDA}
\label{KDA methods}
To obtain distances we first apply the two methods, HISA and HI absorption, separately and then combine the results. This is illustrated as flow chart in Fig. \ref{flow-chart}. We summarize here the most important steps.

Before we look for HISA we divide the ATLASGAL sample into groups of sources, which are located close together and have similar velocities (see Sect. \ref{grouping}). Then, we derive maps of average HI intensity to investigate HISA toward the groups. These are overlaid with molecular line emission as contours tracing small- and large-scale cloud structure, which allows us to assign ATLASGAL sources without known velocities to the molecular cloud complexes and thus to determine distances to them as well. In addition, we look for absorption lines in the HI spectrum at source positions with known velocities. This is the first step of the flow chart in Fig. \ref{flow-chart} together with the investigation of absorption lines in the 21 cm continuum extracted as described in Sect. \ref{continuum absorption method}. As next step in Fig. \ref{flow-chart} we use the method from \cite{2003ApJ...582..756K} for sources showing an absorption feature, explained in Sect. \ref{continuum absorption method}, to distinguish between near and far distances. For clumps not associated with HI absorption we use only HISA.

To combine the two methods we distinguish between complexes that contain one or several continuum sources. The third step in the flow chart shows the case of only one continuum source in a complex. If the KDA resolutions from HISA and HI absorption agree, we will use the assigned distance, indicated as dashed red arrows in Fig. \ref{flow-chart} for the near distance and solid green arrows for the far distance. If the far distance is determined from HISA and the near distance from HI absorption as shown by the solid black arrows, the HI intensity map might not show self-absorption because star formation can be ongoing in the cloud and emission of protostars heats their environment, which can be detected at 8 $\mu$m by the GLIMPSE survey. To identify these cases we look for 8 $\mu$m emission in the surroundings of our source and will assign the near distance if we find GLIMPSE counterparts for sources with a high S/N ratio of the absorption line in the 21 cm continuum. If there are no associations with GLIMPSE sources, the far distance is chosen. For the near distance determined from HISA and the far distance from HI absorption as shown by the dashed blue arrows, we use the far distance because the self-absorption might not result from absorption of the HI background, but from the absorption of the HII region embedded in our source. 

For complexes that have more than one 21 cm continuum source one can distinguish between near and far distance if the KDA resolutions from absorption features in the 21 cm continuum agree. If they do not agree and e.g. two HI spectra give a near and far distance from HI absorption, while HISA determines the far distance, one has to examine the S/N ratio of the HI spectra. If one of those is favoured because of a higher S/N ratio, one can resolve the KDA, otherwise no distance is assigned.

\begin{figure*}
\centering
\includegraphics[angle=0,width=9.0cm]{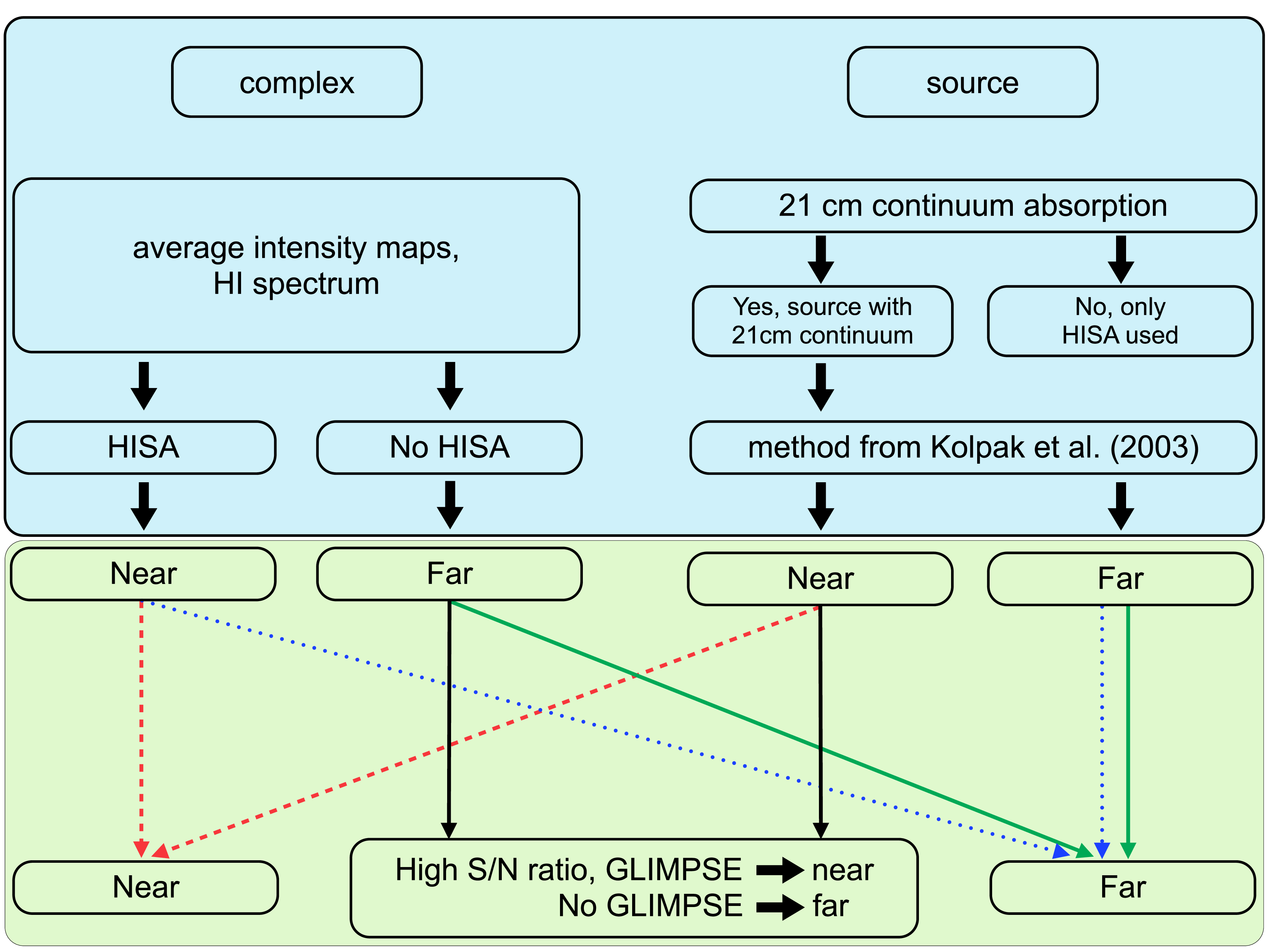}
\caption[histogram of the size of a complex]{Flow chart illustrating HI self-absorption and HI absorption to resolve the KDA. The upper panel shows the main points of the two methods and the lower panel combines their KDA resolutions.}\label{flow-chart}
\end{figure*}

\subsection{Resolved kinematic distances}
\label{resolved distances}
As described in Sect. \ref{KDA methods} we first arrange our sample with known velocities of 749 sources in the first quadrant and 1065 clumps in the fourth quadrant into groups that are coherent in space and velocity (see Table \ref{molecular}). Comparison of these complexes with line and dust continuum emission reveals molecular cloud structure, which is important to associate ATLASGAL sources without known velocities with the complexes as well. Using HISA we distinguish between near and far distances for the whole ATLASGAL sample including 1489 sources in the first quadrant and 2069 clumps in the fourth quadrant with and without known velocities. We arrange 165 ATLASGAL sources in the first quadrant and 184 clumps in the fourth quadrant in more than one complex and without velocity information we cannot determine distances to them.

For sources identified as HII regions with strong continuum emission we obtain additional indication for near and far distances from our analysis of absorption in the 21 cm continuum following the work by \cite{2003ApJ...582..756K}. While we locate all sources with $|\rm v_{\rm t} - \rm v_{\rm s}| < 8.28$ km~s$^{-1}$ at the tangent point taking non-circular motions into account (see Sect. \ref{continuum absorption method}), \cite{2003ApJ...582..756K} place only HII regions with $\rm v_{\rm t} - \rm v_{\rm s} < 0$ at the tangent point. Based on their criterium we find 36 sources with a 21 cm continuum in the first quadrant from our sample with known velocities at the near distance, 23 at the far distance, and 6 at the tangent point, 107 HII regions in the fourth quadrant are at the near distance, 38 at the far distance, and 1 at the tangent point. Considering non-circular motions the number of sources at the tangent point is increased to 17 in the first quadrant, where we obtain 31 near and 17 far HII regions. We assign a near distance to 110 sources with a 21 cm continuum in the fourth quadrant, a far distance to 29, and 7 are placed at the tangent point.

A comparison of near and far distances to HII regions determined from HISA as well as HI absorption using the method from \cite{2003ApJ...582..756K} yields an agreement in the results of the two methods of 58\% in the first quadrant and 71\% in the fourth quadrant. The relatively low agreement is expected because HISA gives better results for the near distance and HI absorption for the far distance (see Sect. \ref{KDA methods}).

We combine our KDA resolutions obtained from HISA and HI absorption as described in Sect. \ref{KDA methods}. Sources with velocities within 8.28 km~s$^{-1}$ of the tangent point velocity are placed at the tangent point. The distance assignments of clumps with measured velocities in the first and fourth quadrant are summarized in Table \ref{kda}. These ATLASGAL sources are divided into complexes, which we locate at the tangent point if the difference of the mean radial velocity and the tangent point velocity is smaller than the amount of non-circular motions estimated by equation \ref{delta v}. The resulting distances to complexes are also shown in Table \ref{kda}. A subsample of 27 ATLASGAL sources with resolved kinematic distances has also been observed by the BeSSeL survey \citep{2014ApJ...783..130R} to derive trigonometric parallax distances. These are compared with the kinematic distances in Appendix \ref{comp}.

Using flux limited ATLASGAL subsamples in the first and fourth quadrant, which goes down to a peak brightness of 1.5 Jy/beam, we obtain a ratio of near-to-far kinematic distances of $1.5 \pm 0.21$ in the first quadrant and of $2.4 \pm 0.35$ in the fourth quadrant. The distributions in the first and fourth quadrant agree within a $3\sigma$ error. This trend is similar to the result from \cite{2009ApJ...699.1153R}, who obtain the near distance to 2.3 times more molecular clouds revealed within the GRS survey than the far distances. The KDA resolution for samples of HII regions from \cite{2012MNRAS.420.1656U} and \cite{2009ApJ...690..706A} yields a larger number of sources located at the far distance than at the near distance. Since the submm continuum and $^{13}$CO emission are tracing all molecular clumps, our source sample as well as that from \cite{2009ApJ...699.1153R} contain a larger number of nearby, low mass clumps. In contrast, \cite{2012MNRAS.420.1656U} and \cite{2009ApJ...690..706A} are probing preferably massive star forming clumps that are typically found at larger distances.

Recently, \cite{2013ApJ...770...39E} determined a distance probability density function (DPDF) from a kinematic distance as the likelihood and prior probabilities related to ancillary data sets to resolve the KDA to BGPS clumps. This work is expanded by \cite{2014arXiv1411.2591E}, who introduce two new prior DBDFs relating BGPS clumps to objects with trigonometric parallax distances from the BeSSeL survey \citep{2009ApJ...700..137R} and to HII regions with resolved kinematic distances using HI absorption methods from the HII Region Discovery Surveys \citep[HRDS,][]{2010ApJ...718L.106B,2012ApJ...759...96B}. A comparison of the kinematic distances to ATLASGAL sources with the distances presented in \cite{2014arXiv1411.2591E} reveals that the agreement is $\sim 70$\% within 2 kpc. The remaining fraction of inconsistent distances do not affect our statistical analysis because distance errors average out for our large ATLASGAL sample.

\begin{table*}
\caption[]{Summary of the KDA resolution of ATLASGAL sources and complexes.}
\label{kda}
\centering
\begin{tabular}{l c c c c c c c c}
\hline\hline
 &  number of sources at &  & tangent & &  number of complexes at & &  tangent & \\
Sample & near distance & far distance &  point & Total & near distance & far distance & point & Total \\ \hline 
first quadrant  &   383   &  332  &  34 &  749  & 143 & 148  & 5 & 296 \\
fourth quadrant   &   737   &  290 & 38  & 1065 & 234  & 141 & 18  & 393 \\\hline 
\end{tabular}
\end{table*}

\section{Results}
\label{results}
With the resolution of the kinematic distance ambiguity of a large sample of 3558 ATLASGAL sources we study their Galactic distribution and analyse the large-scale structure of the Milky Way. In addition, we use the distances to calculate important physical properties such as masses and sizes.

\subsection{Galactocentric distribution of ATLASGAL sources in the Milky Way}
\label{galradius distribution}
We show the galactocentric distribution of the ATLASGAL clumps in the first and fourth quadrant in the upper and lower panel of Fig. \ref{galactocentric radius}. 

\subsubsection{Galactocentric distribution in the first quadrant}
The top panel indicates two peaks at a galactocentric radius, R$_{\mbox{\tiny Gal}}$, between 4 and 5 kpc and at $\sim 6$ kpc. A comparison of these peaks with the large-scale structure of the Milky Way using a model with four spiral arms \citep[e.g.][]{1976A&A....49...57G} reveals that the enhancement of sources at R$_{\mbox{\tiny Gal}} = 4.5$ kpc can be attributed to the intersection of the Scutum-Centaurus arm and the Galactic bar. The largest fraction of sources contributing to that peak, 19\%, are assigned to complex 172 ($l = 30.767^{\circ}, b = -0.050^{\circ}$), which contains the star forming region W43 (see Appendix \ref{gmc}). The enhancement at 6 kpc coincides with the galactocentric radius of the Sagittarius arm. Most sources from that peak are assigned to two complexes, $\sim 27$\% are located in complex 50 ($l = 14.833^{\circ}, b = -0.602^{\circ}$), the high mass star forming region M16, and 19\% are in complex 261 ($l = 49.2^{\circ}, b = -0.7^{\circ}$), which is the massive giant molecular cloud W51 (see Appendix \ref{gmc}).

The two peaks in the Galactocentric distribution of the first quadrant are also detected in HII regions by \cite{2009ApJ...690..706A}, in Bolocam Galactic Plane survey (BGPS) sources observed in the 1.1 mm dust continuum \citep{2011ApJ...741..110D}, and in a sample of $6.7-$GHz methanol masers by \cite{2011MNRAS.417.2500G}. However, an additional peak at a galactocentric radius of $\sim 8$ kpc is also obtained by previous studies, e.g. in a sample of RMS sources by \cite{2011MNRAS.410.1237U}, in IRDCs by \cite{2008ApJ...680..349J}, in high mass star forming regions with FIR colours of UCHIIRs \citep{2000A&A...358..521B}, and in molecular clouds observed in $^{13}$CO by the GRS survey \citep{2009ApJS..182..131R}. This peak results from the building up of sources on the solar circle, which are primarily nearby. \cite{2004MNRAS.347..237P} mention that the high concentration of HII regions detected at R$_{\mbox{\tiny Gal}} = 8$ kpc is removed in the galactocentric distribution of sources with LSR velocities larger than $\arrowvert 10 \arrowvert$ km~s$^{-1}$. The comparison of the number distribution of RMS sources with radial velocities greater than $\arrowvert 12 \arrowvert$ km~s$^{-1}$ \citep{2011MNRAS.410.1237U} to that of the whole sample also reveals a lack of the peak at $\sim 8$ kpc. The small number of ATLASGAL sources in the first quadrant at this galactocentric radius might consequently result from a few observations of clumps with small velocities.\\

\subsubsection{Galactocentric distribution in the fourth quadrant}
The histogram including the whole ATLASGAL sample in the fourth quadrant reveals three peaks at galactocentric radii of $\sim 3$ kpc, between 5 and 6 kpc and at approximately 7 kpc. The enhancement of sources at 3 kpc can be attributed to the near 3-kpc arm and the southern end of the Galactic bar. Most clumps at 3 kpc, $\sim 20$\%, are in complex 363 ($l = 351.653^{\circ}, b = 0.176^{\circ}$), where IRAS 17200$-$3550 is located. It was observed at 5 GHz with the VLA by \cite{1994ApJS...91..347B} and is likely a UCHIIR. The enhancement between 5 and 6 kpc corresponds to the Norma arm and Scutum-Centaurus arm. A large subsample, 31\% of the number of clumps between 5 and 6 kpc, clustering at the Scutum-Centaurus arm is in complex 210 ($l = 333.19^{\circ}, b = -0.36^{\circ}$), which contains the G333 giant molecular clouds (see Appendix \ref{gmc}). The peak at $\sim 7$ kpc can be attributed to the Scutum-Centaurus arm as well. Most sources, 42\%, are assigned to complex 377 ($l = 353.115^{\circ},b = 0.720^{\circ}$), which harbours the HII region G353.186+0.8 located at the near kinematic distance from HI absorption by \cite{2006ApJ...653.1226Q}. This is in agreement with our KDA resolution using HI self-absorption and HI absorption. Another large contribution to the number of clumps, 32\%, originates from complex 356 ($l = 351.317^{\circ}, b = 0.661^{\circ}$), which is the NGC 6334 molecular cloud complex. Some of the peaks in the galactocentric distribution of the ATLASGAL clumps in the fourth quadrant are similar to features obtained by previous studies. A sample of $6.7-$GHz methanol masers in the fourth quadrant \citep{2011MNRAS.417.2500G} also revealed the two peaks at $3-3.5$ kpc and $6-6.5$ kpc.

\subsubsection{Differences in the Galactocentric distribution between the subsamples in the first and fourth quadrant}
The RMS survey did not observe the region within $\arrowvert l \arrowvert < 10^{\circ}$ and is thus not sensitive to sources within 4 kpc. Consequently, \cite{2012MNRAS.420.1656U} obtain only one strong peak between 5.5 and 6 kpc for a sample of compact HII regions from the RMS survey in the fourth quadrant in contrast to three peaks for their sample in the first quadrant \citep{2011MNRAS.410.1237U}. This led them to deduce that HII regions are not located symmetrically in the first and fourth quadrant. Taking the ATLASGAL sample in the fourth quadrant into account we detect three peaks, while there are two concentrations of sources in the first quadrant at $\sim 4$ and 6 kpc. The comparison of the histograms for the subsamples in the first and fourth quadrant therefore indicates a difference in the overall galactocentric distributions as well, which is consistent with spiral arm structure rather than a molecular ring.

\begin{figure}[h]
\centering
\includegraphics[angle=-90,width=9.0cm]{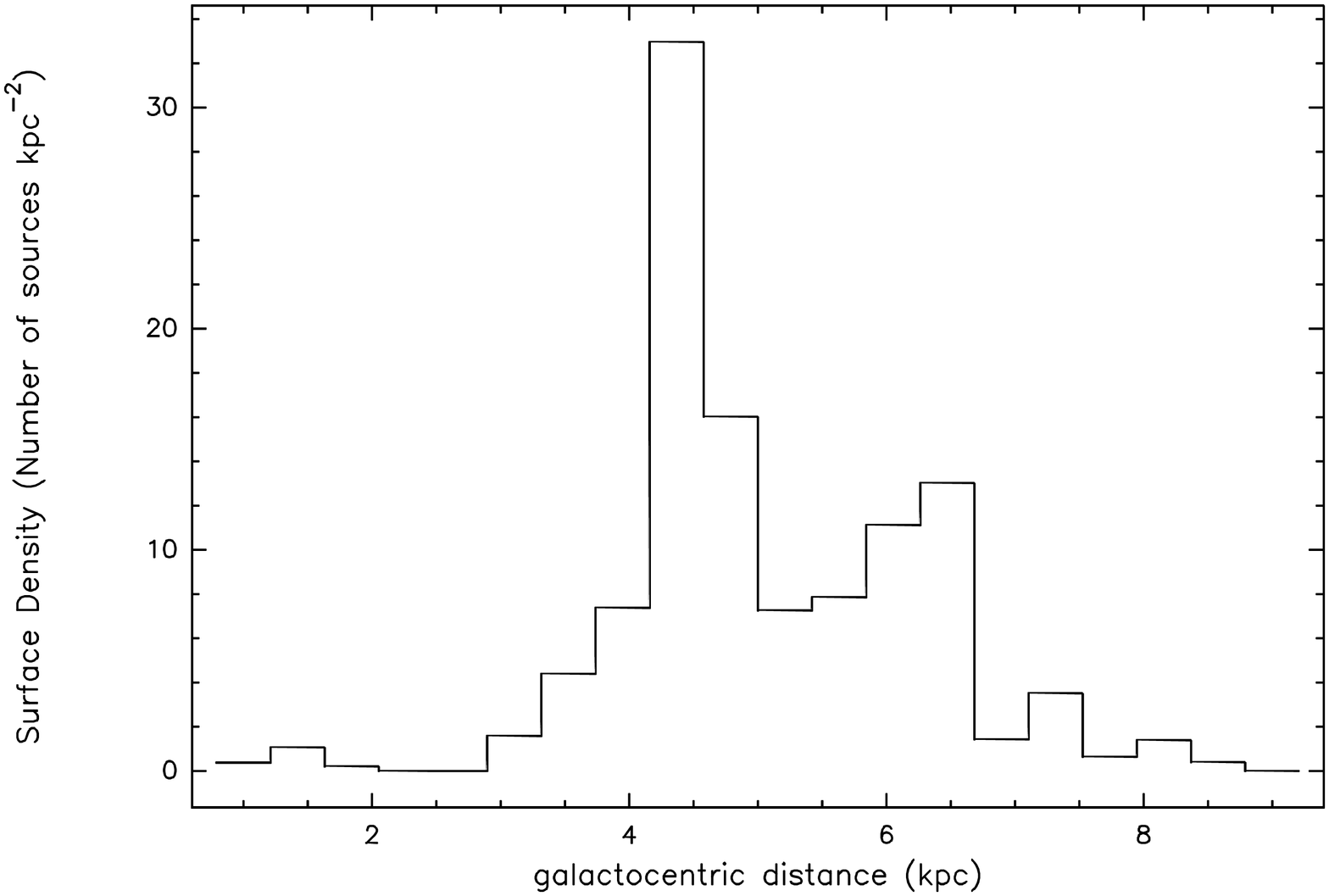}\vspace*{0.5cm}
\includegraphics[angle=-90,width=9.0cm]{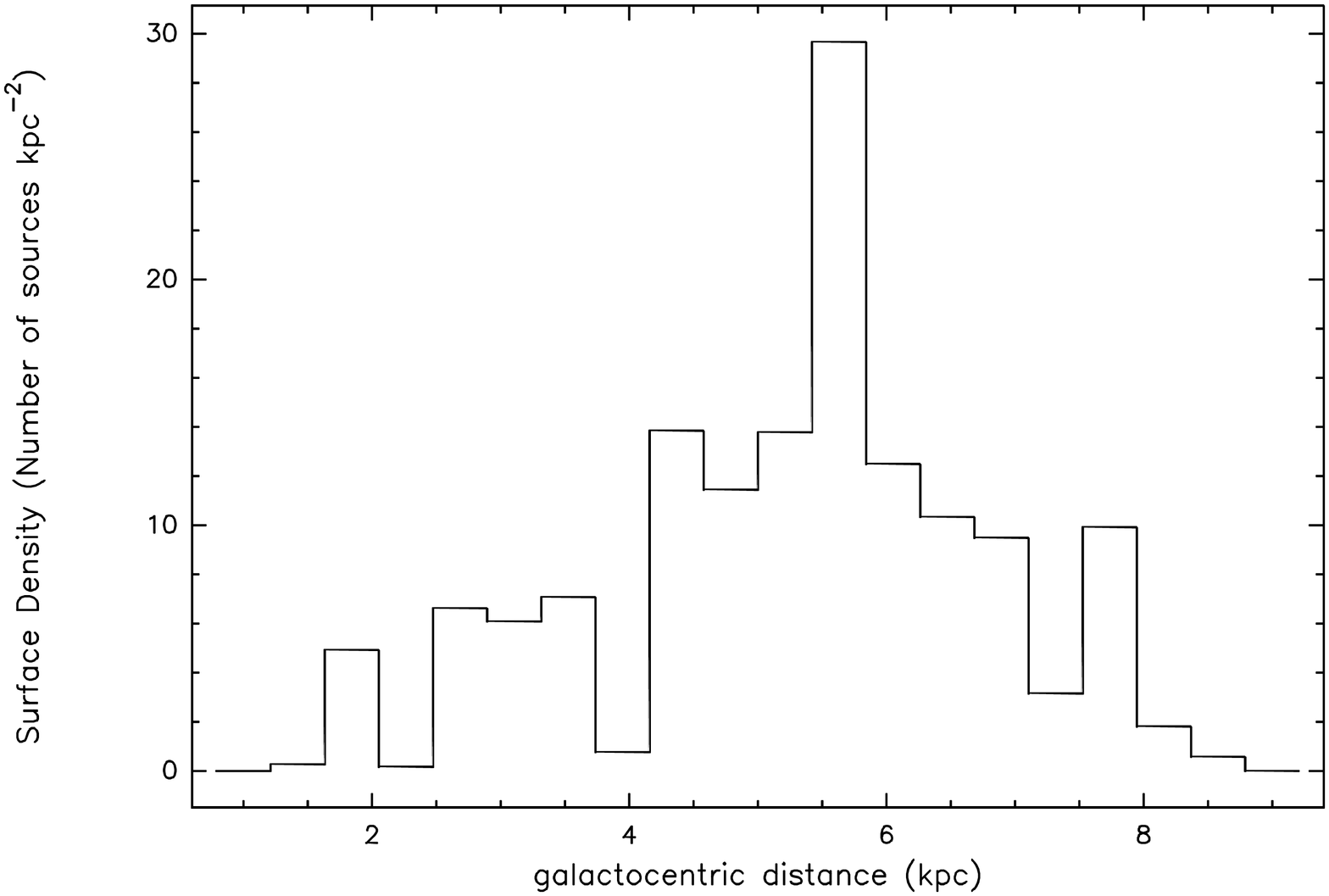}
\caption[Galactocentric distribution]{Number distribution of ATLASGAL sources with galactocentric radii, the clumps in the first quadrant are plotted at the top and the sources in the fourth quadrant at the bottom.}\label{galactocentric radius}
\end{figure}

\subsection{Scale height and distance from the Galactic midplane}
We calculate the distance from the Galactic plane in pc from the Galactic latitude at the distance of the clumps and show the number distribution of the whole ATLASGAL sample in the first and fourth quadrant as a black histogram in Fig. \ref{height}. An exponential fit to the data gives a scale height of $27.93 \pm 1.97$ pc. We fit the height distributions of the subsamples in the first and fourth quadrant as well, which results in scale heights of $23.62 \pm 1.05$ pc in the first quadrant and $30.63 \pm 3.32$ pc in the fourth quadrant. The comparison of the two samples shows that their scale heights are consistent within a $3 \sigma$ uncertainty. The red histogram in Fig. \ref{height} gives the number distribution with Galactic latitude of ATLASGAL sources with masses above the completeness limit of 1000 M$_{\odot}$ (see Sect. \ref{mass size}). A fit to those data yields a scale height of $32.99 \pm 2.6$ pc, similar to that of the whole ATLASGAL sample. The scale height derived from the latitude distribution of high mass star forming clumps is therefore not affected by their masses. The comparison of our scale height to other samples in the inner Galaxy reveals that we obtain a similar scale height as derived from $6.7-$GHz methanol masers of $27 \pm 1$ pc by \cite{2011MNRAS.417.2500G} and $\sim 30$ pc determined from a sample from the Arecibo Methanol Maser Survey \citep{2009ApJ...706.1609P}. The scale height of the whole ATLASGAL sample also agrees with that determined for extended radio sources in the first quadrant, which are likely compact and ultracompact HII regions, of 25 pc \citep{1990ApJ...358..485B}. \cite{2012ApJ...747...43B} estimate an upper limit of the scale height of 46 pc for ATLASGAL submm clumps, which is higher than our measure, because they do not have distances and only approximate them using a sample of YSOs. A similar scale height of 47 pc is determined for an ATLASGAL sample of embedded, compact clumps by \cite{2014A&A...565A..75C}. This was calculated
without distance information of the individual sources and therefore resulted in a larger scale height than our estimate. The scale height of ATLASGAL clumps above 1000 M$_{\odot}$ (see Sect. \ref{mass size}) agrees with that derived from a RMS sample in the first quadrant of $30.2 \pm 2.8$ pc \citep{2011MNRAS.410.1237U}. A fit to the latitude distribution of the whole RMS sample reveals a larger scale height of $37.7 \pm 0.8$ pc \citep{2014MNRAS.437.1791U} because it also includes sources in the outer Galaxy, which can be located at larger distances from the Galactic plane. \cite{2013MNRAS.435..400U} obtain a scale height of $20.7 \pm 1.7$ pc for compact and UCHIIRs, which lies slightly below our value estimated for the whole ATLASGAL sample. A larger scale height of $29.1 \pm 3$ pc is derived by \cite{2013MNRAS.435..400U} for the submm host clumps, in which the compact and UCHIIRs are still deeply embedded, which agrees with our value measured for the whole ATLASGAL sample.

In addition, we obtain the distance of ATLASGAL sources to the Galactic midplane from the exponential fit to the histograms illustrated in Fig. \ref{height}. The peak of the distribution of the whole ATLASGAL sample is below the Galactic midplane at $-6.68 \pm 0.97$ pc. A similar displacement of $-10.34 \pm 0.51$ pc and $-4.09 \pm 1.66$ pc is also derived for the subsamples in the first and fourth quadrant. This is in agreement with the peak of the latitude distribution of ATLASGAL sources in the longitude range from $-30^{\circ}$ to $20^{\circ}$, which is skewed to a negative value \citep{2009A&A...504..415S}. Our results are consistent with previous studies, which also locate their samples below the Galactic midplane using e.g. 1.1 mm clumps from the BGPS \citep{2010ApJS..188..123R}, an ATLASGAL sample consisting of embedded, compact sources \citep{2014A&A...565A..75C}, bubbles in the inner Galactic disk identified with GLIMPSE \citep{2006ApJ...649..759C}, UCHIIRs \citep{2000A&A...358..521B}, and CO observations of molecular clouds \citep{1984ApJ...276..182S,1988ApJ...327..139C}. It is usually assumed that this offset results from the location of the Sun above the Galactic plane \citep[e.g.][]{1995AJ....110.2183H,2009A&A...504..415S}.

\begin{figure}[h]
\centering
\includegraphics[angle=-90,width=9.0cm]{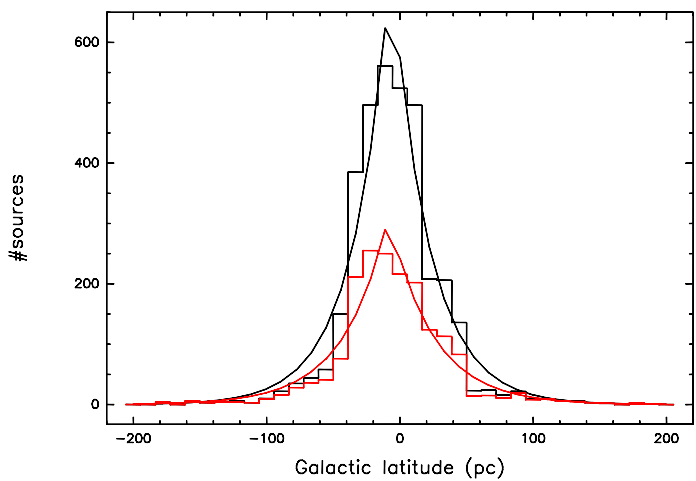}
\caption[scale height]{Number distribution of ATLASGAL subsamples in the first and fourth quadrant with the height above Galactic plane is plotted in black. The histogram of sources with masses above the completeness limit of 1000 M$_{\odot}$ is shown in red.}\label{height}
\end{figure}

\subsection{Mass and size distribution}
\label{mass size}
The gas mass of the ATLASGAL clumps is derived from the dust emission \citep{2008A&A...487..993K}
\begin{eqnarray}
\label{gas mass}
 M_{\mbox{\tiny gas}} (\rm M_{\odot}) = 1.2\times 10^{-14}\frac{S_{870\mu m} \lambda^3 d^2 \left({{\rm exp}} \left(\frac{1.44\times 10^4}{T_{\mbox{\tiny d}} \lambda}\right)-1 \right)}{\kappa (\lambda)} \frac{Z_{\odot}}{Z}
\end{eqnarray}
with the 870 $\mu$m flux density, $S_{870\mu m}$, given in \cite{2013A&A...549A..45C} and \cite{2014arXiv1406.5741U}; the wavelength, $\lambda$, in $\mu$m; the distance to the clump, $d$, in kpc; and the dust temperature, $T_{\mbox{\tiny d}}$, with $T_{\mbox{\tiny d}}$ = $T_{\mbox{\tiny kin}}$ under the assumption of equal gas and dust temperatures. We average the kinetic temperatures obtained from NH$_3$ observations and compute masses with one temperature of 20.8$\pm 2.9$ K for the subsample in the first quadrant and one value of 23.1$\pm 5.5$ K for the sources in the fourth quadrant. An absorption coefficient, $\kappa$, of 1.85 cm$^2$/g at 870 $\mu$m at a gas density n(H) = 10$^6$ cm$^{-3}$ \citep{1994A&A...291..943O} is used resulting from models of dust grains with thick ice mantles and the ratio of metallicity to the solar metallicity, Z/Z$_{\odot}$, is assumed to be 1. These parameters are equivalent to those from \cite{2009A&A...504..415S}. Uncertainties in the gas mass estimates are discussed in Sect. 5.2 in \cite{2012AA...544A.146W}.

The gas masses are plotted against kinematic distances for the ATLASGAL sources in the first quadrant as black triangles and the clumps in the fourth quadrant as red points in the lower panel of Fig. \ref{mass-size}. The error bars indicate the average distance and mass errors. We also added a contour plot by dividing the kinematic distance range into bins of 2 kpc and the logarithm of the mass range into bins of 0.25 and counted the number of sources in each bin (see upper panel of Fig. \ref{mass-size}). The broad distribution of kinematic distances with a peak at 4 kpc and gas masses with a peak at 630 M$_{\odot}$ is similar for the two subsamples. Sources with small masses exhibit a velocity close to 0 km~s$^{-1}$ and are therefore located at near distances. However, the variation of the velocity about 8.28 km~s$^{-1}$ (see Sect. \ref{dist error}) results in large errors in the distance and in the mass of these sources. Fig. \ref{mass-size} reveals a trend of increasing mass with distance resulting in part from an enhanced number of clumps, which may fall within the beam at larger distances. The curve shown in Fig. \ref{mass-size} indicates the mass corresponding to a minimum 5$\sigma$ detection for point sources. Within the whole distance range the effective completeness of the survey to compact unresolved sources is $\sim  600$ M$_{\odot}$. Approximately 65\% of the ATLASGAL sources in the first and fourth quadrant with derived gas masses are above 600 M$_{\odot}$, which is approximately the limit for a clump to form high mass stars \citep{2014A&A...565A..75C}. However, the majority of sources are extended with respect to the beam, their masses are significantly above the catalogue point source sensitivity limit and therefore the turnover in the mass distribution, which is $\sim 1000$ M$_{\odot}$ (see Fig. \ref{mass distribution}) \citep{2013MNRAS.431.1752U}, is a better estimate of the true limit of the mass sensitivity of ATLASGAL.

The rotation curve by \cite{1993A&A...275...67B} yields large distances of $\sim 20$ kpc for one source with $l = 348.551^{\circ}, b = -0.339^{\circ}$ and a velocity of 10.4 km~s$^{-1}$ and 24 kpc for another clump with $l = 349.721^{\circ}, b = 0.121^{\circ}$ and a velocity of 17.6 km~s$^{-1}$. Because these velocities also allow us to locate the sources in the Far 3 kpc arm, we give a more realistic distance of 11 kpc.

Some extreme sources in Fig. \ref{mass-size} are e.g. the two clumps ($l = 338.374^{\circ}, b = -0.152^{\circ}$ and $l = 338.406^{\circ}, b = -0.205^{\circ}$) located far away at a distance of 16.47 kpc and with gas masses of 8.86$\times 10^3$ and $2.12 \times 10^3$ M$_{\odot}$. They are in the same complex, for which we can use HI self-absorption as well as HI absorption toward the more massive clump associated with an HII region. We obtain the far distance, which is in agreement with the kinematic distance given by \cite{2003A&A...397..133R}. Details about other sources, which have extremely high masses, are discussed in Appendix \ref{extreme sources}.

To estimate the size distribution of the ATLASGAL sample we take the source radius, which is not deconvolved from the beam, from \cite{2013A&A...549A..45C} and \cite{2014arXiv1406.5741U} and converted it into pc with the distance of the source. The distribution of radii of the clumps in the first and fourth quadrant is similar, ranging from 0.01 to 2.7 pc with a peak at $\sim 0.3$ pc. They are compared with kinematic distances for the ATLASGAL sample in the first quadrant as black triangles and the sources in the fourth quadrant as red points in Fig. \ref{radius-size}. The upper curve indicates radii calculated using 1.3$\arcmin$, above which large-scale emission is filtered out \citep{2009A&A...504..415S}, while the lower curve shows radii, which correspond to the APEX beam radius of 9$\arcsec$ at the wavelength of 870 $\mu$m. ATLASGAL sources are extended relative to the beam and we obtain a trend of increasing source radius with rising distance. As already revealed in the mass distribution, Fig. \ref{radius-size} shows that the ATLASGAL sample basically consists of molecular clumps, which exhibit radii between $\sim 0.15$ and 1.5 pc \citep{2007ARA&A..45..339B}, and clusters containing several clumps.

\begin{figure}
\centering
\includegraphics[angle=0,width=9.0cm]{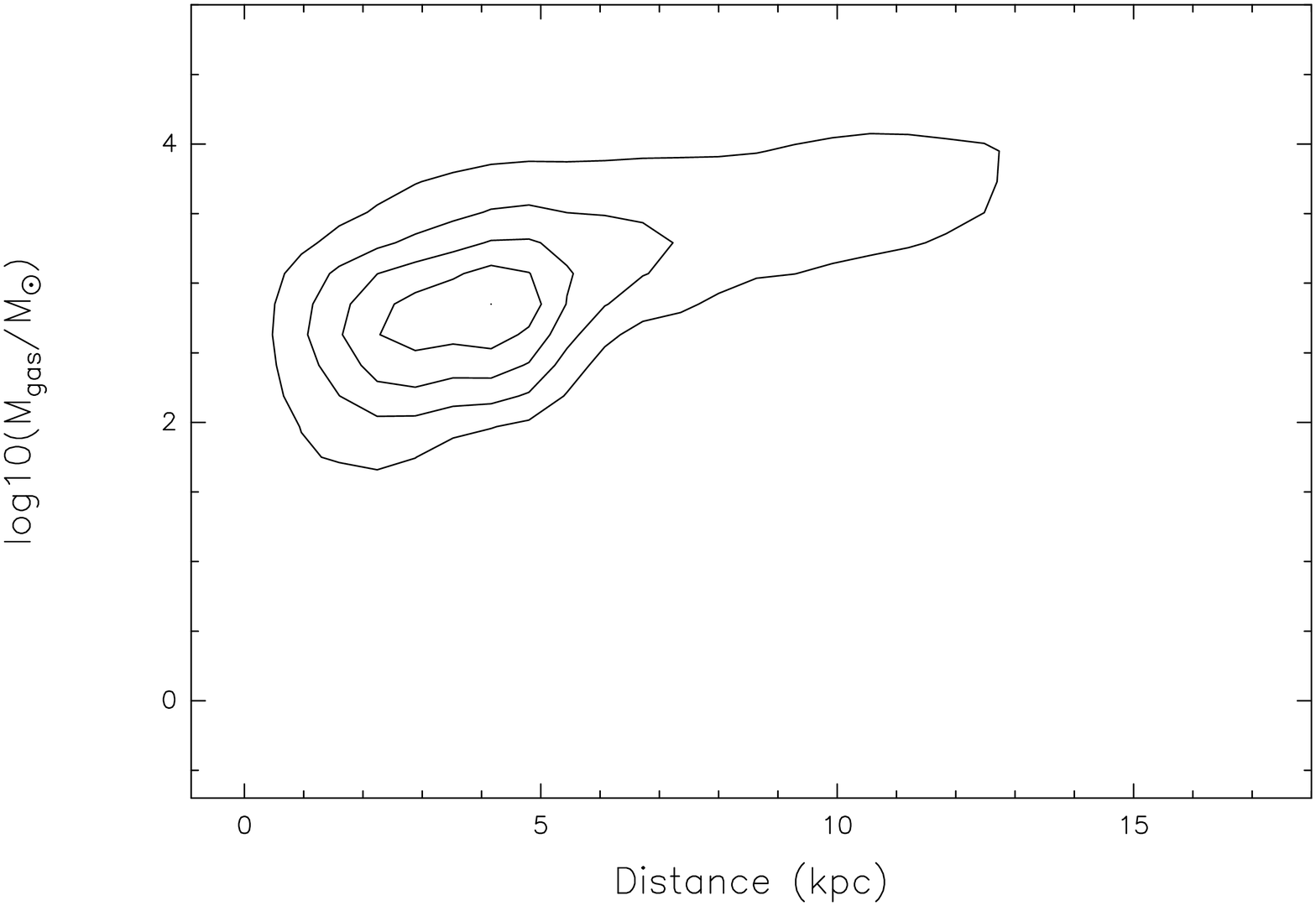}\vspace*{0.5cm}
\includegraphics[angle=0,width=9.0cm]{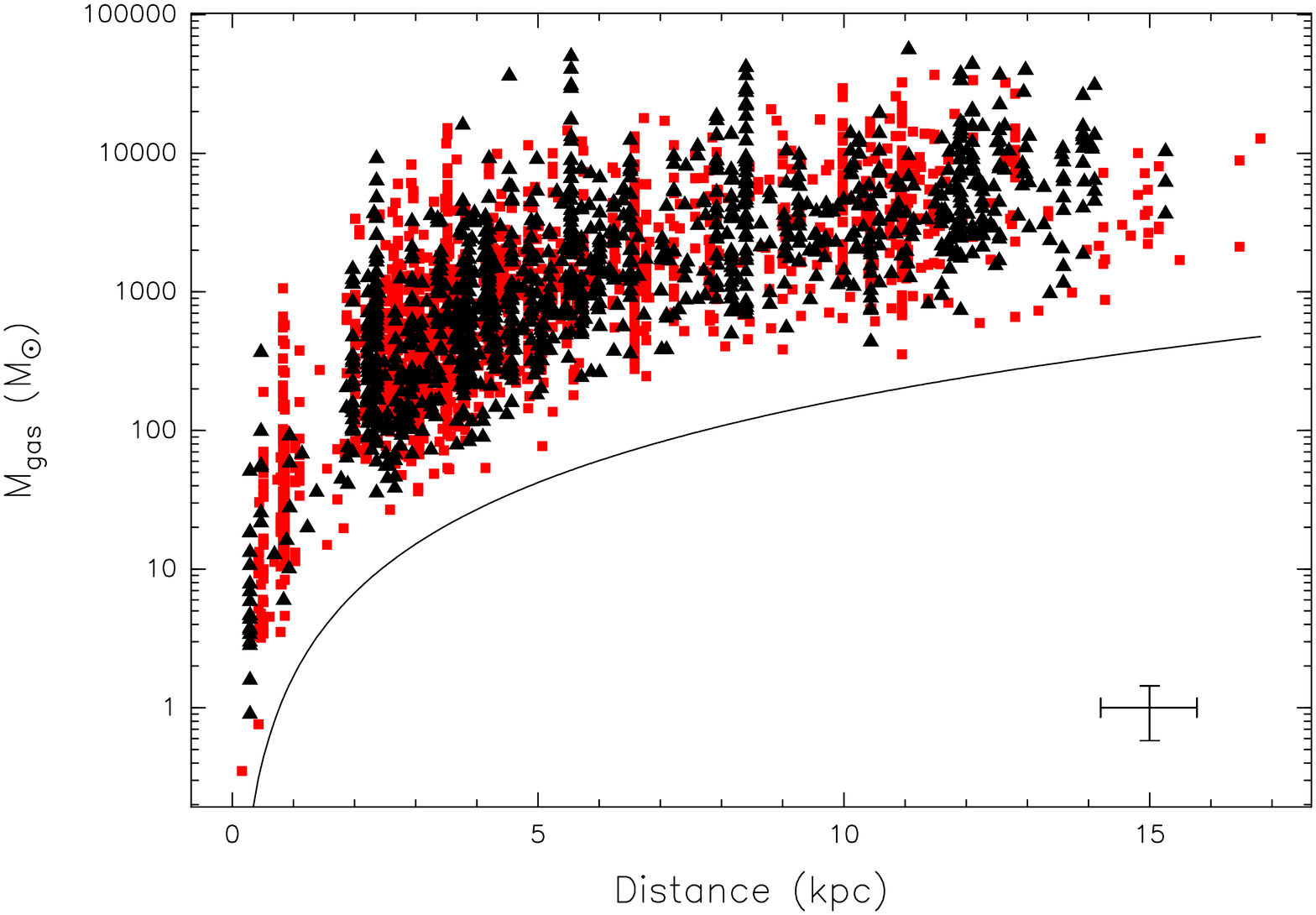}
\caption[comparison of kinematic distance and gas mass]{Correlation plot of the kinematic distances and gas masses for the ATLASGAL sources in the first quadrant as black triangles and the clumps in the fourth quadrant as red points. The curve shows a minimum 5$\sigma$ detection. The upper panel illustrates the contour plot, for which we counted the number of sources in each kinematic distance bin of 2 kpc and each logarithmic gas mass bin of 0.25. The contours give 10 to 90\% in steps of 20\% of the peak source number per bin. The error bar displays the average distance and mass errors.}\label{mass-size}
\end{figure} 

\begin{figure}
\centering
\includegraphics[angle=0,width=9.0cm]{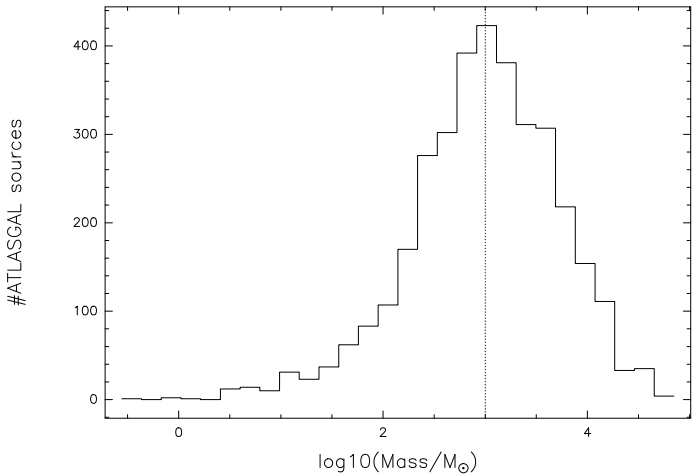}
\caption[histogram of gas mass]{Histogram of the logarithm of the gas mass. The vertical line displays the turnover in the mass distribution.}\label{mass distribution}
\end{figure} 

\begin{figure}
\centering
\includegraphics[angle=-90,width=9.0cm]{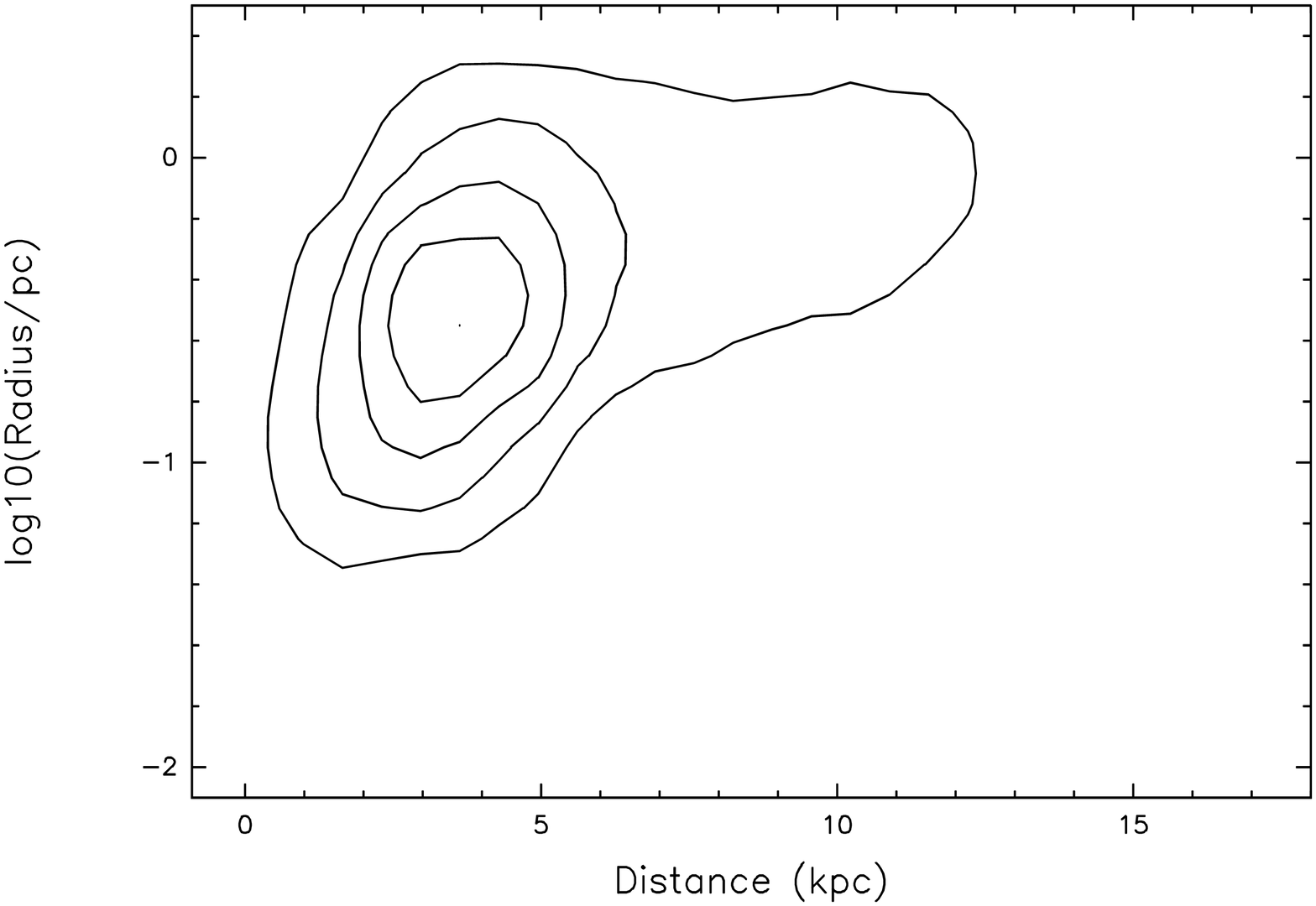}\vspace*{0.5cm}
\includegraphics[angle=0,width=9.0cm]{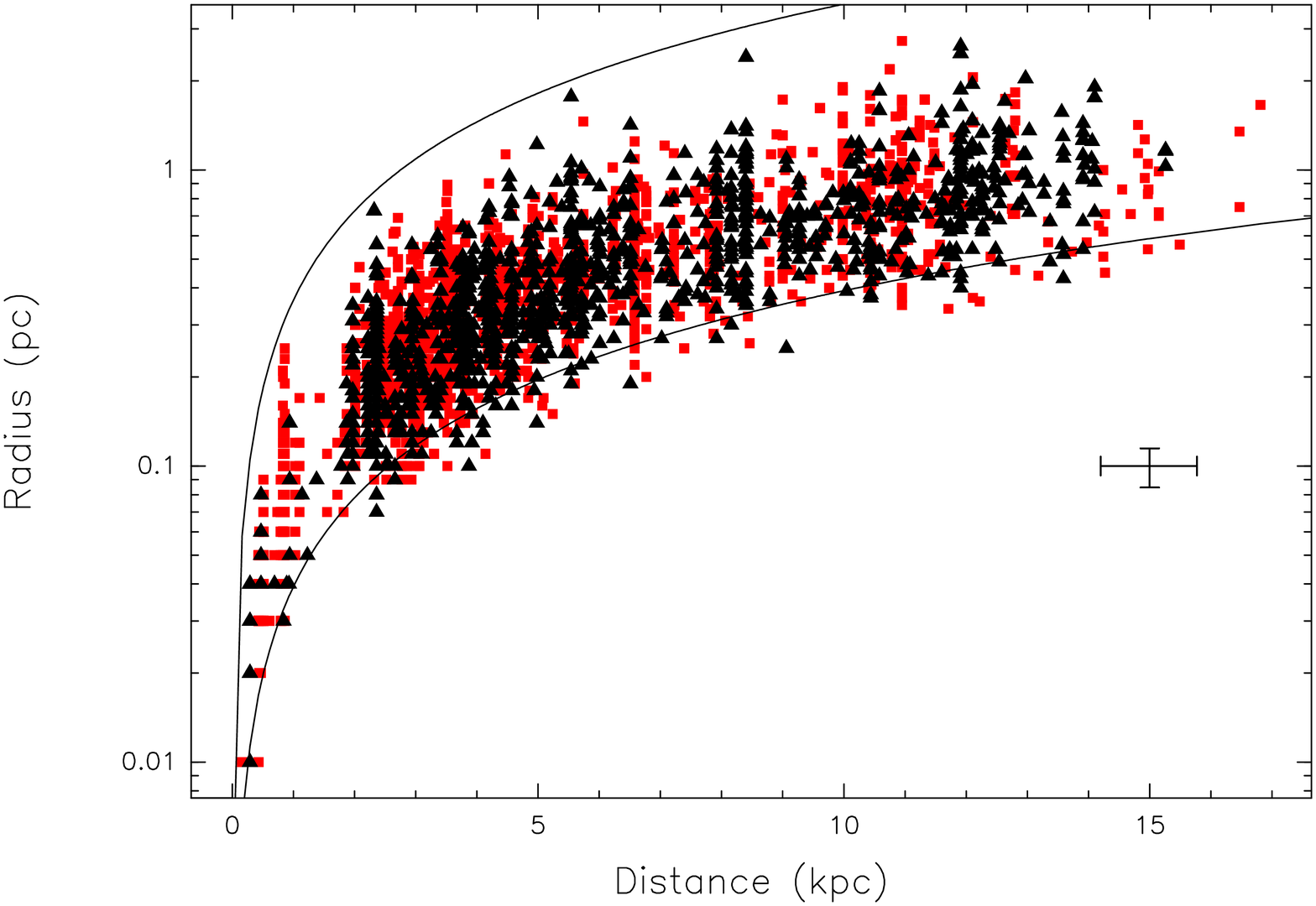}
\caption[comparison of kinematic distance and radius]{Dependence of the source radius not deconvolved from the beam on the kinematic distance to the ATLASGAL sample in the first quadrant as black triangles and the clumps in the fourth quadrant as red points is shown as scatter plot in the lower panel. The lower curve indicates radii corresponding to the beam radius of 9$\arcsec$, the upper curve illustrates radii computed with 1.3$\arcmin$. The binning of the kinematic distance in the contour plot in the upper panel is 1.5 kpc and that of the logarithm of the radius is 0.5. The contour levels from 10 to 90\% are shown in intervals of 20\% of the peak source number per bin. The error bar indicates the average distance and radius errors.}\label{radius-size}
\end{figure}

\subsection{Clump mass function calculation}
\label{clump mass}
With the gas masses calculated in Sect. \ref{mass size} we can examine the statistics of clump masses. We first analyse statistically the influence of the assumption of a constant kinetic temperature, used in the mass calculation, on the slope of the clump mass function. The distribution of gas masses resulting from the observed kinetic temperature of each source is thus compared with masses computed with the mean kinetic temperature of the sample. To investigate if the two are statistically different, we perform a Kolmogorov-Smirnov (KS) test using the ATLASGAL subsample in the first quadrant, which was observed in NH$_3$. The KS test does not contradict that the two clump mass distributions are the same. Cumulative distribution plots in Fig. \ref{KS-mass} reveal similar mass distributions, which are calculated using observed kinetic temperatures and a constant temperature. To derive the clump mass function we can therefore use gas masses computed with an average kinetic temperature (see Sect. \ref{mass size}).

Two approaches are usually used to characterize the clump mass distribution, which are the differential and cumulative clump mass function. The upper panel of Fig. \ref{diff mass spectrum} illustrates the differential mass distribution of sources with distances between 2 and 5 kpc, which is fitted by a power law as a function of mass bins having equal logarithmic widths \citep{2012ARA&A..50..531K}
\begin{eqnarray}
 dN/d \mathrm{log_{10}}(M) = A M^{\gamma}
\end{eqnarray}
with the number of clumps, $dN$, in a mass bin of width $d \mathrm{log_{10}} (M)$ and constants $A$ and $\gamma$. The power-law exponent $\gamma$ is related to the index $\alpha$ in the power law fitting the number distribution against mass, which is often used to describe structures in gas, \citep{2012ARA&A..50..531K}
\begin{eqnarray}
 dN/dM = AM^{\alpha} 
\end{eqnarray}
by $\alpha = \gamma -1$. We divide the mass range of the whole ATLASGAL sample at any distance into different numbers of bins and fit differential clump mass functions to the mass range above 1000 M$_{\odot}$, for which we are complete (see Sect. \ref{mass size}). The best fit to the data gives similar values for $\alpha$ ranging from $-1.75$ to $-1.89$ for varying bin widths with an average of $-1.82$. We also determined the slope using a method based on maximum likelihood \citep{2009SIAMR..51..661C}, which is independent of the histogram binning and results in $\alpha = -1.82 \pm 0.02$ consistent within the errors with the steeper values obtained by our analysis. We also used different bin numbers to plot the mass distribution of ATLASGAL sources with heliocentric distances from 2 to 5 kpc, which is selected around the peak in the heliocentric distance histogram. As illustrated in Fig. \ref{mass-size} we probe a large range in masses over a small distance range. Fitting the differential clump mass function above 1000 M$_{\odot}$, which indicates the turnover for sources within 2 to 5 kpc, leads again to a narrow range of $\alpha$ between $-2.19$ and $-2.38$ for different bin widths with an average of $-2.29$, which is similar to the values for $\alpha$ obtained for the whole ATLASGAL sample at any distance. The method based on maximum likelihood \citep{2009SIAMR..51..661C} gives a slope of $-2.26 \pm 0.05$, which is in agreement with the power-law exponents from our investigation. The upper panel of Fig. \ref{diff mass spectrum} gives an example of the differential clump mass function fitted to masses of the sources from 2 to 5 kpc resulting in a power-law exponent of $-2.19 \pm 0.11$.

We also fit the mass distribution of the ATLASGAL subsample between 2 and 5 kpc by a lognormal function, which is given by equation 5 in \cite{2010ApJ...723..555P}
\begin{eqnarray}\label{logn}
 \frac{dN}{d \mathrm{log_{10}}(M)} = A {\rm exp} \left(- \left(\mathrm{log_{10}} (M) - \mathrm{log_{10}} \left(M_{\mbox{\tiny peak}}\right) \right)^2/2 \sigma^2 \right)
\end{eqnarray}
with the constant $A$, the peak mass $M_{\mbox{\tiny peak}}$, and the dispersion, $\sigma$. All masses are given in units of M$_{\odot}$, hence $\frac{dN}{d \mathrm{log_{10}}(M)}$ is in $(\mathrm{ log_{10} \ M_{\odot}})^{-1}$. The best fit to our data reveals $A = 188 \ (\mathrm{ log_{10} \ M_{\odot}})^{-1}$, $M_{\mbox{\tiny peak}} = 700$ M$_{\odot}$, $\sigma = 0.49$ (see lower panel of Fig. \ref{diff mass spectrum}), and is a better expression of the mass distribution than the power law.

The cumulative mass function describes the fraction of clumps with masses greater than a specific value: it is shown in Fig. \ref{cmf mass spectrum}. The distribution hints at a steep power law for the clumps with high masses and a flatter power law for sources with lower masses. We use a double power law for the cumulative mass function with a break mass, $M_{\mbox{\tiny break}}$, of 3000 M$_{\odot}$ \citep{2006ApJ...650..970R}
\begin{eqnarray}\label{cmf-dpl}
 N (>M) = AM_{\mbox{\tiny break}}^{(\alpha_{\mbox{\tiny high}}-\alpha_{\mbox{\tiny low}})}M^{\alpha_{\mbox{\tiny low}}+1}, 
\end{eqnarray}
where $\alpha_{\mbox{\tiny low}}$ and $\alpha_{\mbox{\tiny high}}$ are the power-law exponents below and above $M_{\mbox{\tiny break}}$. Equation \ref{cmf-dpl} is used for masses smaller than the break mass and 
\begin{eqnarray}
 N (>M) = AM^{\alpha_{\mbox{\tiny high}}+1} 
\end{eqnarray}
for masses equal to and greater than the break mass. Our fit to the mass distribution of sources within 2 to 5 kpc results in $\alpha_{\mbox{\tiny low}} = -1.83$ and $\alpha_{\mbox{\tiny high}} = -3.28$. The power-law exponent of $-3.28$, which describes the high mass portion of the mass spectrum, is steeper than the values of $\alpha$, obtained from the fit of the differential clump mass function.

The cumulative mass function of ATLASGAL clumps with distances between 2 and 5 kpc is better described by a lognormal distribution than by the double power law. A relation for the lognormal function is given by equation 10 in \cite{2006ApJ...650..970R}:
\begin{eqnarray}\label{cmf-logn}
 N (>M) = \frac{1}{2} \left( 1- {\rm erf} \left((\mathrm{log_{10}} (M) - A_0)/\sqrt{2} A_1\right)\right).
\end{eqnarray}
Its fit to our data is illustrated in Fig. \ref{cmf mass spectrum} and results in $A_0 = 6.44$ and $A_1 = 1.15$.

To summarize, our fit of the differential clump mass function to the whole ATLASGAL sample as well as to sources within 2 to 5 kpc leads to consistent power-law exponents, while high masses have a steeper index of the cumulative mass function. The masses of the ATLASGAL clumps can be fitted best by lognormal distributions.

\begin{figure}[h]
\centering
\includegraphics[angle=0,width=9.0cm]{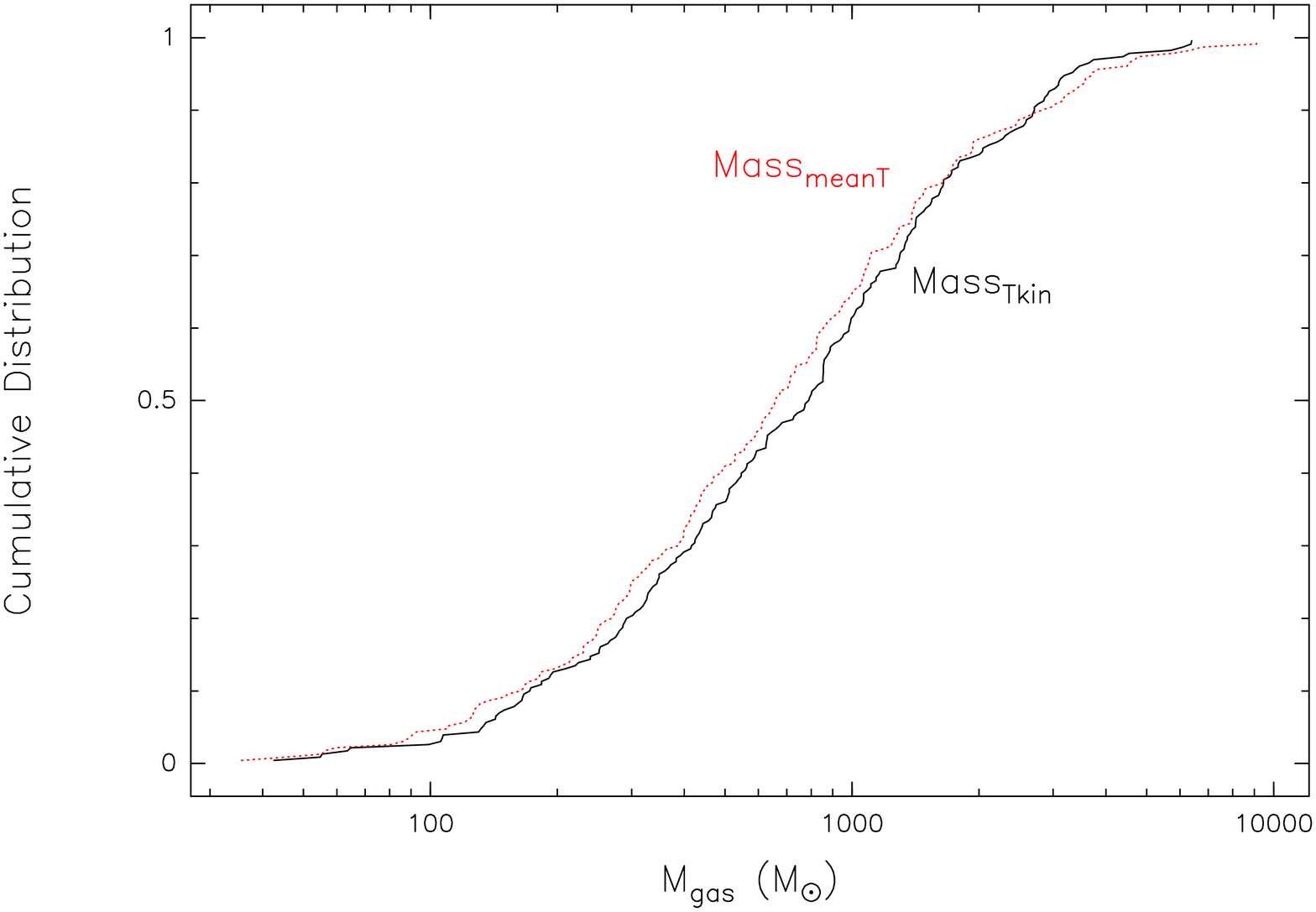}
\caption[mass spectrum]{Cumulative distribution function of the gas mass, which is computed using the observed kinetic temperature of each source shown as solid black line and using the average kinetic temperature of the sample illustrated as dotted red line.}\label{KS-mass}
\end{figure}

\begin{figure}[h]
\centering
\includegraphics[angle=0,width=9.0cm]{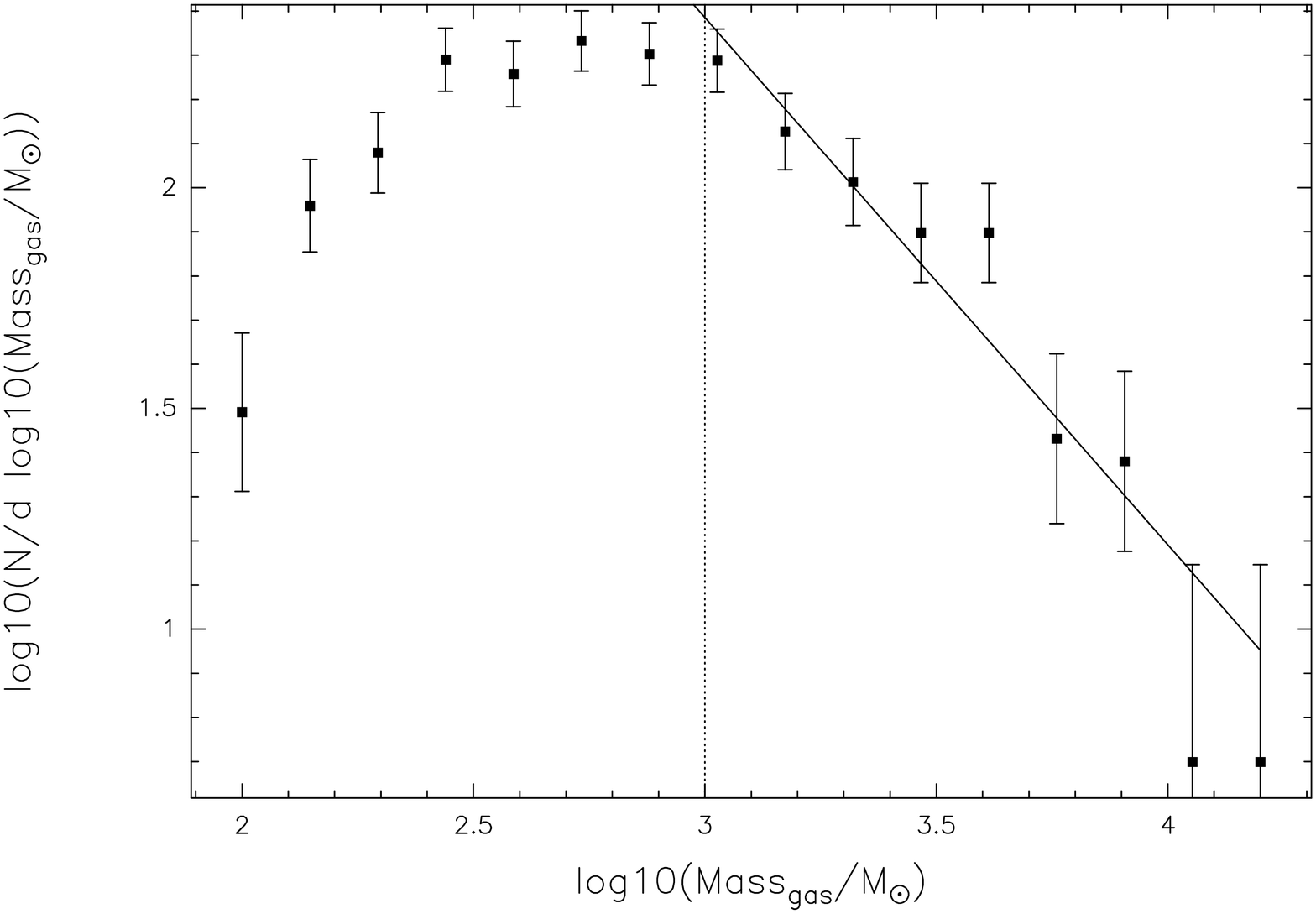}\\
\includegraphics[angle=0,width=9.0cm]{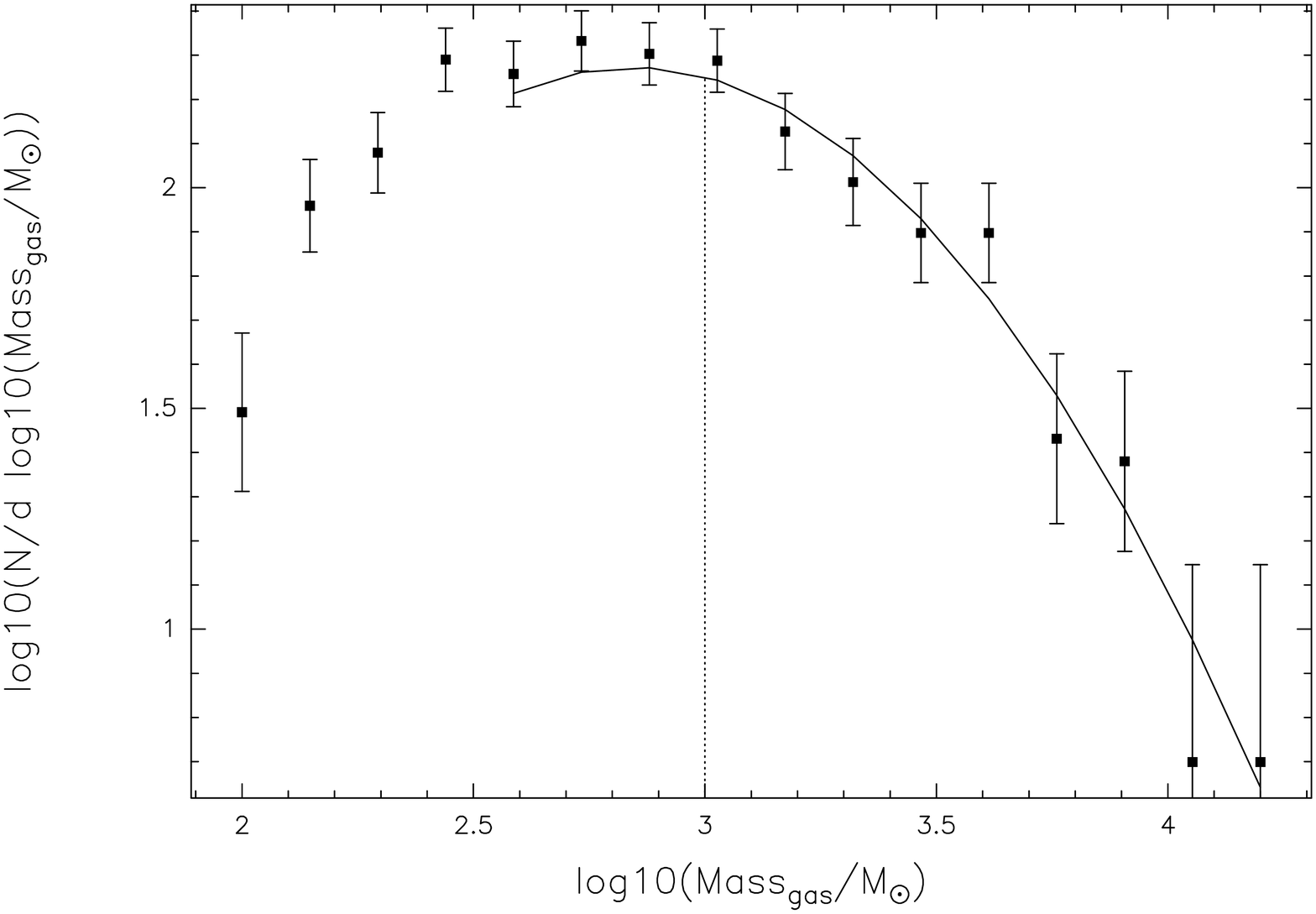}
\caption[mass spectrum]{Differential clump mass function of ATLASGAL sources with distances from 2 to 5 kpc fitted to masses above 1000 M$_{\odot}$ in the upper panel yields a power-law exponent of $-2.19 \pm 0.11$. The vertical line displays the completeness limit of 1000 M$_{\odot}$. A lognormal function is fitted to the mass distribution of this ATLASGAL subsample resulting in $A = 188$, $M_{\mbox{\tiny peak}} = 700$ M$_{\odot}$, and $\sigma = 0.49$. $M_{\mbox{\tiny peak}}$ is below the completeness limit and the peak itself is thus not well constrained.}\label{diff mass spectrum}
\end{figure}

\begin{figure}[h]
\centering
\includegraphics[angle=0,width=9.0cm]{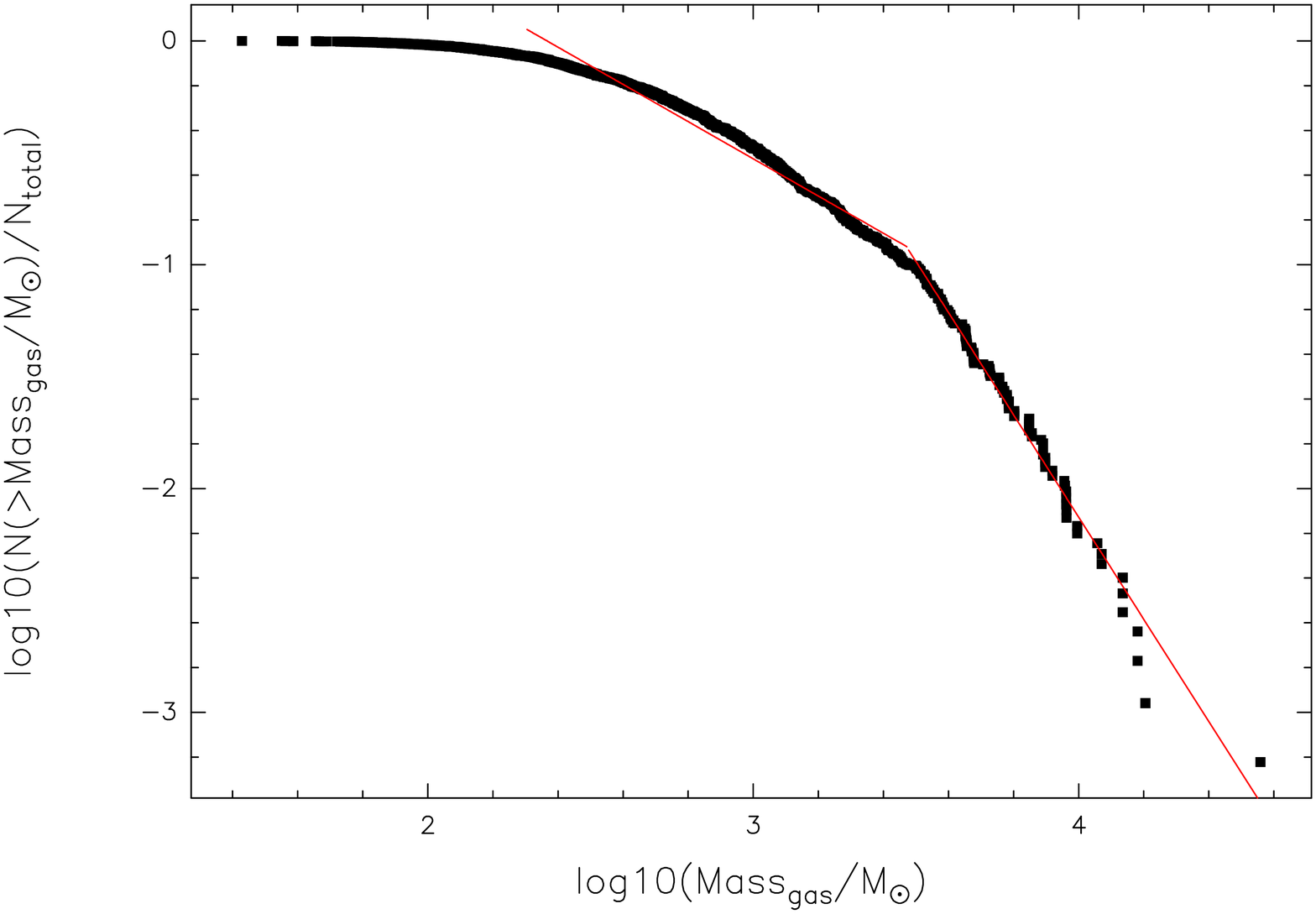}\\
\includegraphics[angle=0,width=10.0cm]{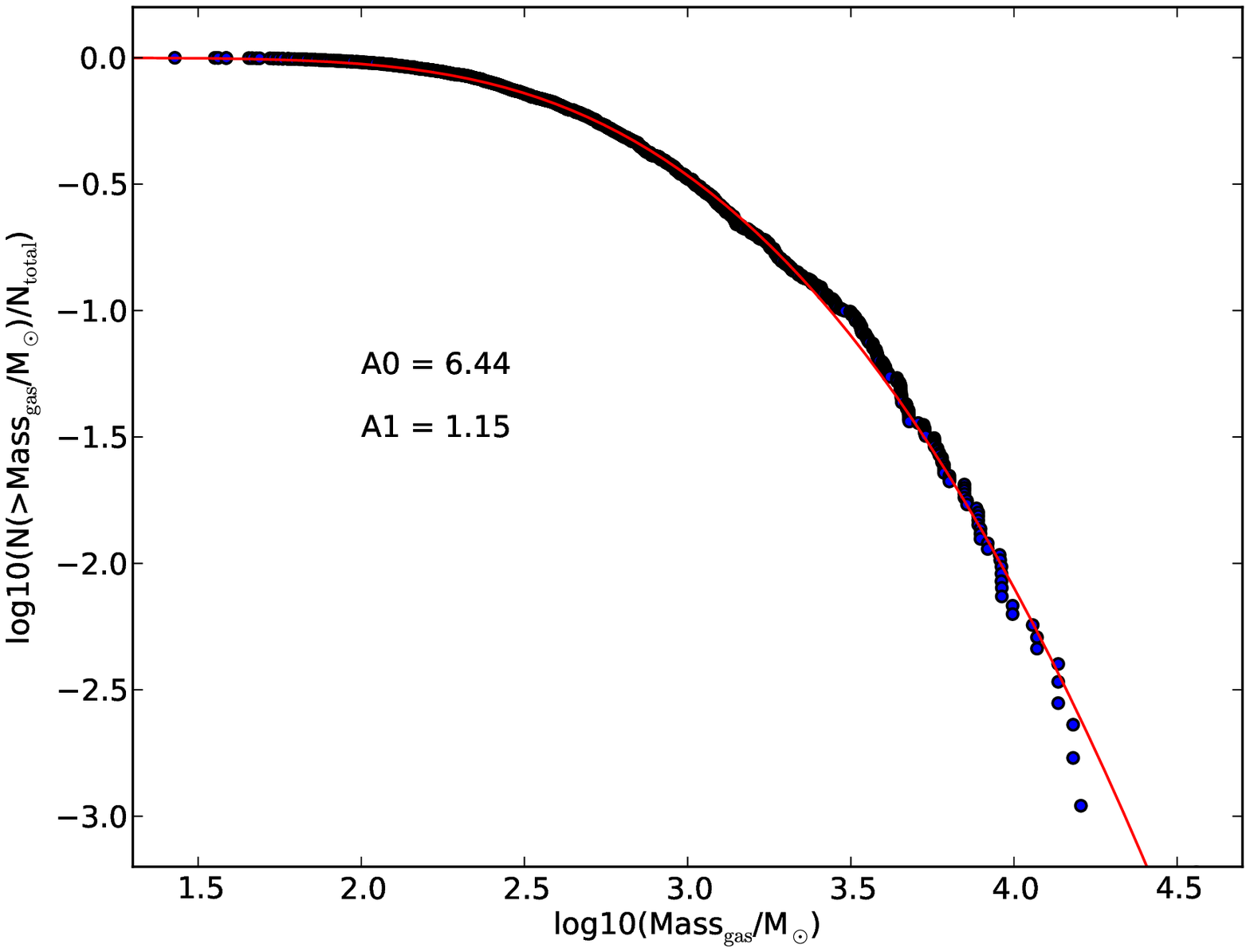}
\caption[mass spectrum]{A double power law fit to the cumulative mass function of the subsample between 2 and 5 kpc gives an exponent of $-1.83$ for masses below 3000 M$_{\odot}$ and of $-3.28$ for masses above that in the upper panel. A lognormal mass distribution is fitted in the lower panel, which leads to $A_0$=6.44 and $A_1$=1.15.}\label{cmf mass spectrum}
\end{figure}

\subsection{Volume density}
\label{density}
The H$_2$ volume density is calculated from the relation $n = 3 M_{\mbox{\tiny gas}}/(4 \pi R_{\mbox{\tiny eff}}^3)$ with the gas mass (see Sect. \ref{mass size}), $M_{\mbox{\tiny gas}}$, and the effective radius, $R_{\mbox{\tiny eff}}$, as given in \cite{2013A&A...549A..45C} and \cite{2014arXiv1406.5741U}. A correction of the mean molecular weight per H$_2$ molecule for helium abundance of m$_{\mbox{\tiny H}_2}$ = 2.72m$_{\mbox{\tiny H}}$ \citep{1973asqu.book.....A} is applied. The ATLASGAL clumps exhibit volume densities between 226 cm$^{-3}$ and $3.4 \times 10^6$ cm$^{-3}$ with a median of 7842 cm$^{-3}$. The number distribution of the volume density is shown in Fig. \ref{volume density}. Our values are similar to those of \cite{2011ApJ...741..110D}, who calculate volume densities between 120 cm$^{-3}$ and $8 \times 10^5$ cm$^{-3}$ but obtain a lower median of 1072 cm$^{-3}$ from 1.1 mm continuum observations towards BGPS sources. We are tracing a smaller range of volume densities than the dust clumps observed in the 1.2 mm continuum by \cite{2006AA...447..221B}, who derive values between $10^3$ and $3 \times 10^8$ cm$^{-3}$ with a median of $4\times 10^4$ cm$^{-3}$. In addition, our volume densities are larger than those of giant molecular clouds in the first and fourth quadrant \citep{1986ApJ...305..892D,2014ApJS..212....2G}, which lie between 5 and $\sim 300$ cm$^{-3}$, because objects identified by the ATLASGAL survey are more compact than molecular clouds.

The ATLASGAL sample consists of high mass clumps in different evolutionary stages from cold cores in an early phase such as IRDCs to more evolved objects with higher temperatures, e.g. UCHIIRs, as revealed by a comparison of their NH$_3$ (1,1) linewidths and rotational temperatures \citep{2012AA...544A.146W}. To investigate any differences in the volume density distribution we divide the ATLASGAL sources in two subsamples: Precluster clumps, which have a faint or no mid-infrared or far-infrared counterpart, do not harbour an embedded protostar or protocluster, while ATLASGAL clumps with embedded or associated luminous infrared sources emit at 21 $\mu$m. Because the ratio of the 21 $\mu$m flux from the MSX survey to the 870 $\mu$m peak flux density as shown in \cite{2013A&A...549A..45C} estimates the heated part of a source, we use this measure to distinguish between cold sources in an early phase or warm clumps in a later phase. Comparison with more sensitive and higher resolution WISE data \citep{2010AJ....140.1868W} is presented in \cite{2014A&A...565A..75C}. A histogram of the 21 $\mu$m flux to the 870 $\mu$m peak flux density ratio reveals a peak at 0.8, which divides the ATLASGAL sources into a cold sample with a ratio smaller than 0.8 and a warm sample with a ratio greater than 0.8 corresponding to UCHIIs and YSOs. A comparison of the volume density distribution of the two samples indicates no difference between the two with similar median values of 8334 cm$^{-3}$ for the cold clumps and of 7753 cm$^{-3}$ for the warm sources. This is in agreement with the analyis of \cite{2006AA...447..221B}, who derived similar H$_2$ volume density distributions of 1.2 mm dust clumps showing MSX emission and sources without MSX counterpart.\\

\begin{figure}
\centering
\includegraphics[angle=0,width=9.0cm]{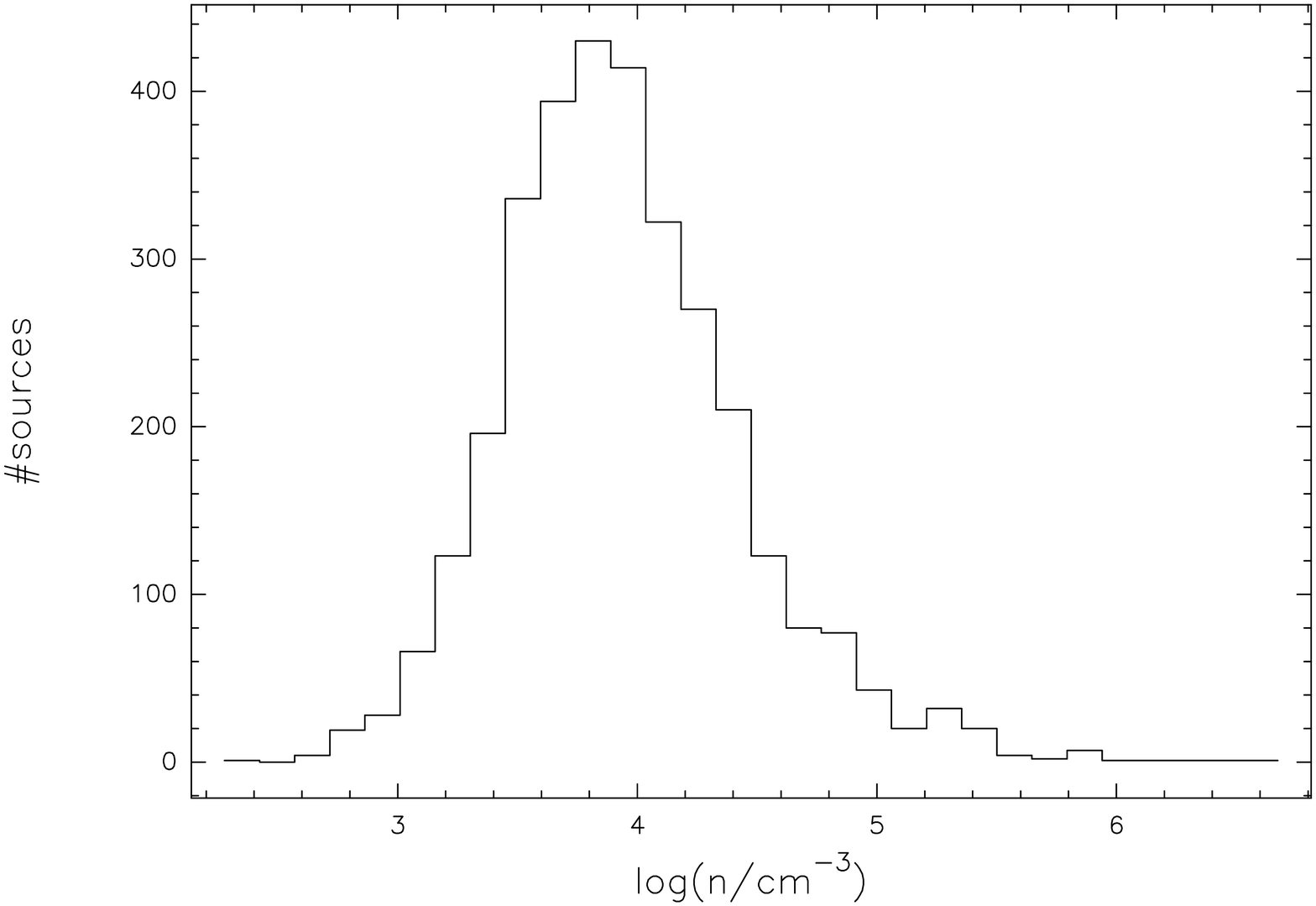}
\caption[volume density]{Histogram of the H$_2$ volume density of the ATLASGAL sample. Most sources exhibit a volume density of 7842 cm$^{-3}$.}\label{volume density}
\end{figure}

\section{Discussion}
\label{discussion}

\subsection{Identification of complexes with known giant molecular clouds}
\label{literature-complex}
We searched in the literature for the complexes with the largest numbers of ATLASGAL sources with measured velocities (see Sect. \ref{s:obs} and \ref{molecular data}) given in Table \ref{v11-dist}. Those had already been identified as giant molecular cloud complexes in previous studies, from which we obtain additional information. Analysis of the source distribution as a function of Galactic longitude \citep{2014A&A...565A..75C} already reveals that giant molecular clouds exist as statistically significant concentrations of compact sources. We summarize our associations in Table \ref{large-complex}. More details can be found in the Appendix.

\begin{figure}[h]
\centering
\includegraphics[angle=0,width=9.0cm]{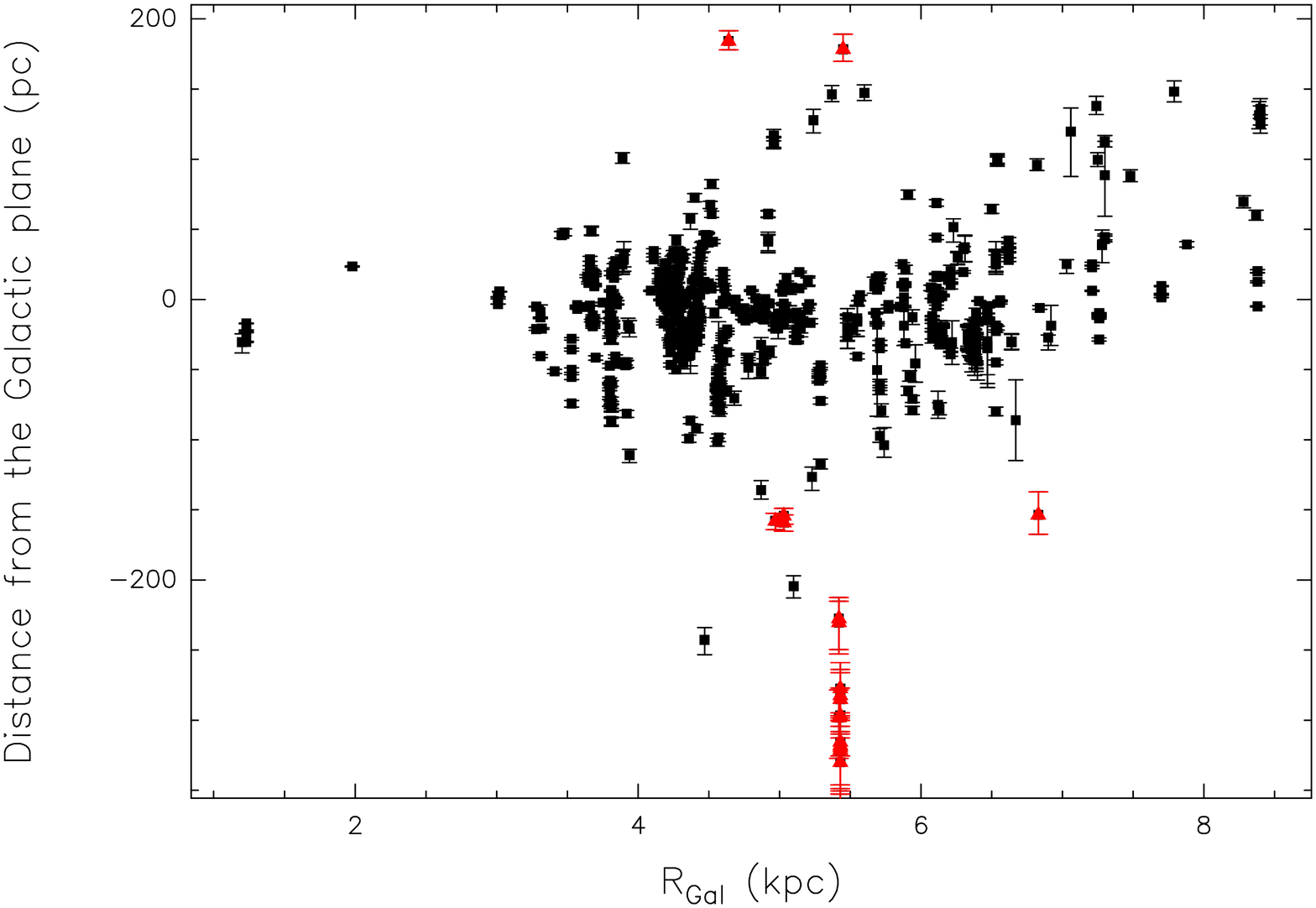}\vspace*{0.5cm}
\includegraphics[angle=0,width=9.0cm]{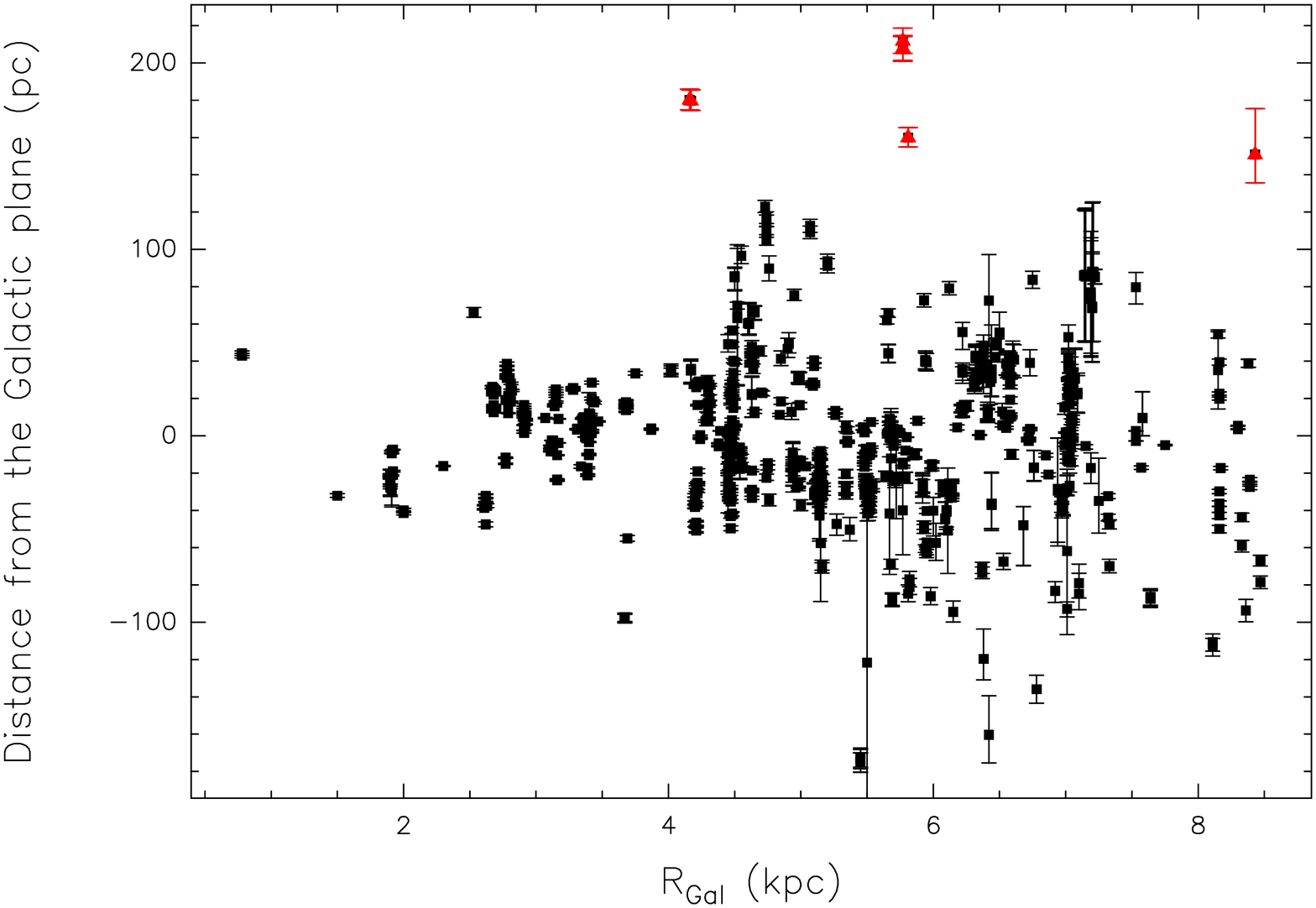}
\caption[scale height]{Distance from the Galactic mid-plane compared with the galactocentric radius for ATLAGAL sources in the first quadrant in the top panel and for clumps in the fourth quadrant in the bottom panel. Sources with a large displacement from the mid-plane are indicated as red triangles.}\label{height samples}
\end{figure}

\begin{figure}[h]
\centering
\includegraphics[angle=0,width=9.0cm]{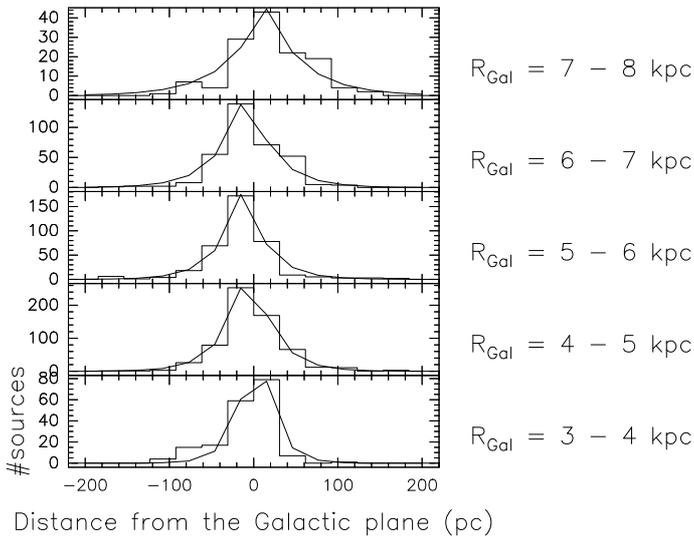}
\caption[scale height]{Histograms of the height above Galactic plane for every bin in galactocentric radius of the subsamples in the first and fourth quadrant. The distributions are fitted by an exponential.}\label{height galradius}
\end{figure}

\begin{figure}[h]
\centering
\includegraphics[angle=0,width=9.0cm]{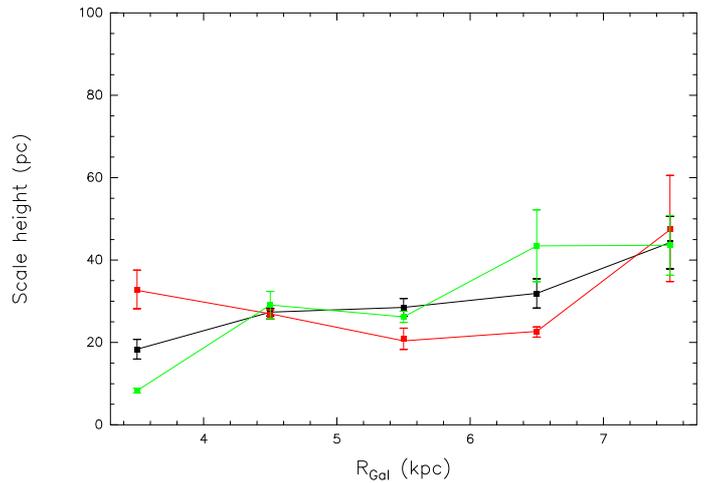}\vspace*{0.5cm}
\includegraphics[angle=0,width=9.0cm]{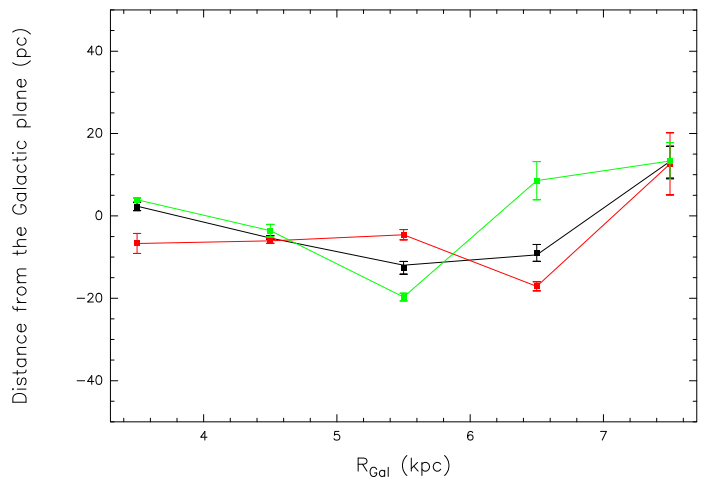}
\caption[scale height]{Dependence of scale height on the galactocentric radius in the top panel and the distance from the disk of the Milky Way shown against galactocentric radii in the lower panel. The subsample in the first quadrant is indicated in red, sources in the fourth quadrant in green, and the two combined in black.}\label{scale height galradius}
\end{figure}

\subsection{Galactic distribution of scale height and molecular gas midplane}
We use the ATLASGAL sample in the first and fourth quadrant to determine the radial distribution of molecular clumps. We plot the distance from the Galactic plane against galactocentric radii of the ATLASGAL subsamples in Fig. \ref{height samples}, clumps in the first quadrant are illustrated in the top panel and sources in the fourth quadrant in the lower panel. Individual complexes with extremely high distances from the Galactic mid-plane are indicated as red triangles and discussed in the Appendix \ref{extreme sources}.

We divide the range in galactocentric radius from 3 to 8 kpc of the ATLASGAL subsamples in the first and fourth quadrant with masses larger than the completeness limit of 1000 M$_{\odot}$ (see Sect. \ref{mass size}) as well as the galactocentric radii of the two combined into bins of width 1 kpc and plot the number distribution of the whole ATLASGAL sample with the height above/below Galactic plane for every bin in galactocentric radius in Fig. \ref{height galradius}. 
An exponential function is fitted to the distributions, which yields the scale height and the distance from the mid-plane in each bin with the error given by the rms value. We show the scale height distribution against galactocentric radius, taken as the mean of each bin, in the upper panel of Fig. \ref{scale height galradius}, where the subsample in the first quadrant is presented in red, the sources in the fourth quadrant in green and the two combined in black. The scale height of the clumps in the first quadrant is approximately constant between $21 \pm 3$ pc and $33 \pm 5$ pc from a galactocentric radius of 3 to 7 kpc and rising to $\sim 48 \pm 13$ pc at 8 kpc. The sample in the fourth quadrant exhibits a scale height distribution varying between $\sim 8 \pm 1$ pc and $\sim 29 \pm 3$ pc at 6 kpc and increasing up to $44 \pm 7$ pc at larger galactocentric radii.

The whole ATLASGAL sample shows a slightly rising scale height distribution from $\sim 18 \pm 2$ pc to $\sim 45 \pm 6$ pc between 3 and 8 kpc with approximately constant scale heights from 4.5 to 6.5 kpc, which are mainly determined by a few complexes with large numbers of sources. These are discussed in the Appendix \ref{extreme sources}. The variation of the scale height with galactocentric radius has already been studied by previous work. Observations of the CO (1$-$0) line in giant molecular clouds \citep{1984ApJ...276..182S} also show a trend of a slightly increasing scale height between R$_{\mbox{\tiny Gal}} \approx 2.5$ and 10 kpc. 
The scale height distribution of the subsamples in the first and fourth quadrant are not significantly different, but reveal an approximately constant scale height from 3 to 8 kpc in agreement with the scale height resulting from CO surveys in the first and fourth quadrant \citep{1988ApJ...324..248B}. Using HI data in the inner Galaxy \cite{1984ApJ...283...90L} also obtain a narrow range in the scale height compared to the flaring of the Galactic midplane in the outer Galaxy at galactocentric radii larger than 9 to 10 kpc \citep{1990A&A...230...21W}. This is also consistent with the analysis of \cite{2014MNRAS.437.1791U}, who investigated $^{13}$CO, CS, and NH$_3$ lines of a sample of RMS sources. Their scale height is also slightly rising from about 20 pc to $\sim 30$ pc between 4 and 8 kpc, while the increase is larger, up to 200 pc, at a galactocentric radius of 11 kpc.

\subsubsection{Variation of the distance from the mid-plane within the inner Galaxy}
The lower panel of Fig. \ref{scale height galradius} shows the displacement from the mid-plane derived from the exponential fit to the distributions plotted in Fig. \ref{height galradius} against the galactocentric radius of ATLASGAL clumps. The clumps in the first and fourth quadrant have a similar distribution of the distance from the midplane within 3 kpc to 8 kpc. Figure \ref{scale height galradius} reveals that the ATLASGAL sample lies close to the Galactic plane, with a distance varying between $-20$ pc and 20 pc. This is a flat distribution from 3 kpc to 8 kpc compared with the increasing mid-plane displacement of the RMS sample \citep{2014MNRAS.437.1791U} and of OB stars associated with UCHIIRs \citep{2000A&A...358..521B} up to 200 pc between 9 kpc and 14 kpc. CO lines in giant molecular clouds \citep{1984ApJ...276..182S} also show an approximately constant distribution of the displacement around $-20$ pc, although it is restricted to negative values. Molecular clouds observed in CO surveys \citep{1988ApJ...324..248B} in the first and fourth quadrant are bounded to the Galactic plane as well, slightly varying between $-40$ pc and 60 pc, similar to distances of the ATLASGAL sources.

Moroever, we analyse if there is a dependence of the gas mass on the distance from the plane. The height above/below Galactic plane of ATLASGAL sources with masses above the completeness limit of 1000 M$_{\odot}$ is divided into bins of 20 pc and the gas mass is averaged within these bins. The distribution of the mean mass in the first and fourth quadrant is approximately constant over the range of distances from the mid-plane. Some clumps with high masses are located at large height above or below the Galactic plane. These are in complexes, for which we could only use the HI self-absorption method to resolve the KDA, because they do not contain any HII region. The distance assignment is therefore uncertain and as a consequence their masses and displacement from the Galactic plane as well. The most massive clumps are expected to be located close to the Galactic plane because of the Galactic gravitational potential. The largest masses are indeed distributed at small distances, but many objects of lower masses between 1000 M$_{\odot}$ and 10$^4$ M$_{\odot}$ are also located near the mid-plane, which lowers the mean value of the gas mass.

\subsection{Mass distribution of ATLASGAL sources in the Milky Way}
With resolved kinematic distances we are able to study the 3D distribution of a large sample of ATLASGAL clumps. We plot the logarithm of the galactocentric radius of sources above the completeness limit of 1000 M$_{\odot}$ against the azimuth, $\theta$, in Fig. \ref{spiral-ns}. Their location is analysed with respect to the large-scale structure of the Milky Way illustrated by the model by \cite{1995ApJ...454..119V}. From a statistical examination of various spiral arm models using different methods such as HI emission, HII regions, magnetic fields, CO emission, thermal electron gas \cite{1995ApJ...454..119V} built a logarithmic four-arm model of the Milky Way with a mean pitch angle of 12.5$^{\circ}$. Using this value we illustrate spiral arms given by the \cite{1995ApJ...454..119V} model as coloured straight lines in Fig. \ref{spiral-ns}. They appear twice on the map because they wrap around, and ATLASGAL sources within 2 kpc of the Galactic centre are excluded where the rotation curve is unreliable. The left panel indicates that ATLASGAL clumps are located close to spiral arms as well as in the interarm region, the source density is high around $\theta = 0^{\circ}$ with clumps at the near distance. The right panel of Fig. \ref{spiral-ns} shows the contour plot, for which the azimuth range is divided into bins of 10$^{\circ}$ and the logarithm of the galactocentric radius into bins of 0.2 and the clump mass is summed in each bin. The peaks in the mass correlate well with the Scutum-Crux arm, Sagittarius and Perseus arms. The map reveals that the mass in the spiral arms is clumpy. The distribution of the ATLASGAL sources in the first quadrant is consistent with the $^{13}$CO surface brightness of GRS clouds in the first quadrant shown in Fig. 12 of \cite{2009ApJ...699.1153R}. In addition, our complexes with the largest number of sources (see Table \ref{large-complex}), which are known molecular cloud complexes, are presented as red dots in Fig. \ref{spiral-ns} and mostly coincide with the enhancements of the summed mass distribution. Two complexes, NGC6334 and N49, do not correlate well with the mass peaks. They have the smallest summed clump mass of the complexes with the largest numbers of sources in the first and fourth quadrant, given in Table \ref{large-complex}. In addition, two mass peaks are not associated with any of the complexes containing the largest numbers of clumps from Table \ref{large-complex}. One of them is at $\theta \approx -127^{\circ}$ and R$_{\mbox{\tiny gal}} \approx 4.7$ kpc and coincides with complex 78 ($l = 18.197^{\circ}, b = -0.308^{\circ}$), associated with the HII region G18.2$-$0.3. Although it harbours fewer sources, 11 clumps with observed velocities, it is located at the far distance of 11.9 kpc and has a large summed clump mass of $4.64 \times 10^5$ M$_{\odot}$. The other mass enhancement at $\theta \approx 110^{\circ}$ and R$_{\mbox{\tiny gal}} \approx 4.6$ kpc is associated with complex 242 ($l = 336.97^{\circ}, b = -0.01^{\circ}$), which contains 15 sources with observed velocities and coincides with the HII region G336.84+0.047 \citep{1987A&A...171..261C}. Our assignement to the far kinematic distance of 10.9 kpc is consistent with previous studies: \cite{2012ApJ...753...62J} and \cite{2014ApJS..212....2G} also place the HII region at the far distance using HI absorption. The far distance leads to a large summed clump mass of $5.07 \times 10^5$ M$_{\odot}$ of complex 242.

The correlation between the location of a large number of high mass star forming clumps and spiral arms as provided by a four-armed model of the Galaxy (see Fig. \ref{spiral-ns}) hints at star formation activity ongoing inside the arms. Because we expect that this trend is also reflected by the distribution of the gas mass, we investigate if the largest amount of gas mass is confined to spiral arms. We therefore study the variation of the gas mass with the galactocentric radius in Fig. \ref{galactocentric radius mass}. The upper panel shows the mass distribution of the ATLASGAL sources in the first quadrant and the lower panel that of the sample in the fourth quadrant plotted against galactocentric radii. The black points indicate masses of the whole sample above the completeness limit of 1000 M$_{\odot}$ (see Sect. \ref{mass size}) and the red points show the summed mass surface density in each bin of 1 kpc in galactocentric radii with the error (1$\sigma$) calculated from Gaussian error propagation.

In the first quadrant there is a large number of masses between $\sim 1000$ and $10^4$ M$_{\odot}$ in addition to large mass values up to $4 \times 10^4$ M$_{\odot}$ at R$_{\mbox{\tiny Gal}} \sim 4$ kpc resulting in a rising sum of the mass surface density per bin to $9 \times 10^4$ M$_{\odot}$~kpc$^{-2}$, which agrees with the increase in the number of ATLASGAL sources at approximately 4 kpc in Fig. \ref{galactocentric radius}. After a decrease to $2.1 \times 10^4$ M$_{\odot}$~kpc$^{-2}$ at 5 kpc, the summed mass surface density is increasing to $2.5 \times 10^4$ M$_{\odot}$~kpc$^{-2}$, which is consistent with the distribution of galactocentric radii showing another peak at approximately 6 kpc in the upper plot of Fig. \ref{galactocentric radius}.

The sum of the mass surface density in the fourth quadrant (see lower panel of Fig. \ref{galactocentric radius mass}) is rising to $2.1 \times 10^4$ M$_{\odot}$~kpc$^{-2}$ at a galactocentric radius of $\sim 3$ kpc, which is reflected by the slightly increasing number surface density in the lower histogram of Fig. \ref{galactocentric radius}. After the summed mass surface density is slightly decreasing up to 4 kpc, it enhances to $4.6 \times 10^4$ M$_{\odot}$~kpc$^{-2}$ at 5 kpc. This trend is also seen by the galactocentric distribution in Sect. \ref{galradius distribution}. There is a decrease of the summed mass surface density at galactocentric radii larger than 6 kpc with a slight rise at 7.5 kpc, which is consistent with the distribution of the number surface density. This analysis consequently reveals an agreement of the gas mass with spiral arms, which are also correlated with the number of ATLASGAL clumps as shown in Sect. \ref{galradius distribution}.

We also investigate the variation of the mean gas mass estimated in bins of 1 kpc in galactocentric radii. However, the peaks of the surface density are not revealed by the average gas mass, which is constant within the errors within the inner Galaxy. This reveals that the clumps in the spiral arms are not more massive, but that the enhancement of the mass in the spiral arms results from a larger number of clumps. In addition, \cite{2012MNRAS.422.3178E} and \cite{2014ApJ...780..173B} revealed no changes in the clump formation efficiency with different environments within the Galactic plane. A recent study of differences in the clump formation efficiency along the line of sight within $37.83^{\circ} \leq l \leq 42.5^{\circ}$ \citep{2013MNRAS.431.1587E} results in no variation in the clump formation efficiency between arm and interam regions. The rise of the mass in the spiral arms is therefore not caused by an increased clump formation efficiency, but likely by a flow of the gas of the interstellar medium and molecular clouds in high density regions resulting in source crowding and an increased probability of collisions in the spiral arms.

\begin{figure*}
\centering
\includegraphics[angle=-90,width=9.0cm]{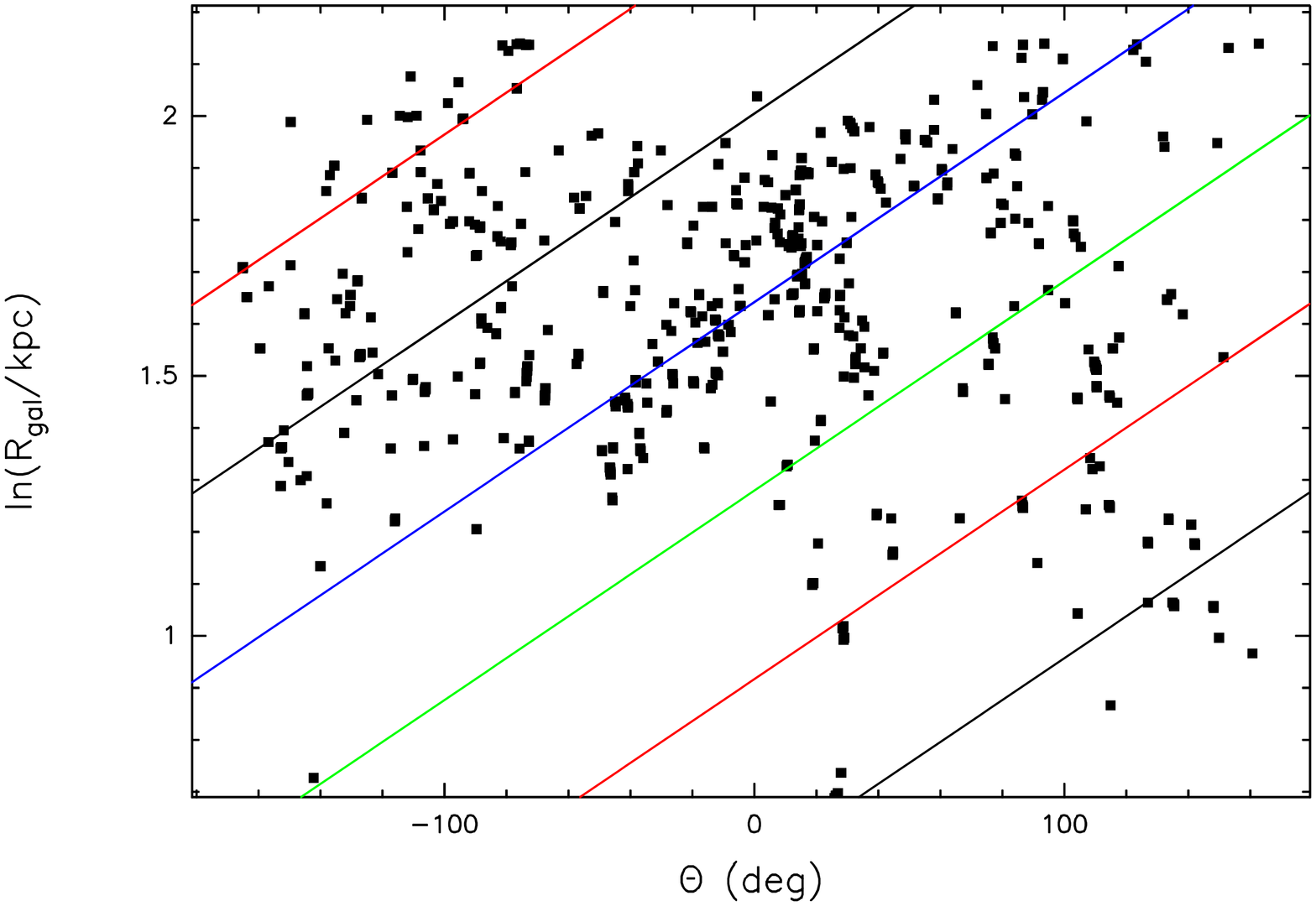}\vspace*{0.5cm}
\includegraphics[angle=-90,width=9.0cm]{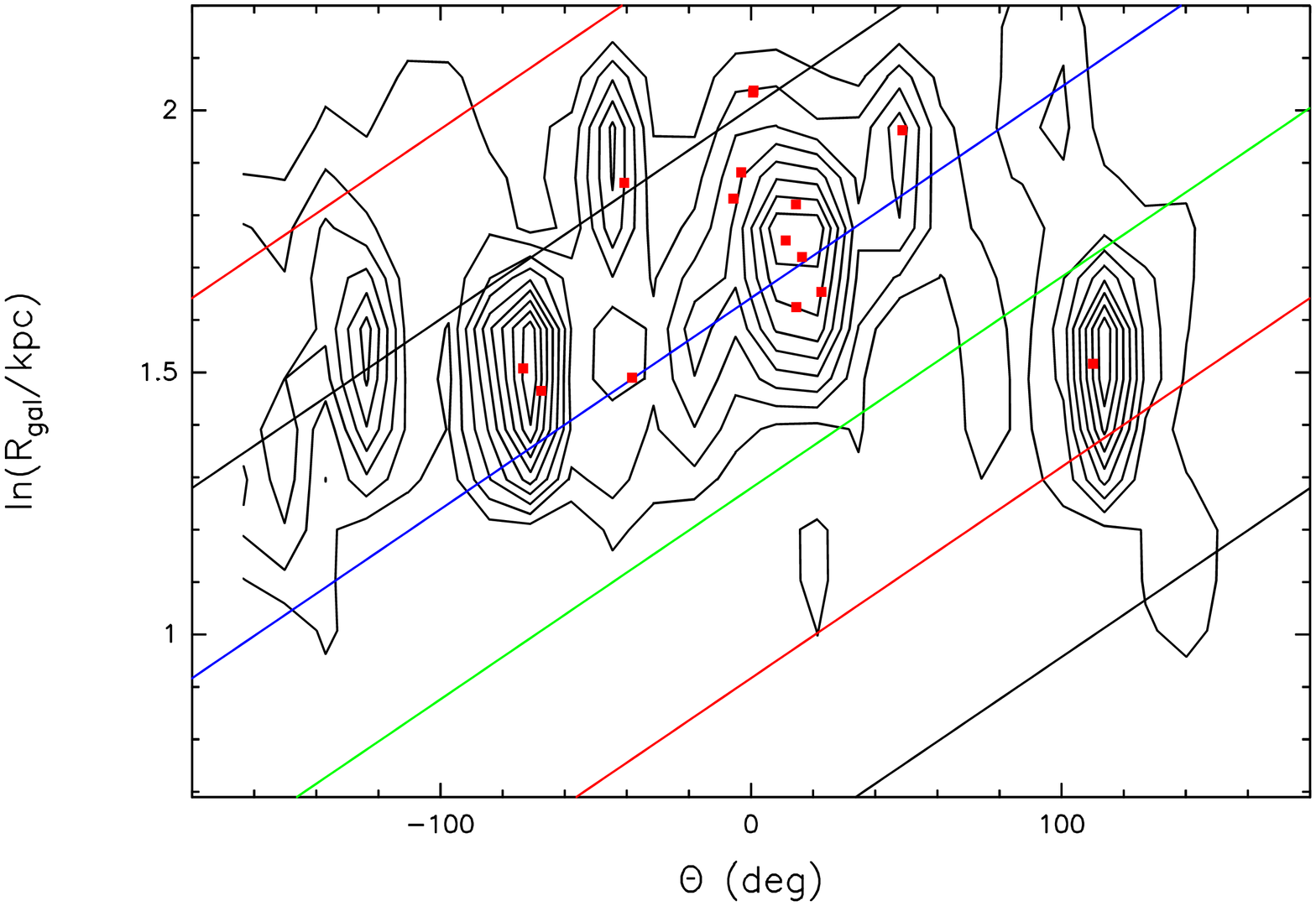}
\caption[Galactocentric distribution]{The logarithm of the galactocentric radius plotted against the azimuth, $\theta$, of the ATLASGAL sources in the first and fourth quadrant above 1000 M$_{\odot}$ as scatter plot in the left panel, clumps within 2 kpc of the galactic centre are excluded. The angle $\theta$ to the Galactic centre-Sun axis is counted anticlockwise as viewed from the north Galactic pole. The right panel displays the contour plot, for which we summed the clump mass in each azimuth bin of 10$^{\circ}$ and each logarithmic galactocentric radius bin of 0.2. The contours give 10 to 90\% in steps of 10\% of the peak summed clump mass per bin. The spiral arms originate from the model by \cite{1995ApJ...454..119V}. The black, red, green, and blue lines indicate the Sagittarius arm, the Perseus arm, the 3-kpc arm, and the Scutum-Crux arm. Complexes with the largest number of sources are illustrated as red dots, most of them agree well with the enhancements of the summed mass.}\label{spiral-ns}
\end{figure*}

\subsection{Clump mass function}
\label{cmf}
\subsubsection{Mass distribution of ATLASGAL clumps and clusters}
The derivation of the power-law exponent of the clump mass function (see Sect. \ref{clump mass}) allows us to analyse if it is consistent with the stellar initial mass function (IMF). This would favour theoretical models, which explain high mass star formation as extension of the theory of low mass stars forming from a molecular cloud collapse via disk accretion. Our $\alpha$ average of $-2.29$ for the differential mass functions agrees with the power-law exponent of the stellar IMF given by \cite{2008LNP...760..181K} of $-2.3 \pm 0.5$. This hints at an origin of the IMF from the clump mass function. However, the confirmation of models of high mass star formation requires that clumps are converted one-to-one into stars or binaries, which cannot be revealed from our data because one clump is likely to form a whole cluster of stars. The similarity between the slope of the clump mass function and the IMF shows that the star formation efficiency in clumps does not depend on the mass of the clump.

\begin{figure}
\centering
\includegraphics[angle=0,width=9.0cm]{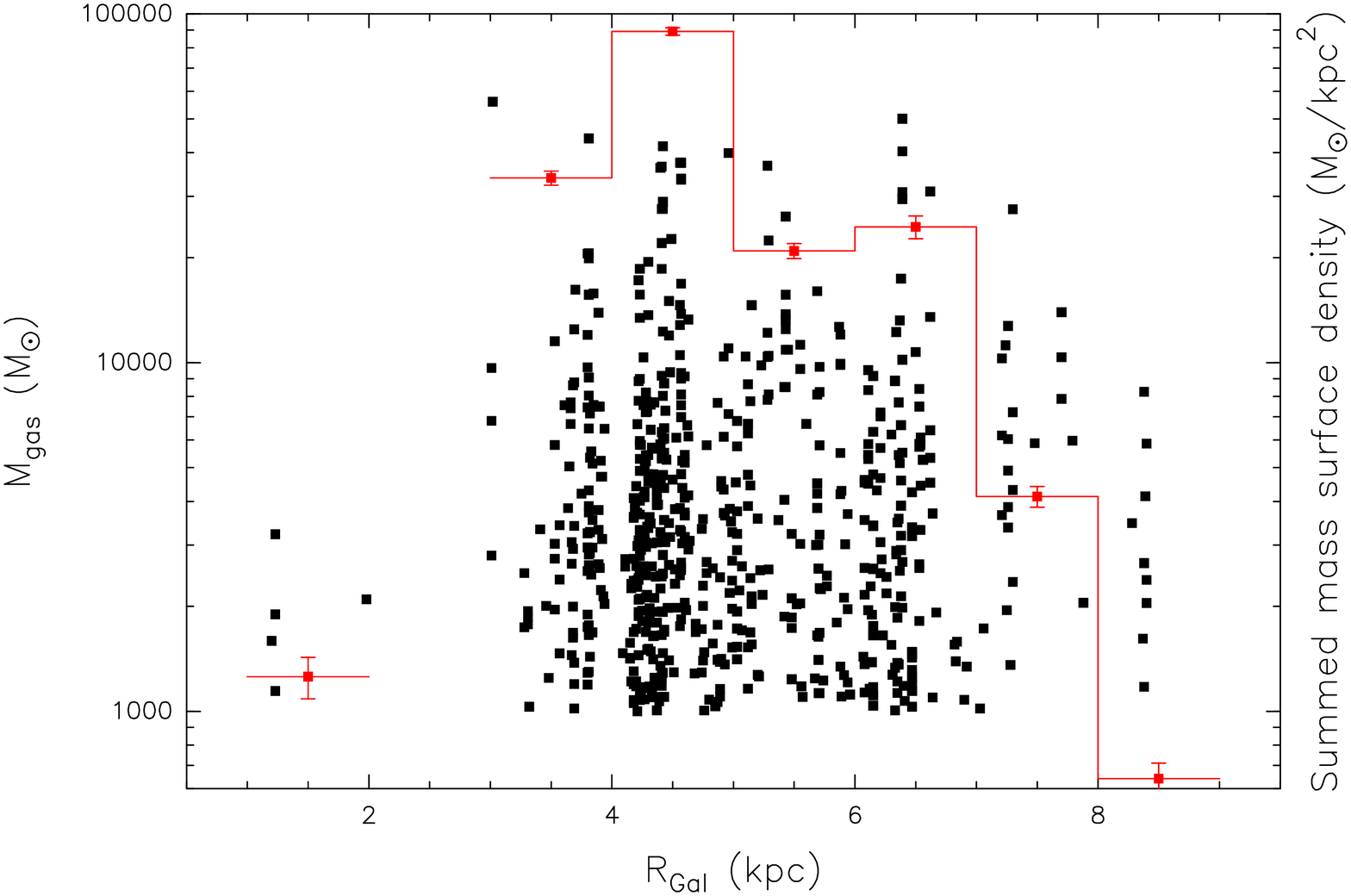}\vspace*{0.5cm}
\includegraphics[angle=0,width=9.0cm]{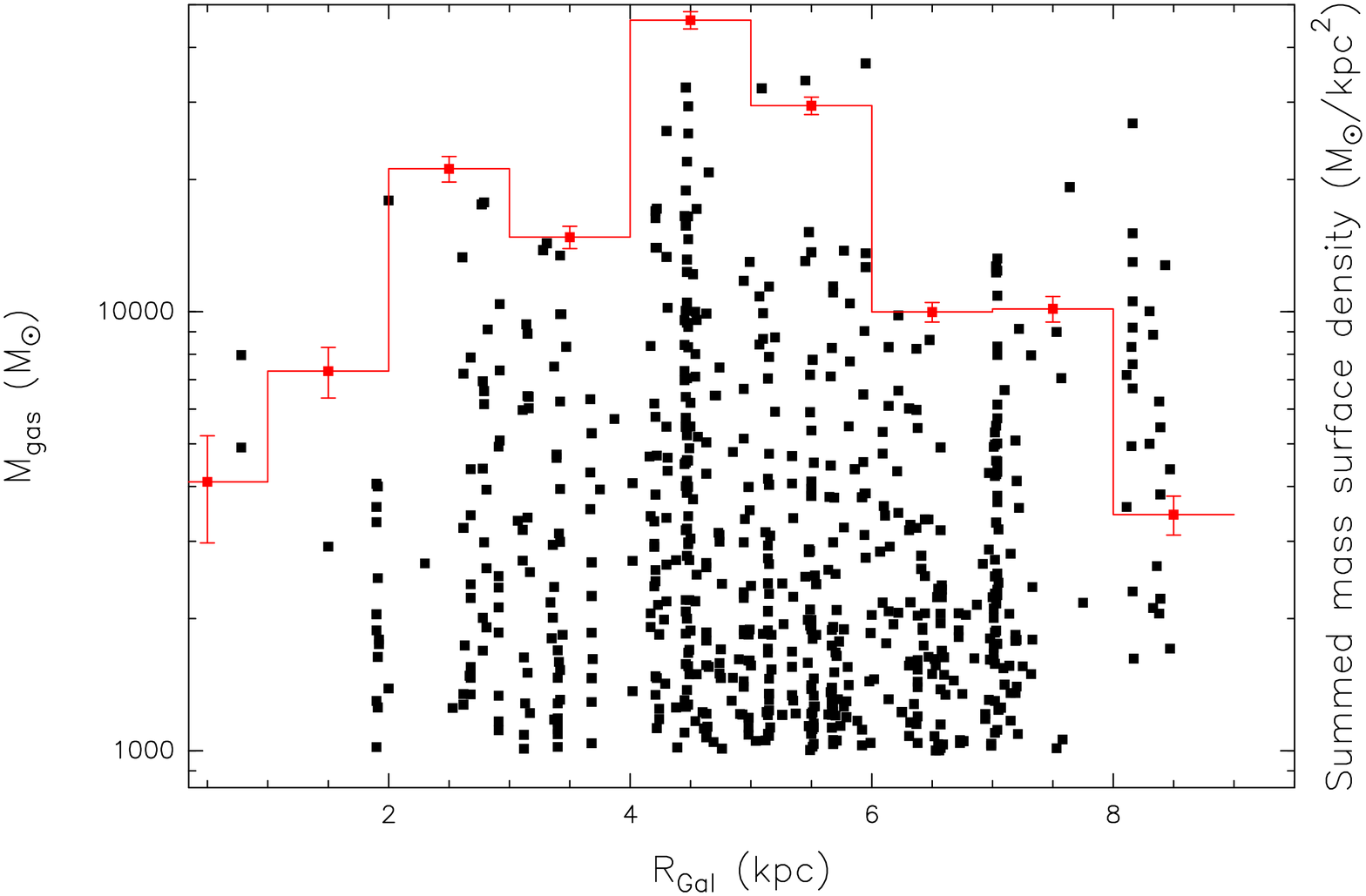}
\caption[Galactocentric distribution]{Distribution of gas masses above 1000 M$_{\odot}$ is illustrated as black points in the upper panel for the ATLASGAL sources in the first quadrant and in the lower panel for clumps in the fourth quadrant. Red points indicate the sum of the gas mass surface density estimated in bins of 1 kpc in galactocentric radii.}\label{galactocentric radius mass}
\end{figure}

Studies of embedded young clusters \citep{2003ARA&A..41...57L}, young and old globular clusters, massive open clusters, and associations as well as interstellar clouds \citep{1997ApJ...480..235E,2010ARA&A..48..431P} yield a power-law exponent $\alpha$ of $\sim -2$, which lies in the $\alpha$ range determined from our clump mass function fits. This consequently reveals a consistency of the mass function of clusters and ATLASGAL clumps. The mass spectrum is therefore not affected by fragmentation. This result is also obtained by other studies using dust continuum emission as well as of high-density gas tracers. \cite{2006AA...447..221B} measured the 1.2 mm dust continuum of clumps in high mass star forming regions; a fit to the mass spectrum above 92 M$_{\odot}$ of sources within 6 kpc yields a slope of $-2.1 \pm 0.02$, which is similar to that derived for pre- and protocluster masses and also agrees with our $\alpha$ values of the ATLASGAL subsample from 2 to 5 kpc. \cite{2003ApJS..149..375S} observed high mass cores, which will form clusters or associations, in the high-density probe CS and calculated virial masses. Fitting the cumulative mass function to their distribution above 1000 M$_{\odot}$ gives a power-law exponent of $-1.91 \pm 0.17$, slightly smaller than our $\alpha$ value while still similar to our power-law exponent and also consistent with the IMF slope of stars within clusters \citep{2003ApJS..149..375S}. This again, shows the similarity of the clump and cluster mass function.

\subsubsection{Comparison with literature values}
\cite{2006ApJ...650..970R} investigated the mass distribution of star forming regions with different median masses. They fitted double power laws to the cumulative mass functions of clumps or cores in low mass objects such as found in the $\rho$ Oph region, as well as in complexes of high mass star formation, e.g. W43 and RCW 106. Their $\alpha_{\mbox{\tiny high}}$ values lie between $-1.8$ and $-3.1$ with a mean of $-2.4 \pm 0.1$. They find no dependence of $\alpha_{\mbox{\tiny high}}$ on the median clump mass for a mass range from 0.4 to 1000 M$_{\odot}$. The ATLASGAL sample covers the same mass range as well as sources of still higher masses, from 0.4 M$_{\odot}$ up to $5.6 \times 10^4$ M$_{\odot}$. Fits to its differential and cumulative mass functions yield power-law exponents, which are consistent with the mean given by \cite{2006ApJ...650..970R}.

We also study the dependence of the power-law exponents on the mass obtained by more recent studies of high mass star forming regions. Table \ref{mass spectrum} gives the name of the source sample, the mass range, the distance range, the minimum mass, $M_{\mbox{\tiny min}}$, above which the distribution is fitted, the power-law exponent, and the reference. We determine the minimum mass from the differential mass functions shown in the listed articles. We consider source samples covering different mass ranges from cores in IRDCs \citep{2006ApJ...641..389R} with masses up to $\sim 2000$ M$_{\odot}$, clumps in IRDCs \citep{2009ApJ...698..324R} exhibiting up to 3000 M$_{\odot}$, high mass clumps detected in the 1.2 mm dust continuum with masses up to 4100 M$_{\odot}$ \citep{2006AA...447..221B}, clumps in molecular cloud complexes from the GRS \citep{2001ApJ...551..747S} having up to 10$^4$ M$_{\odot}$ as well as IRDCs \citep{2010ApJ...723..555P} with masses up to 10$^5$ M$_{\odot}$. Comparing the different mass distributions reveals that the IRDCs and the ATLASGAL sources have the broadest mass ranges with a similar minimum mass. We plot the power-law exponents of the objects listed in Table \ref{mass spectrum} against their minimum masses in Fig. \ref{alpha samples}. The study of GRS molecular clouds (see Table \ref{mass spectrum}) with varying star formation activity and located at various distances \citep{2001ApJ...551..747S} revealed no difference in the slope of the mass functions. Figure \ref{alpha samples} shows that it is also in agreement with the power-law exponent of clumps in IRDCs and the ATLASGAL sample, while the slope of the mass spectrum of 1.2 mm dust clumps and of cores in IRDCs are slightly steeper. This comparison indicates that the mass distribution of cores, clumps, and clusters consisting of several clumps can be fitted by power laws with similar slopes. We also add four high mass star forming regions studied by \cite{2010PASP..122..224S}, who estimated the slope of the mass spectra using a method based on maximum likelihood \citep{2009SIAMR..51..661C}. These samples consist of M17, NGC 7538, Cygnus X, and NGC 6334, which are all located nearby and exhibit various mass ranges: Sources in M17 only have masses up to 120 M$_{\odot}$, while Cygnus X contains clumps with up to 949 M$_{\odot}$, still larger mass ranges are probed by NGC 7538 exhibiting masses up to 2700 M$_{\odot}$ and NGC 6334 with masses up to 6000 M$_{\odot}$. Although \cite{2010PASP..122..224S} used a method, which is independent of the histogram binning, their power-law exponents are similar to those of the other samples in Table \ref{mass spectrum}. The slope of NGC 6334 is consistent with that of clumps in IRDCs, of ATLASGAL sources, and of clumps in GRS molecular clouds. The power-law index of M17, NGC 7538, and Cygnus X are the same as obtained for cores in IRDCs and of 1.2 mm dust clumps. Figure \ref{alpha samples} shows that there is no trend of $\alpha$ with the minimum mass. Table \ref{mass spectrum} reveals that the power-law exponents of the mass function derived from $^{13}$CO emission, observed towards GRS molecular clouds, and from dust emission are consistent.

Using CO emission of molecular clouds in the inner Galaxy \cite{2005PASP..117.1403R} fitted a slope of $-1.5$ to the mass spectrum, which is flatter than our range of $\alpha$ values. This difference results from a larger amount of material probed by CO, which traces low-density regions. High-density tracers such as NH$_3$ detected in the ATLASGAL sample are more appropriate to study the clump mass distribution because stars form only in molecular clouds that exceed a gas surface density threshold of $\Sigma_{\mbox{\tiny gas}} \approx 116$ M$_{\odot}$ pc$^{-2}$ \citep{2012ApJ...745..190L}.

To investigate any differences between the mass distributions of sources in different evolutionary phases, we distinguish between clumps that have the ratio of the MSX 21 $\mu$m flux to the 870 $\mu$m peak flux density smaller or larger than 0.8 as described in Sect. \ref{density}. This leads to a cold sample with a mean gas mass of 2980 M$_{\odot}$ and a warm sample with a mean mass of 3325 M$_{\odot}$. Figure \ref{msx mass spectrum} compares the mass spectrum of the cold sample in red with that of the whole ATLASGAL sample in black, which shows that the two have the same distribution. From a power-law fit to the differential clump mass function of the whole ATLASGAL sample above the completeness limit of 1000 M$_{\odot}$ we derive a slope of $- 1.83 \pm 0.1$. The fit to the mass spectrum of the whole ATLASGAL sample in Fig. \ref{msx mass spectrum} indicates that it has a lognormal distribution (see Sect. \ref{clump mass}). We obtain a power-law exponent of $-1.86 \pm 0.09$ from the fit of the mass distribution of the cold sources above 1000 M$_{\odot}$, which agrees within the errors with the slope of the whole ATLASGAL sample. The clump mass distribution is therefore already set in IRDCs and does not change in later evolutionary phases. This is also supported by the analysis of \cite{2013MNRAS.431.1752U}, who derived a similar power-law exponent of $-2.0 \pm 0.1$ for ATLASGAL sources associated with methanol masers. A comparison of their clumps with other samples also yields a similar clump mass function for the embedded phases of high mass star formation. \cite{2013A&A...551A.111O} divided Hi-GAL sources at a longitude of 30$^{\circ}$ and 59$^{\circ}$ in samples of starless and proto-stellar clumps. They fitted the differential clump mass function to the distribution of the different samples and also obtained a similar power-law index for the whole sample, the starless clumps, and proto-stellar clumps. In addition, we compare our result with the mass spectra of massive dust clumps in the fourth quadrant from \cite{2006AA...447..221B} who investigated if these are associated with point-like, diffuse or no MSX emission. Their study of different clump properties revealed that their subsample without MSX counterpart has a lower mean mass of 96 M$_{\odot}$ than sources associated with point-like or diffuse MSX emission, which has a mean mass of 336 M$_{\odot}$ similar to the trend of average masses of our subsamples. We cannot fit the mass distribution of clumps not associated with mid-IR emission from the sample of \cite{2006AA...447..221B} because of the small source number, which shows that large samples such as ATLASGAL are required for a statistically significant analysis. The power-law fit to the differential mass distribution of massive dust clumps with MSX counterparts from \cite{2006AA...447..221B} gives a slope of $- 1.84 \pm 0.06$, which is consistent within the errors with our fit result of the whole ATLASGAL sample.

\subsubsection{Variation of the clump mass function with Galactocentric radii}
To investigate the change in the slope of the mass distribution in the inner Galaxy we divide ATLASGAL sources with galactocentric radii between 3 and 8 kpc into equally spaced bins of 1 kpc and derive their mass spectra above 1000 M$_{\odot}$, which is the turnover in the complete mass distribution as seen in Fig. \ref{mass distribution}. Fitting of differential clump mass functions results in power-law exponents between $-1.32$ and $-1.86$. They are plotted against galactocentric radii in Fig. \ref{mass spectrum galradius}, which shows a slightly decreasing slope with increasing galactocentric radius out to $\sim$ 7 kpc after which the slope increases to $\alpha = -1.3$ at 8 kpc. This indicates that there is generally a higher proportion of lower mass clumps as a function of galactocentric radius. To compare the mass function of inner and outer Galaxy data we use observations of 132 massive protoclusters in the outer Galaxy \citep{2005ApJS..161..361K}. Their gas masses range from $\sim 0.2$ to 5200 M$_{\odot}$ with an average of 290 M$_{\odot}$. We fit a power law to the differential mass spectrum of the protoclusters and derive an exponent of $-1.48$, which is flatter than measured for the ATLASGAL sample. However, the outer Galaxy sample consists of fewer sources than the observed ATLASGAL clumps, which results in a larger uncertainty of the protocluster mass spectrum.\\

\begin{figure}[h]
\centering
\includegraphics[angle=0,width=9.0cm]{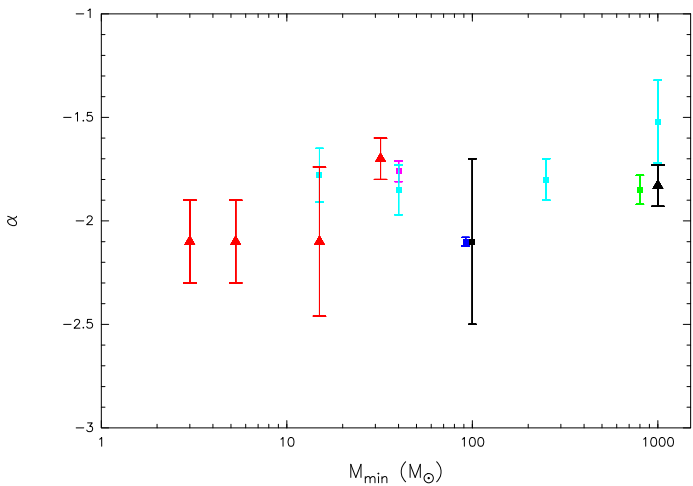}
\caption[mass spectrum]{Slope of the mass functions is compared with the minimum mass for different high mass star forming samples from Table \ref{mass spectrum}: Cores within IRDCs are indicated as black point, clumps within IRDCs as purple point, IRDCs as green point, 1.2 mm dust clumps as blue point, clumps within GRS molecular clouds and W49 as light blue points, ATLASGAL sources as black triangle, M17, NGC 7538, Cygnus X, and NGC 6334 as red triangles.}\label{alpha samples}
\end{figure}

\begin{figure}[h]
\centering
\includegraphics[angle=0,width=9.0cm]{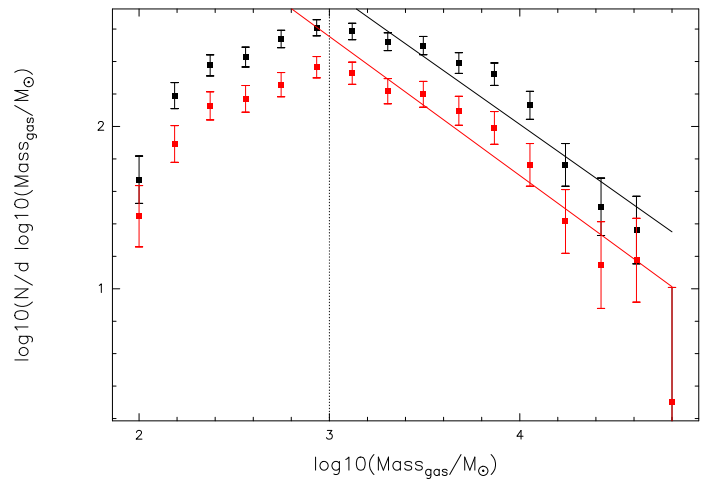}
\caption[mass spectrum]{Similar slopes result from power-law fits to the differential mass function of the whole ATLASGAL sample in black and of a cold subsample in red with the ratio of the MSX 21 $\mu$m flux to the 870 $\mu$m peak flux density smaller than 0.8. The vertical line displays the completeness limit of 1000 M$_{\odot}$, above which the mass spectra are fitted.}\label{msx mass spectrum}
\end{figure}

\begin{figure}[h]
\centering
\includegraphics[angle=0,width=9.0cm]{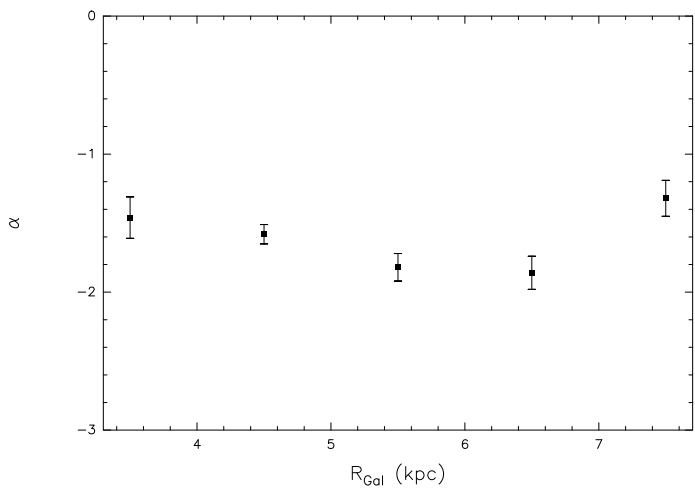}
\caption[mass spectrum]{Fitting of the differential mass distributions of ATLASGAL clumps above 1000 M$_{\odot}$ within 1 kpc over the range of galactocentric radii from 3 to 8 kpc gives similar power-law exponents, $\alpha$. There is a slightly decreasing slope of $\alpha$ with increasing galactocentric radius out to $\sim$ 7 kpc and an increasing slope to $\alpha = -1.3$ at 8 kpc.}\label{mass spectrum galradius}
\end{figure}

\begin{table*}[htbp]
\caption[]{Fit results of the mass spectra of different high mass star forming samples.}
\label{mass spectrum}
\centering
\begin{tabular}{l l l l c l}
\hline\hline
Sample & mass range (M$_{\odot}$) & distance range (kpc) & $M_{\mbox{\tiny min}}$\tablefootmark{1} (M$_{\odot}$) & $\alpha$ & Reference \\ \hline 
IRDC cores   &   $10 - 2100$ & $ 0.4 - 7.8$ &  100 &  $-2.1 \pm 0.4$ &  \cite{2006ApJ...641..389R} \\
IRDC clumps   &   $30 - 3000$ & $2 - 5$ &  40\tablefootmark{2} & $-1.76 \pm 0.05$  & \cite{2009ApJ...698..324R} \\
IRDCs  &  $0.1 - 10^5$ & $2.5 - 7.5$ & 800  & $-1.85 \pm 0.07$   & \cite{2010ApJ...723..555P} \\
1.2 mm dust clumps & $10 - 4100$ & $0.01 - 4$ & 92 & $-2.1 \pm 0.02$ & \cite{2006AA...447..221B} \\
GRSMC 45.60+0.30 clumps  & $4 - 1000$ & 1.8   &  15   & $-1.78 \pm 0.13$ & \cite{2001ApJ...551..747S} \\
GRSMC 43.30$-$0.33 clumps  & $10 - 1600$ & 3 & 40 & $-1.85 \pm 0.12$ & \cite{2001ApJ...551..747S} \\
GRSMC 45.46+0.05 clumps &  $25 -10^4$  & 6 & 250  & $-1.80 \pm 0.10$ & \cite{2001ApJ...551..747S} \\
W49  &  $250 - 4 \times 10^4$  & 11.4 &  1000   & $-1.52 \pm 0.20$  &  \cite{2001ApJ...551..747S}  \\ 
ATLASGAL  &  $0.4 - 5.6 \times 10^4$  & $0.2 - 17$ & 1000  & $-1.83 \pm 0.1$  & Sect. \ref{clump mass} \\
M17  &  $0.8 - 120$  &  1.6 & 3\tablefootmark{3}  & $-2.1 \pm 0.2$  & \cite{2010PASP..122..224S} \\ 
NGC 7538  &  $1.4 - 2700$  & 2.8 &  15\tablefootmark{3}  & $-2.1 \pm 0.36$  & \cite{2010PASP..122..224S} \\ 
Cygnus X  &  $4 - 949$  & 1.7 &  5.3\tablefootmark{3}  & $-2.1 \pm 0.2$  & \cite{2010PASP..122..224S} \\ 
NGC 6334  &  $3 - 6000$  & 1.7 &  32\tablefootmark{3}  & $-1.7 \pm 0.1$  & \cite{2010PASP..122..224S} \\ \hline 
\end{tabular}
\tablefoot{
\tablefoottext{1}{$M_{\mbox{\tiny min}}$ is the minimum mass, above which the distribution is fitted.}\\
\tablefoottext{2}{We use the mass $m_{\mbox{\tiny lim}}$ given in Table 1 in \cite{2009ApJ...698..324R} as the minimum mass.}\\
\tablefoottext{3}{\cite{2010PASP..122..224S} fit a double power law to the mass spectrum of IRDC clumps, we use the break mass as the minimum mass.}
}
\end{table*}

\subsection{How many ATLASGAL sources form high mass stars?}
\label{mass radius}
The derivation of the kinematic distances allows us to examine the range of gas masses and sizes of the ATLASGAL sample (see Sect. \ref{mass size}), which consists of different kinds of sources. To examine their star formation activity we have to connect masses and sizes, relations between the two have been obtained by previous studies. \cite{2010ApJ...723.1019H} fitted the star formation rate against surface densities of 20 molecular clouds, associated mainly with low mass star formation, as well as high mass dense clumps. The changing slope of the fit from a steep to a linear function yields a star formation threshold of $129 \pm 14$ M$_{\odot}$pc$^{-2}$. \cite{2010ApJ...724..687L} analysed the relation between the star formation rate of molecular clouds within 500 pc and the mass of the cloud. This revealed that the star formation activity does not depend on the total mass of the cloud, but on the extinction of the molecular gas and therefore volume density. \cite{2010ApJ...724..687L} obtained a linear relation between the star formation rate and the mass of a cloud above a threshold in visual extinction of $\sim 7$ mag, which is a gas surface density threshold of $\sim 116$ M$_{\odot}$pc$^{-2}$. 
The analysis of molecular clouds, which do not form high mass stars such as Taurus, Perseus, Ophiuchus, the Pipe Nebula, by \cite{2010ApJ...716..433K} revealed that their masses are limited by
\begin{eqnarray}\label{Kauffmann mass}
 M(r) = 580 {\rm M}_{\odot} (R_{\mbox{\tiny eff}} {\rm pc}^{-1})^{1.33}
\end{eqnarray}
with the gas mass, $M(r)$, and the effective radius, $R_{\mbox{\tiny eff}}$, as given in \cite{2013A&A...549A..45C} and \cite{2014arXiv1406.5741U}. \cite{2010ApJ...716..433K} decreased the dust opacities of \cite{1994A&A...291..943O} by a factor of 1.5, while we use the unmodified opacities in the calculation of the gas mass. This change is taken into account in equation \ref{Kauffmann mass}, where we reduce the factor 870 given in \cite{2010ApJ...716..433K} to 580. In contrast to the nearby clouds, two massive star forming complexes have masses larger than derived in equation \ref{Kauffmann mass}, which led \cite{2010ApJ...716..433K} to consider the relation as threshold for high mass star formation. However, this must still be confirmed by the study of larger samples of clouds that form high mass stars. Recent work by \cite{2013MNRAS.431.1752U} and \cite{2013MNRAS.435..400U} has analysed the mass-size relation of a large number of ATLASGAL sources, but it traces only the more evolved phases of massive star formation from methanol masers to UCHIIRs. In contrast, ATLASGAL clumps studied in this article are selected from the unbiased ATLASGAL sample and therefore include cold sources in an early, even prestellar, evolutionary stage as well as objects in later phases.

We show the correlation between the logarithm of the effective radius and of the gas mass of the ATLASGAL sample with derived kinematic distances in Fig. \ref{radius-mass}. The light blue line shows the limits in gas surface density of 116 M$_{\odot}$pc$^{-2}$ \citep{2010ApJ...724..687L} and $129 \pm 14$ M$_{\odot}$pc$^{-2}$ \citep{2010ApJ...723.1019H}, above which sources are likely to form stars. Using an average of the two thresholds, 122.5 M$_{\odot}$pc$^{-2}$, we obtain 3226 ATLASGAL clumps with masses larger than this limit, which is 100\% of the sources with determined gas masses. The comparison to BGPS sources \citep{2011ApJ...741..110D} reveals that a smaller number of them, $\sim 46$\% of the whole sample, satisfy the criteria from \cite{2010ApJ...724..687L} and \cite{2010ApJ...723.1019H} using the surface density of the clumps. However, this fraction will increase to $\sim 87$\% if they refer to smaller regions within the sources by taking the peak column density into account.

The red line in Fig. \ref{radius-mass} indicates the mass-radius relationship derived by \cite{2010ApJ...716..433K}. We obtain 2953 ATLASGAL sources, 92\% of the whole sample, which exhibit larger masses than expected from equation \ref{Kauffmann mass} and therefore might form high mass protostars. This fraction of ATLASGAL clumps is higher than the result from \cite{2011ApJ...741..110D}, who calculated that $\sim 48$\% of their sample have masses larger than the limit given by \cite{2010ApJ...716..433K}, and smaller than $\sim 97$\% of ATLASGAL sources identified as methanol masers \citep{2013MNRAS.431.1752U} that are potentially forming high mass protostars. While those trace later evolutionary phases from hot molecular cores to UCHIIRs, the ATLASGAL clumps analysed in this article and the BGPS sources are unbiased samples covering various evolutionary stages. Because these also include cold sources in an early phase, which show no sign of high mass star formation yet, they can explain the larger portion of clumps below the threshold from \cite{2010ApJ...716..433K} compared to ATLASGAL sources associated with methanol masers. \cite{2013MNRAS.431.1752U} obtain similar results from a fit to the mass-size relation of ATLASGAL sources associated with methanol masers and with compact HII regions. A fit to our data yields $\mathrm{log_{10}} (M) = 3.22 \pm 0.002 + (1.76 \pm 0.01) \times \mathrm{log_{10}} (R_{\mbox{\tiny eff}})$ in Fig. \ref{radius-mass}. Their comparison shows that the slope of our sample, which also contains cold clumps, is consistent within $3 \sigma$ with that of the sources in later evolutionary phases, although Fig. \ref{radius-mass} reveals the trend of a broader mass range of the ATLASGAL sample studied in this article extending down to lower values. The dashed black line in Fig. \ref{radius-mass}, which indicates the sensitivity of the ATLASGAL survey of 5$\sigma$, is a lower boundary to the derived gas masses.

\cite{2014A&A...565A..75C} used a limit of 650 M$_{\odot}$, above which ATLASGAL sources are likely to harbour massive dense cores and high-mass protostars. This is indicated as a dotted green line in Fig. \ref{radius-mass}. We find 2097 ATLASGAL clumps, 71\% of the sources with masses larger than the threshold given by \cite{2010ApJ...716..433K}, that also exhibit masses larger than 650 M$_{\odot}$.

\begin{figure}
\centering
\includegraphics[angle=0,width=9.0cm]{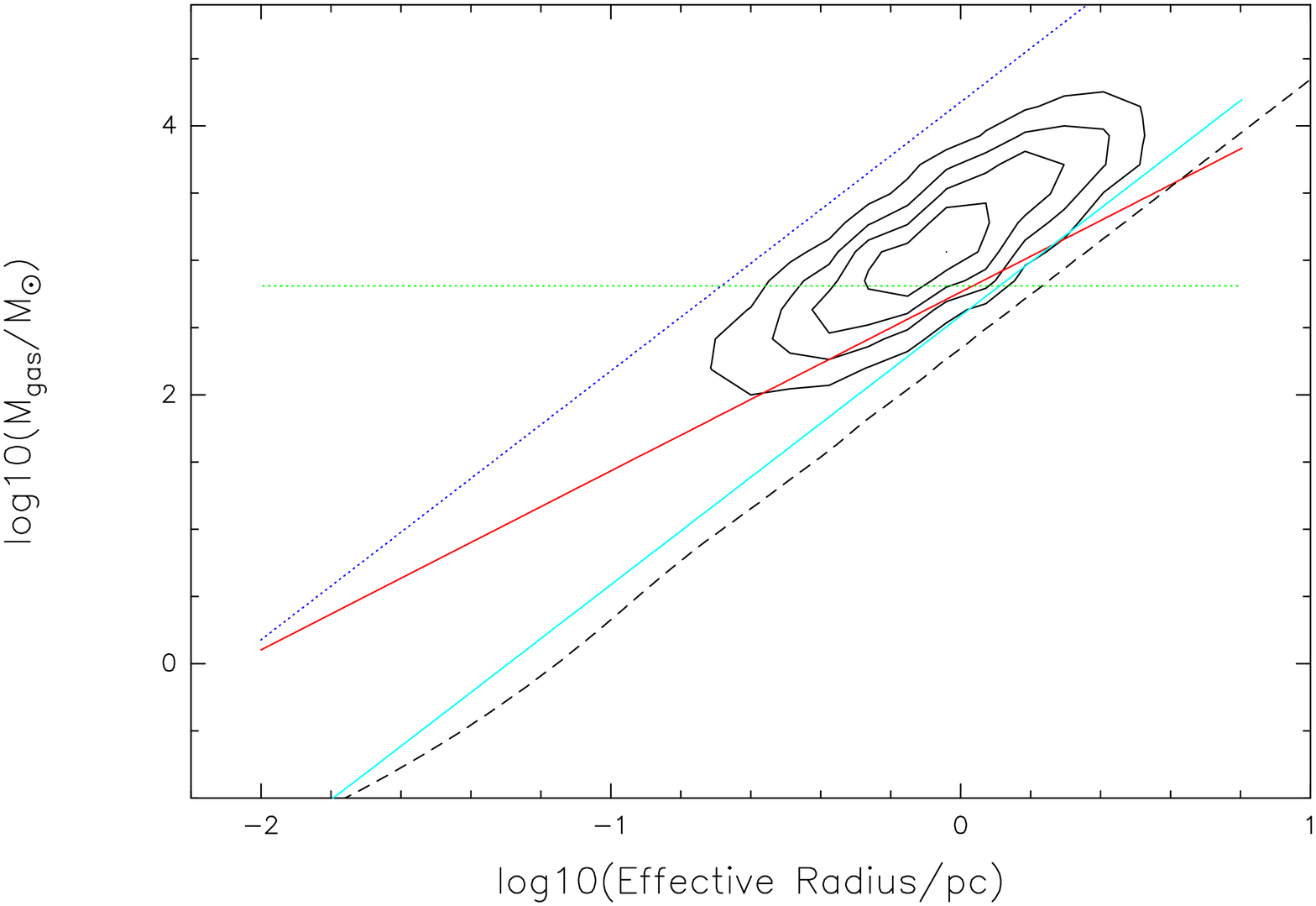}\vspace*{0.5cm}
\includegraphics[angle=0,width=9.0cm]{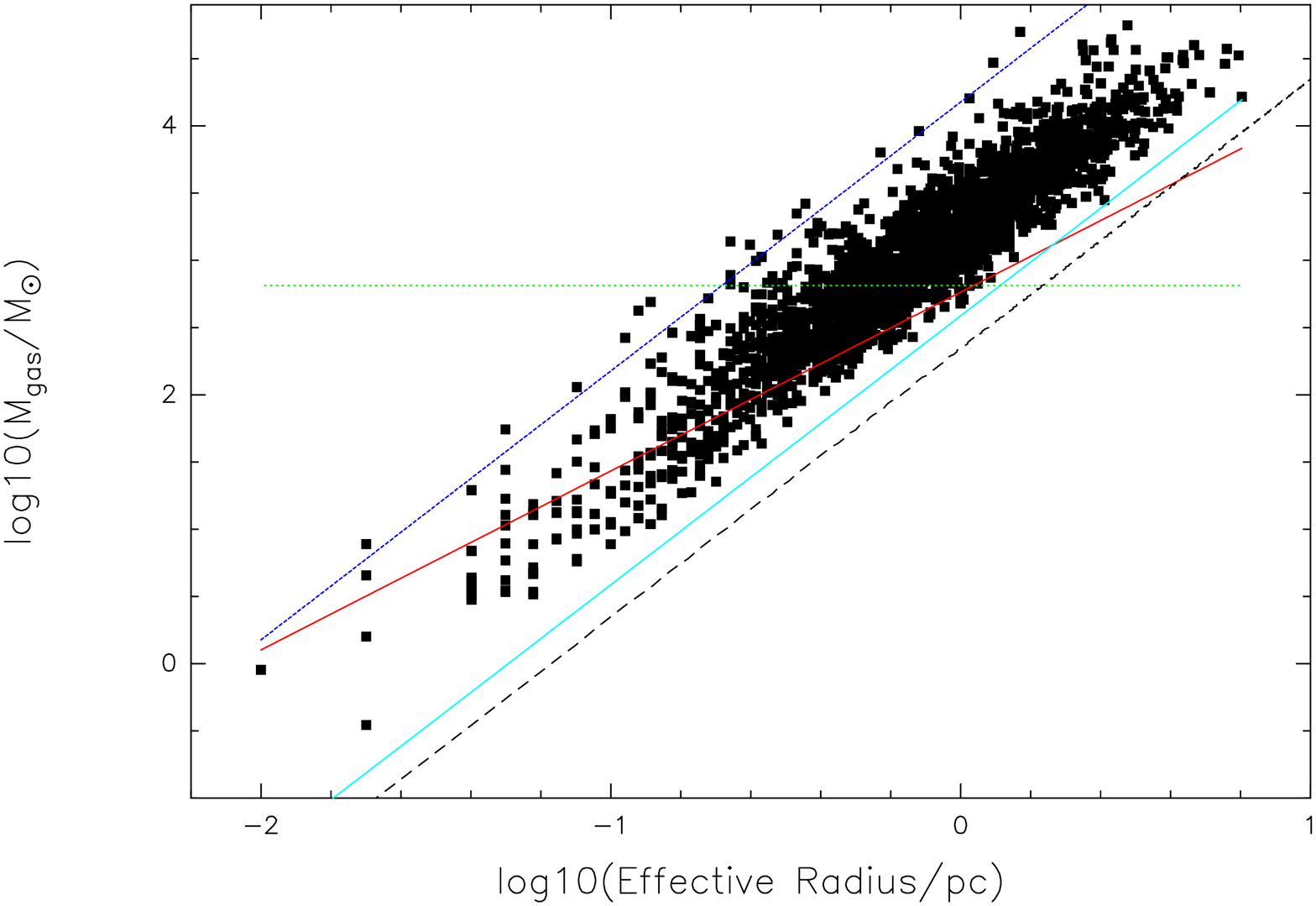}
\caption[mass spectrum]{Logarithm of the gas mass compared with the logarithm of the effective radius of ATLASGAL sources. The red line illustrates the threshold of massive star formation by \cite{2010ApJ...716..433K}. The dotted green line illustrates a threshold of 650 M$_{\odot}$ for the formation of high mass dense cores and massive protostars derived by \cite{2014A&A...565A..75C}. Limit in gas surface density as average of 116 M$_{\odot}$pc$^{-2}$ \citep{2010ApJ...724..687L} and $129 \pm 14$ M$_{\odot}$pc$^{-2}$ \citep{2010ApJ...723.1019H} is displayed as light blue line. A gas surface density of 1 g cm$^{-2}$ is shown as dotted blue line. The sensitivity of the ATLASGAL survey is displayed as dashed black line. The upper panel illustrates the contour plot with a binning of the logarithmic effective radius of 0.2 and the logarithmic gas mass of 0.2.}\label{radius-mass}
\end{figure}

The dotted blue line in Fig. \ref{radius-mass} shows masses corresponding to a constant surface density of 1 g cm$^{-2}$. This value is determined to be the limit above which fragmentation is suppressed and high mass stars form \citep{2008Natur.451.1082K}. The investigation of high mass star forming regions as well as star clusters with massive stars by \cite{2003ApJ...585..850M} revealed that the surface densities of $\sim 1$ g cm$^{-2}$ of these objects vary by a factor of 4. Because this threshold is derived for small-scale structures such as cores, while we are tracing larger objects with clump radii and smaller densities, only a small part of the ATLASGAL sources lie close to the dotted blue line. Details about clumps, which have high gas masses concentrated to small source radii, are given in Appendix \ref{extreme sources}.
 
\begin{figure}[!h]
\centering
\includegraphics[angle=0,width=9.0cm]{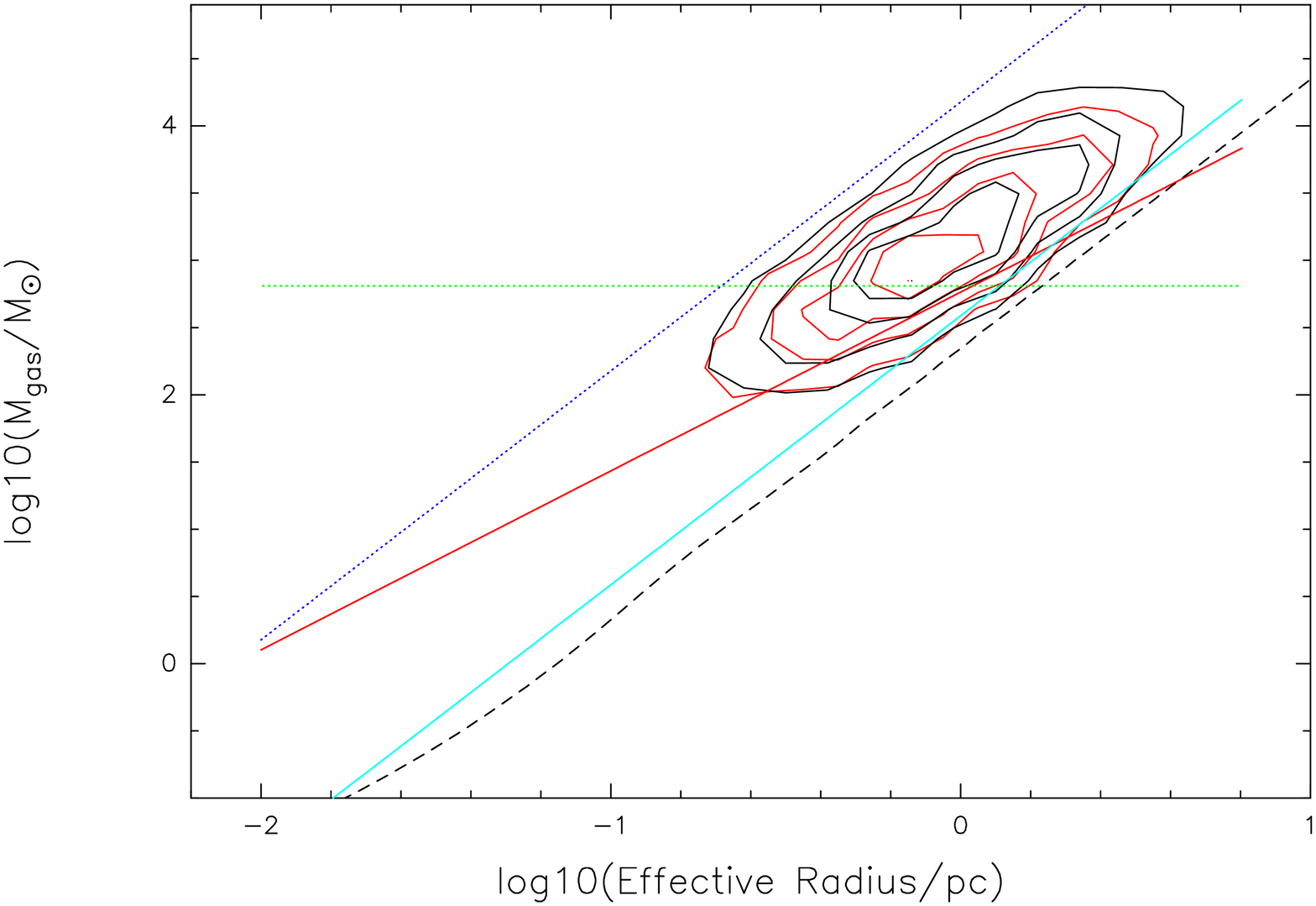}\vspace*{0.5cm}
\includegraphics[angle=0,width=9.0cm]{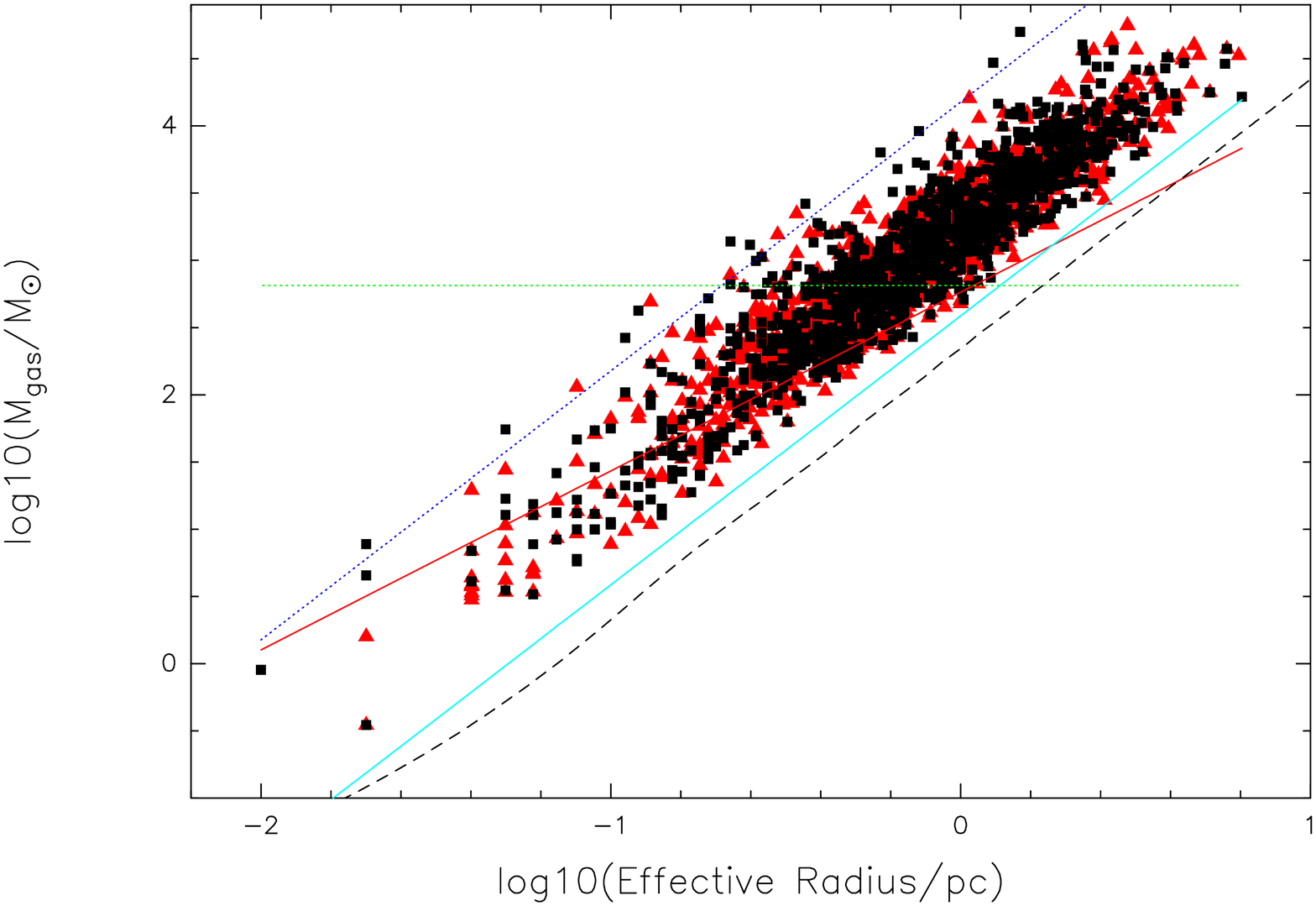}\vspace*{0.5cm}
\includegraphics[angle=0,width=9.0cm]{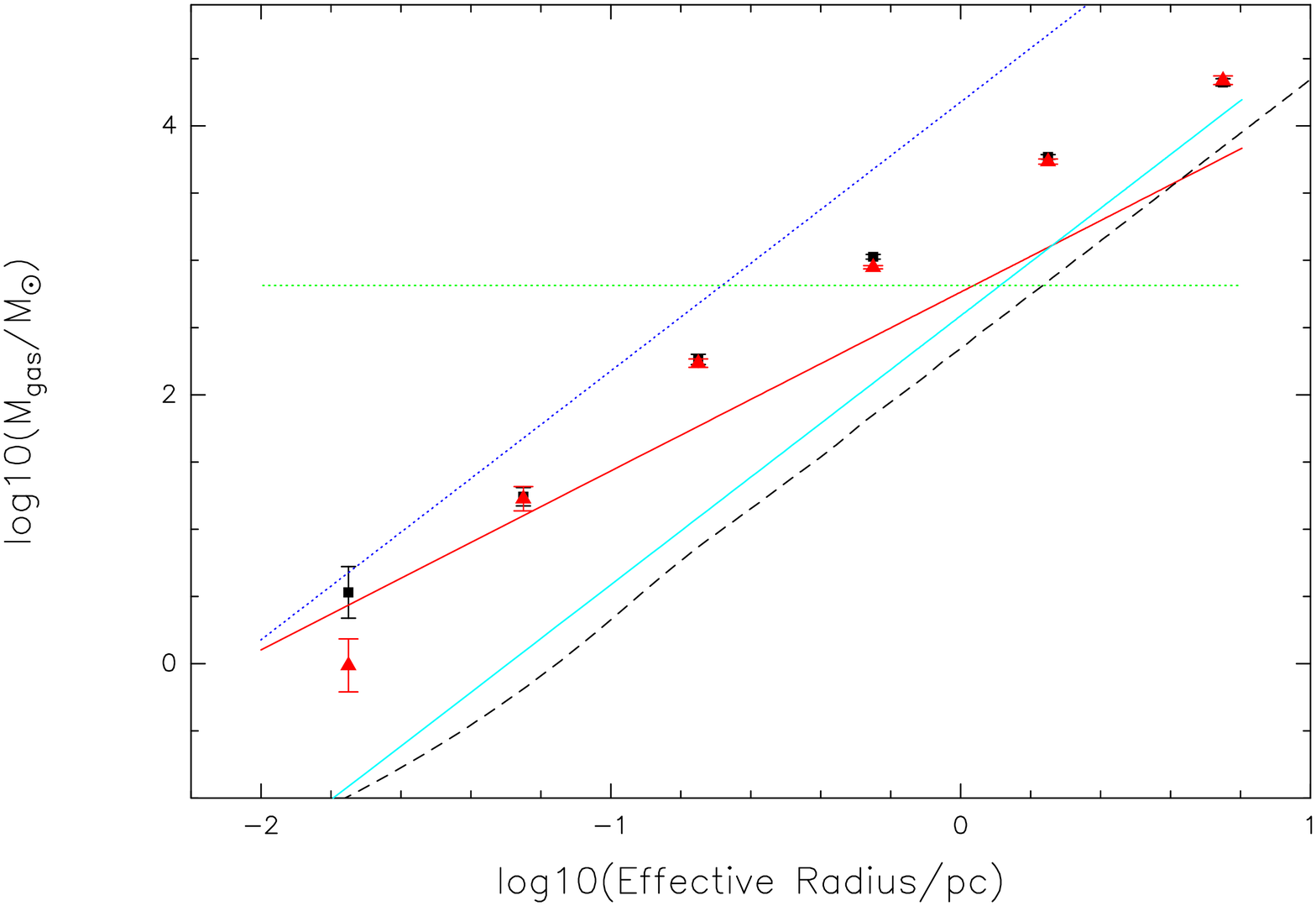}
\caption[mass spectrum]{The mass-size relation for a cold and a warm ATLASGAL subsample (see Sect. \ref{mass radius}) is shown as a red and black contour plot in the top panel, bins of 0.25 are used for the logarithm of the effective radius and bins of 0.2 for the logarithm of the gas mass. The scatter plot with the cold subsample as red triangles and the warm sources as black points is displayed in the middle panel. Mean values of the logarithm of the gas masses and of the effective radii are plotted in the bottom panel. The lines are the same as described in Sect. \ref{mass radius}.}\label{radius mass msx}
\end{figure}

To analyse if there is a trend of star formation activity with the mass-size relation of ATLASGAL clumps with derived kinematic distances we divide them into two samples with the ratio of the MSX 21 $\mu$m flux to the 870 $\mu$m peak flux density smaller or larger than 0.8. As described in Sect. \ref{cmf} this provides us a cold and a warm sample. We plot the logarithm of their gas mass against the logarithm of the effective radius as red triangles and black points in Fig. \ref{radius mass msx}. The two subsamples exhibit similar distributions. The lines illustrate the same relations as given in Fig. \ref{radius-mass}. In addition, we average the gas mass within bins of 0.5 in the logarithm of the effective radii to examine if there is a trend in the mean values, which are shown in the lower panel of Fig. \ref{radius mass msx}. However, we find no difference between the distributions of the two subsamples either.

In conclusion, most ATLASGAL clumps exhibit gas masses larger than the limit above which high mass stars form \citep{2010ApJ...716..433K}. This indicates that the majority of ATLASGAL clumps have the potential to form high mass protostars. In addition, the distribution of the mass-size relation is similar for a cold subsample in an early evolutionary stage without star formation activity as well as for warm clumps harbouring an embedded heating source.

\section{ Conclusions}
\label{conclusions}
We calculated near and far kinematic distances from the radial velocity of various molecular lines, mostly the NH$_3$ (1,1) inversion transition, to a large sample of dust clumps discovered by the ATLASGAL survey using the \cite{1993A&A...275...67B} rotation curve. This section summarizes our results.
\begin{enumerate}
\item To determine distances we first divide the ATLASGAL sample into groups of sources, which are coherent in space and velocity. We associate observed ATLASGAL sources with molecular clouds by comparing their location with $^{12}$CO emission probing molecular clouds on a large scale, with the $^{13}$CO (1$-$0) line in the first quadrant and 870 $\mu$m dust continuum tracing the small-scale structure. This allows us to assign additional ATLASGAL sources without any known velocities to the complexes of measured clumps and therefore the same distance.
\item Near and far kinematic distances to the complexes of ATLASGAL sources are computed using the rotation curve by \cite{1993A&A...275...67B}. To resolve the KDA we investigate HI self-absorption toward all ATLASGAL clumps and HI absorption toward 65 sources in the first quadrant with an embedded HII region and 146 sources in the fourth quadrant. We combine our KDA resolutions derived from the two techniques and obtain distances to 689 complexes. We locate 143 complexes in the first quadrant at the near distance, 148 at the far distance, and 5 at the tangent point, while 234 complexes in the fourth quadrant are assigned to the near distance, 141 to the far distance, and 18 at the tangent point.
\item While Galactocentric distances of the ATLASGAL subsample in the first quadrant show a concentration of sources at $\sim 4$ and 6 kpc, Galactocentric distances in the fourth quadrant reveal an increase of sources at 3, between 5 and 6 kpc, and $\sim 7$ kpc. These enhancements of clumps probe different spiral arms of the Milky Way.
\item An exponential fit to the number distribution of ATLASGAL sources with the distance from the Galactic plane gives a similar scale height of $\sim 28 \pm 2$ pc for the samples in the first and fourth quadrant. In addition, we obtain a similar displacement of ATLASGAL sources below the Galactic midplane in the first and fourth quadrants, at $\sim -10 \pm 0.5$ and $\sim -4 \pm 1.7$ pc. This is consistent with the results derived for other high mass star forming samples \citep{2011MNRAS.417.2500G,1990ApJ...358..485B}. The distribution of the scale height and the distance from the midplane are constant from 3 kpc to 8 kpc with respect to the outer Galaxy.
\item Using kinematic distances we calculate a broad mass and size range of the ATLASGAL sources. Within 2 to 18 kpc we trace clumps with 50 to 500 M$_{\odot}$ and radii from 0.15 to 1.5 pc. A fit of the differential clump mass function to the mass distribution of ATLASGAL sources between 2 and 5 kpc results in a power-law exponent of $-2.2 \pm 0.1$ on average. The cumulative mass function of ATLASGAL clumps is better represented by a lognormal distribution than by a double power law. Our slope of the differential mass distribution agrees with the power-law exponent of the IMF \citep{2008LNP...760..181K}, the mass function of clusters, associations, and that derived from high-density tracers as well as dust continuum emission for massive star forming regions \citep{2003ARA&A..41...57L,2006AA...447..221B,2001ApJ...551..747S}. Our fit to the mass spectrum of different ATLASGAL subsamples reveals that the power-law index does not vary from the early, cold phase to later evolutionary stages. In addition, the slope also remains similar within the inner Galaxy.
\item The comparison of mass and radius shows that the whole ATLASGAL sample has gas surface densities higher than 116 M$_{\odot}$pc$^{-2}$ \citep{2010ApJ...724..687L} and 129 M$_{\odot}$pc$^{-2}$ \citep{2010ApJ...723.1019H}, which is the threshold for star formation. Masses of 92\% of the ATLASGAL clumps are larger than the limit for high mass star formation given by \cite{2010ApJ...716..433K}. Most sources might therefore potentially form high mass protostars. A fit of the mass-size relationship by a power law results in a slope of $1.76 \pm 0.01$, which remains invariant for subsamples in different evolutionary phases.
\end{enumerate}

\noindent
\textit{Acknowledgements.} L.B. acknowledges support from CONICYT through project BASAL PFB-06. M.W. participates in the CSIRO Astronomy and Space Science Student Program and acknowledges support from the ATNF staff within the course of the observations. This work was partially funded by the ERC Advanced Investigator Grant GLOSTAR (247078).

\begin{appendix}
\section{Grouping of ATLASGAL sources}
\label{grouping details}
We give in this section details about our analysis to group sources, which are coherent in space and velocity using the friends-of-friends algorithm \citep{1982ApJ...257..423H,1993MNRAS.261..827M,2006ApJS..167....1B}. As a first step we sort ATLASGAL sources with known velocities (see Sect. \ref{molecular data}) in the first and fourth quadrant according to an increasing galactic longitude range. For each ATLASGAL source we search for related sources in the whole sample according to a predetermined threshold in position and velocity. Associated ATLASGAL sources are added to the same group until no more sources that lie within a maximum distance and velocity interval to the neighbouring sources are found. The group is then considered to be complete and a new search around the next ATLASGAL source of the catalogue begins. All distances between at least two individual sources within one group must not exceed the maximum distance and the velocities of these must be within a given velocity interval. However, the size and the maximum difference of velocities within a group can exceed the input values for the maximum distance and the velocity interval used to search for associated sources. This algorithm results in an arrangement of the ATLASGAL sample into groups and sources that do not belong to any group.

We study the dependency of the grouping on the input parameters by varying the maximum distance and the velocity interval. This results in various divisions of the ATLASGAL sources into groups, which are analysed to improve the matching between our grouping and properties of known molecular cloud complexes (see Appendix \ref{gmc}). For different input values for the maximum distance, $\Delta d_{\rm i}$, and a fixed interval of 10 km~s$^{-1}$ for velocities of neighbouring sources in a group we plot histograms of the size of a group in Fig. \ref{complex-size-atlasgal}. The size is defined as $(\Delta l \times \Delta b)^{1/2}$ with the maximum difference in longitude, $\Delta l$, and in latitude, $\Delta b$, of sources within a group. For $\Delta d_{\rm i} = 0.05^{\circ}$ or $0.1^{\circ}$ the size of most groups is smaller than those values, while the size of a larger number of groups is in the range of the input value $\Delta d_{\rm i} = 0.2^{\circ}$ or 0.3$^{\circ}$. If $\Delta d_{\rm i}$ is larger, 0.5$^{\circ}$, some groups will have sizes larger than $\Delta d_{\rm i}$. This trend is visible for the ATLASGAL samples in the first (upper panel of Fig. \ref{complex-size-atlasgal}) and fourth quadrant (lower panel of Fig. \ref{complex-size-atlasgal}), although it is more obvious for the groups in the fourth quadrant. To avoid gathering separate groups together, we do not take a large input value such as 0.5$^{\circ}$, but choose 0.3$^{\circ}$. In addition, the median size of GRS molecular clouds \citep{2009ApJ...699.1153R} of $\sim 0.3^{\circ}$ is also consistent with the maximum distance of ATLASGAL sources within a group, which corresponds to $\sim 22$ pc at a median distance of $\sim 4$ kpc. Using $^{12}$CO emission \cite{2014ApJS..212....2G} obtained even sizes up to 100 pc for giant molecular clouds in the fourth quadrant.

Figure \ref{complex-dv11-atlasgal} illustrates histograms of the maximum difference in velocities within a group for varying input values of the velocity interval used to search for neighbouring sources within a group, $\Delta \rm v_{\rm i}$, and a fixed maximum distances between individual sources in a group. Figure \ref{complex-dv11-atlasgal} indicates that with an increase in $\Delta \rm v_{\rm i}$ from 10 to 15 km~s$^{-1}$ the maximum velocity difference in a few groups becomes greater than the chosen $\Delta \rm v_{\rm i}$. This is similar to the trend seen in the maximum distance, but less obvious. Since the maximum velocity difference of most groups should not exceed the input value of $\Delta \rm v_{\rm i}$, we take 10 km~s$^{-1}$ for that. This analysis lets us choose the arrangement of ATLASGAL sources into groups resulting from 0.3$^{\circ}$ and 10 km~s$^{-1}$ as input values for the maximum distance and the maximum difference in velocities of neighbouring sources in a group. The maximum velocity difference of ATLASGAL sources in a group also agrees with the velocity dispersion of GRS molecular clouds \citep{2009ApJ...699.1153R} ranging up to 10 km~s$^{-1}$ and with the median linewidth of GMCs in the fourth quadrant \citep{2014ApJS..212....2G} of 9 km~s$^{-1}$.

We call sources that are associated by their position and velocity ''a group'' and refer to the group's spatial structure in a comparison with the ATLASGAL dust continuum emission as ''a complex''. We illustrate an example of a complex within $22.1^{\circ} < l < 22.6^{\circ}$ and $0.25^{\circ} < b < 0.5^{\circ}$, which results from our method, in the top panel of Fig. \ref{HISA}.

This method results in 296 groups in the first quadrant and 393 groups in the fourth quadrant, an uncertainty associated with these identifications might result from gathering separated groups along the line of sight together into one 
group.

\subsection{Properties of groups}
To characterize the groups we study some physical properties such as the velocity dispersion and physical radius. The distribution of the number of sources per group is illustrated in panel a) of Fig. \ref{vdisp radius}, the histogram of the velocity dispersion in panel b) and of the physical radius of the groups in panel c) using distances determined in Sect. \ref{kinematic distance}. We obtain $\sim$ 300 sources in the first and fourth quadrants that within 0.3$^{\circ}$ and 10 km~s$^{-1}$ are not related to another ATLASGAL source, as indicated in panel a) of Fig. \ref{vdisp radius}. The histograms of the velocity dispersion and physical radius of the groups containing $\geq 2$ sources show that most groups exhibit a velocity dispersion of $< 2$ km~s$^{-1}$ and radius of $\sim 5$ pc. The peak in the radius distribution mainly results from a subsample of about 100 groups, which contain only two sources per group. In contrast, the peak of the velocity dispersion is a property of the whole sample. Figure \ref{vdisp radius} reveals a large dispersion of radii, which range to 80 pc. We obtain 12 groups out of the whole sample of 689 with radii larger than 35 pc, most of them consist of more than ten sources. Some of the large radii belong to groups at the far distances. 

The correlation of the physical radius and velocity dispersion of all 709 groups is shown in Fig. \ref{size-linewidth}. A power-law relationship between the two properties is indicated as black line, which is given by \cite{1981MNRAS.194..809L}
\begin{eqnarray}
 \delta \rm \upsilon = 1.43 R^{0.38},
\end{eqnarray}
where $\delta \rm \upsilon$ is the velocity dispersion and $R$ the radius. Some of the groups are located close to the relationship, but many groups lie in a broad distribution. To describe the whole sample we added a relation for a gravitationally bound cloud, resulting from a re-examination of Larson's law by \cite{2011MNRAS.411...65B}
\begin{eqnarray}
\label{vdisp-sigma}
 \delta \rm \upsilon = \sqrt{2G \Sigma R}
\end{eqnarray}
with the H$_2$ column density $\Sigma$. We plot Eq. \ref{vdisp-sigma} for H$_2$ column densities of $10^{21}$ cm$^{-2}$ as green line, $10^{22}$ cm$^{-2}$ as red line, and $5 \times 10^{22}$ cm$^{-2}$ as blue line in Fig. \ref{size-linewidth}. The distribution of radius and velocity dispersion can be fitted by a power law with an exponent of $0.4 \pm 0.02$, which is shown as purple line. The slope is thus consistent with the power-law exponent given in \cite{1981MNRAS.194..809L}. We performed a Spearman correlation test for an ATLASGAL subsample with heliocentric distances between 2 and 5 kpc, which gives a correlation coefficient of 0.57 with a p-value < 0.0013. This indicates that the correlation between radius and velocity dispersion is significant over 3$\sigma$. A few groups are below the green line and exhibit extremely small velocity dispersions, $< 0.1$ km~s$^{-1}$, because they mostly include only fewer than five sources with observed velocities. Because these do not have good statistics, we aim at describing groups associated with a larger number of measured velocities. A contour plot is presented in the upper panel of Fig. \ref{size-linewidth} with a binning of 0.3 for the logarithm of the radius and for the logarithm of the velocity dispersion. It shows a trend of growing radii with increasing velocity dispersions and most groups follow the law of \cite{2011MNRAS.411...65B}. 

\captionsetup[subfigure]{position=top}
\begin{figure}[tbp]
\centering

\subfloat[\hspace*{10cm}]{\includegraphics[angle=-90,width=9.0cm]{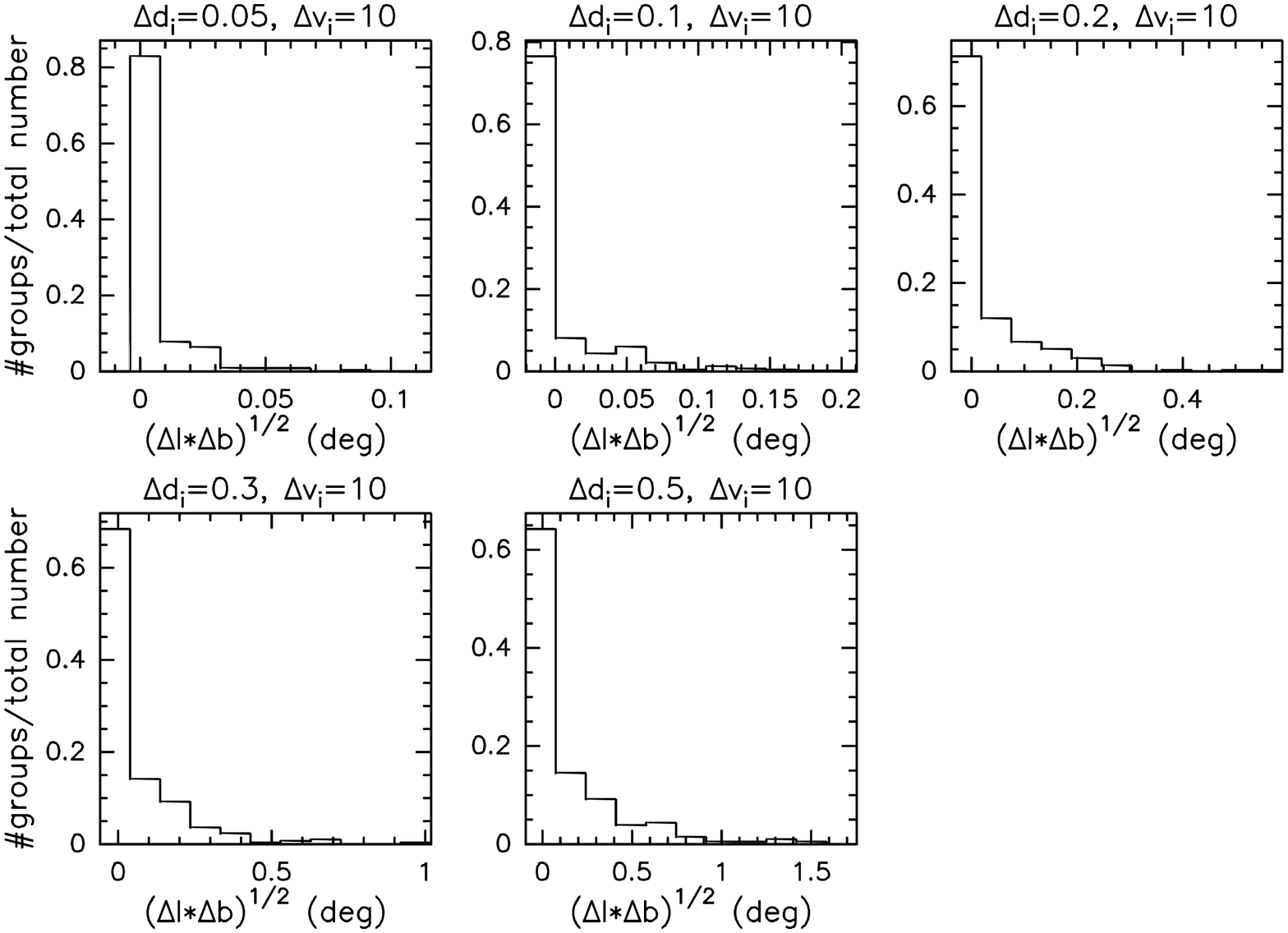}}\vspace*{0.5cm}

\subfloat[\hspace*{10cm}]{\includegraphics[angle=-90,width=9.0cm]{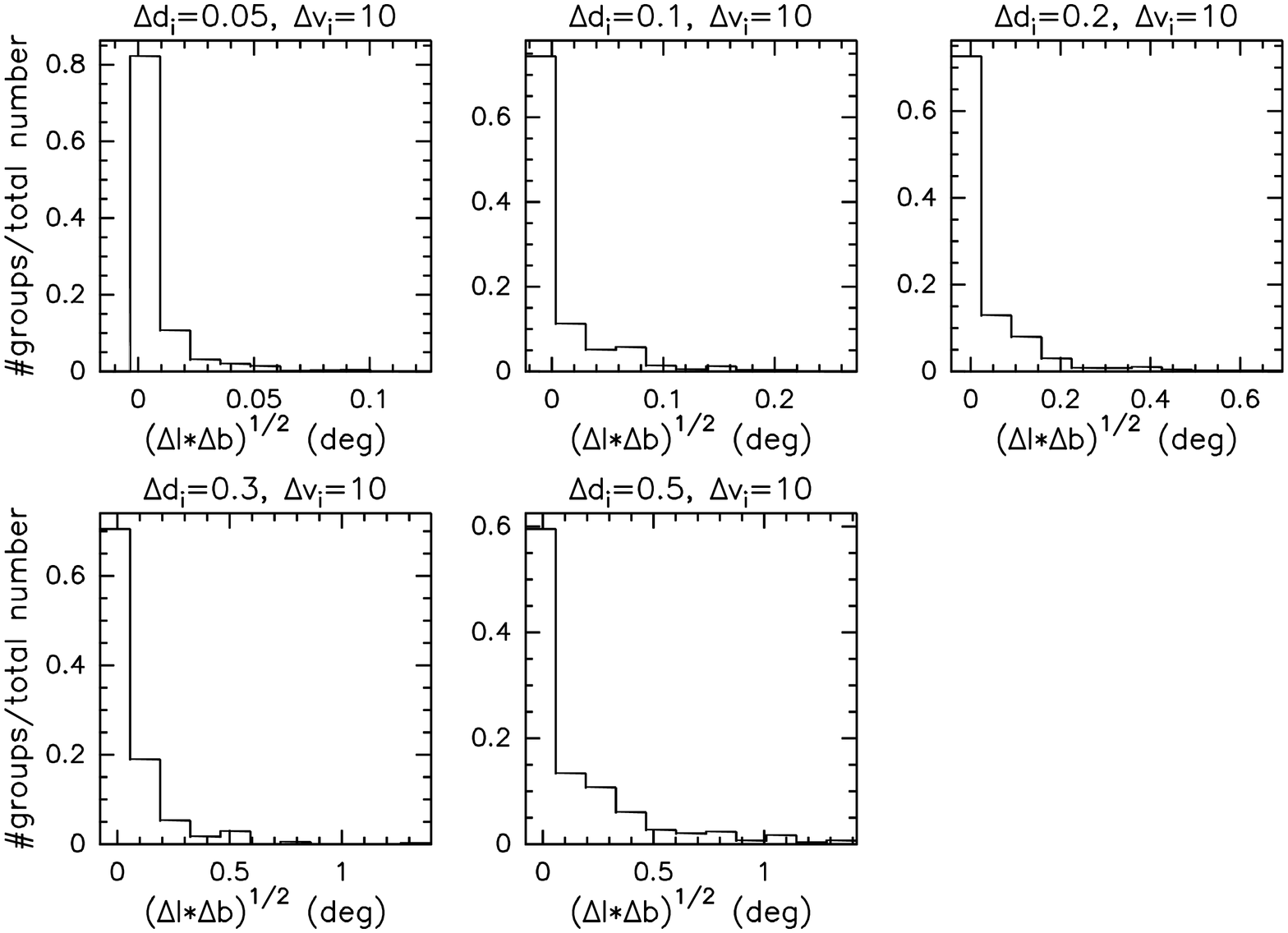}}

\caption[histogram of the size of a complex]{Relative number distribution of ATLASGAL groups with the size of a group for different input values of the maximum distance between individual sources within a group, $\Delta d_{\rm i}$. This is varied between 0.05$^{\circ}$ and 0.5$^{\circ}$ with a fixed interval for velocities of neighbouring sources in a group, $\Delta \rm v_{\rm i}$, of 10 km~s$^{-1}$ for the sample in the first quadrant in panel \textbf{a)} and the groups in the fourth quadrant in panel \textbf{b)}.}\label{complex-size-atlasgal}
\end{figure}

\captionsetup[subfigure]{position=top}
\begin{figure}[tbp]
\centering

\subfloat[\hspace*{10cm}]{\includegraphics[angle=-90,width=9.0cm]{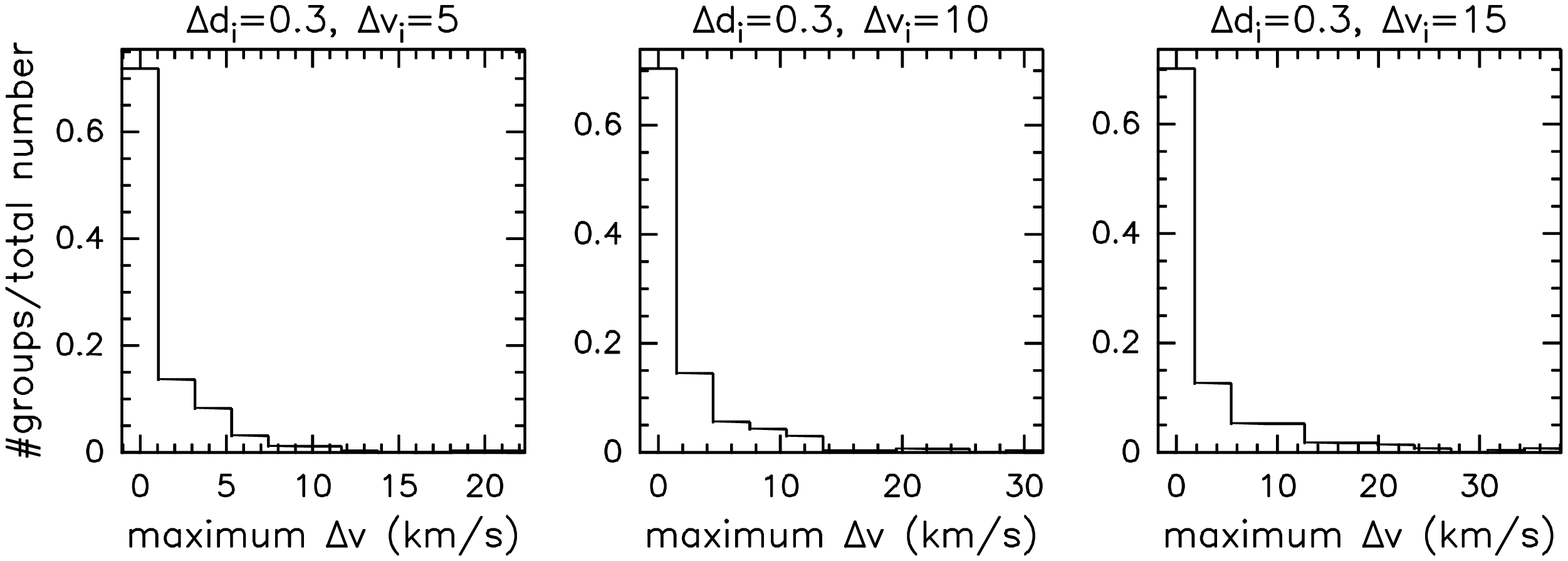}}\vspace*{0.5cm}

\subfloat[\hspace*{10cm}]{\includegraphics[angle=-90,width=9.0cm]{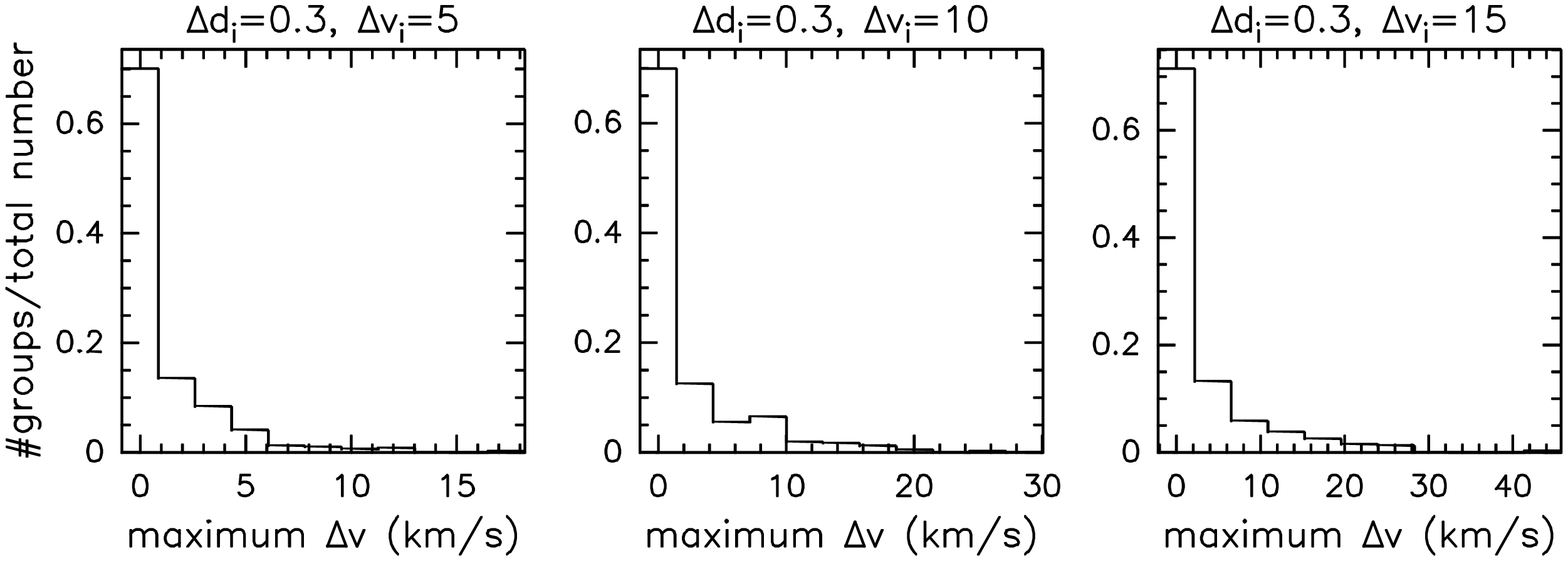}}

\caption[histogram of the maximum velocity difference of a complex]{Relative number distribution of ATLASGAL groups in the first quadrant in panel \textbf{a)} and groups in the fourth quadrant in panel \textbf{b)} with the maximum velocity difference in a group for different input values of the velocity interval to search for associated sources in a group, $\Delta \rm v_{\rm i}$. We use 5 to 15 km~s$^{-1}$ for that and a fixed maximum distance between individual sources in a group, $\Delta d_{\rm i}$, of 0.3$^{\circ}$.}\label{complex-dv11-atlasgal}
\end{figure}

\captionsetup[subfigure]{position=top}
\begin{figure}[tbp]
\centering

\subfloat[\hspace*{10cm}]{\includegraphics[angle=0,width=9.0cm]{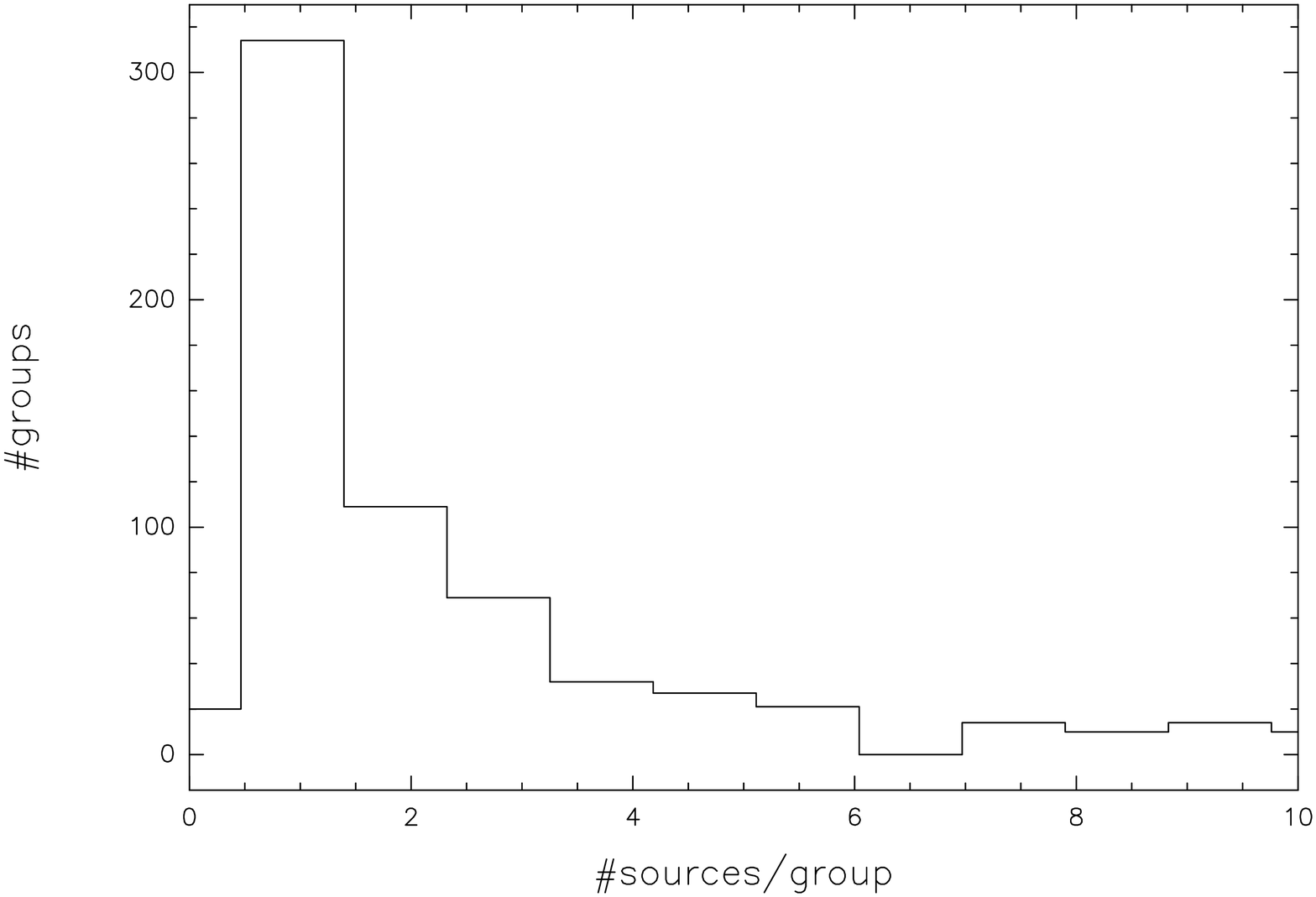}}\vspace*{0.5cm}

\subfloat[\hspace*{10cm}]{\includegraphics[angle=0,width=9.0cm]{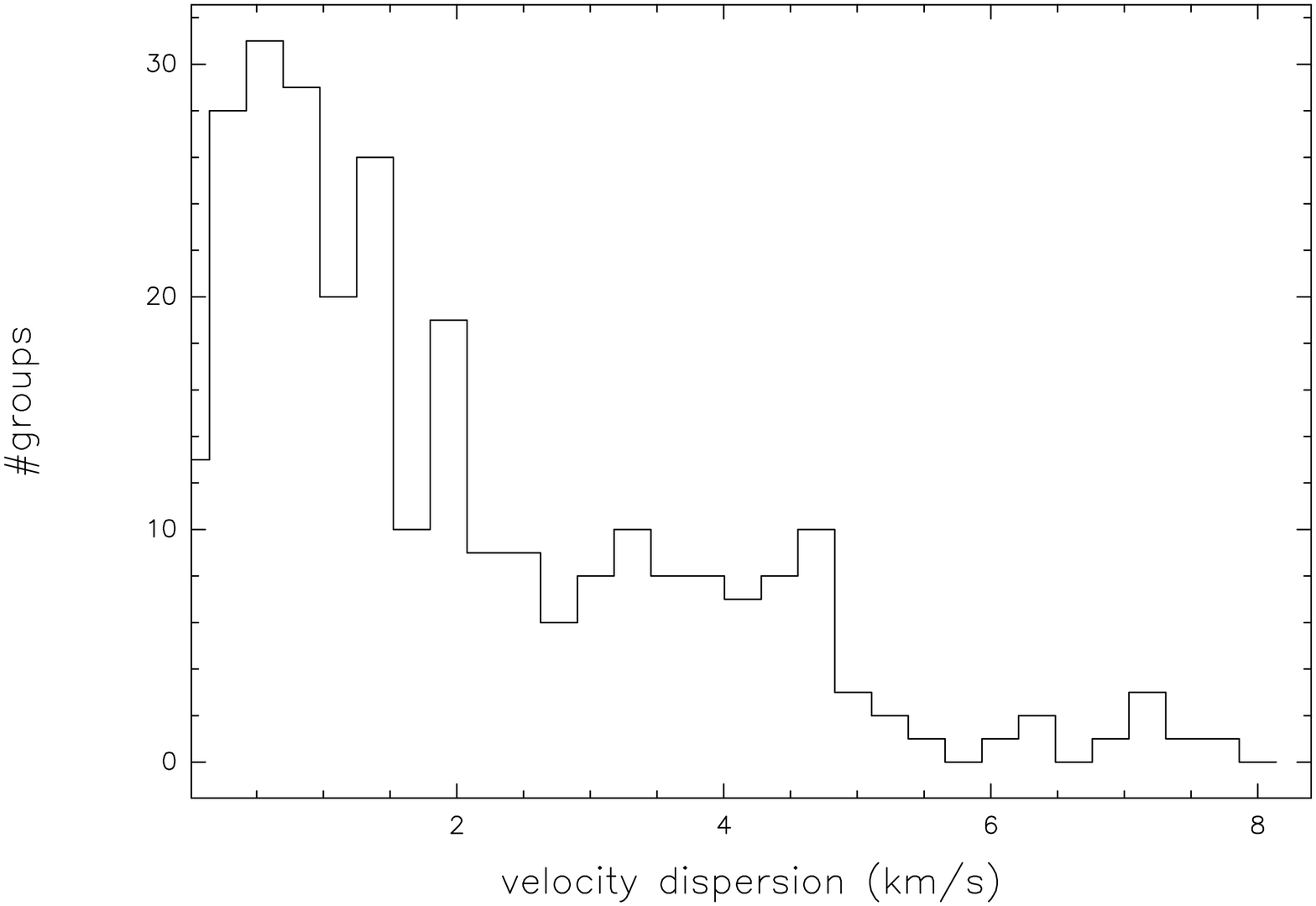}}\vspace*{0.5cm}

\subfloat[\hspace*{7cm}]{\includegraphics[angle=0,width=9.0cm]{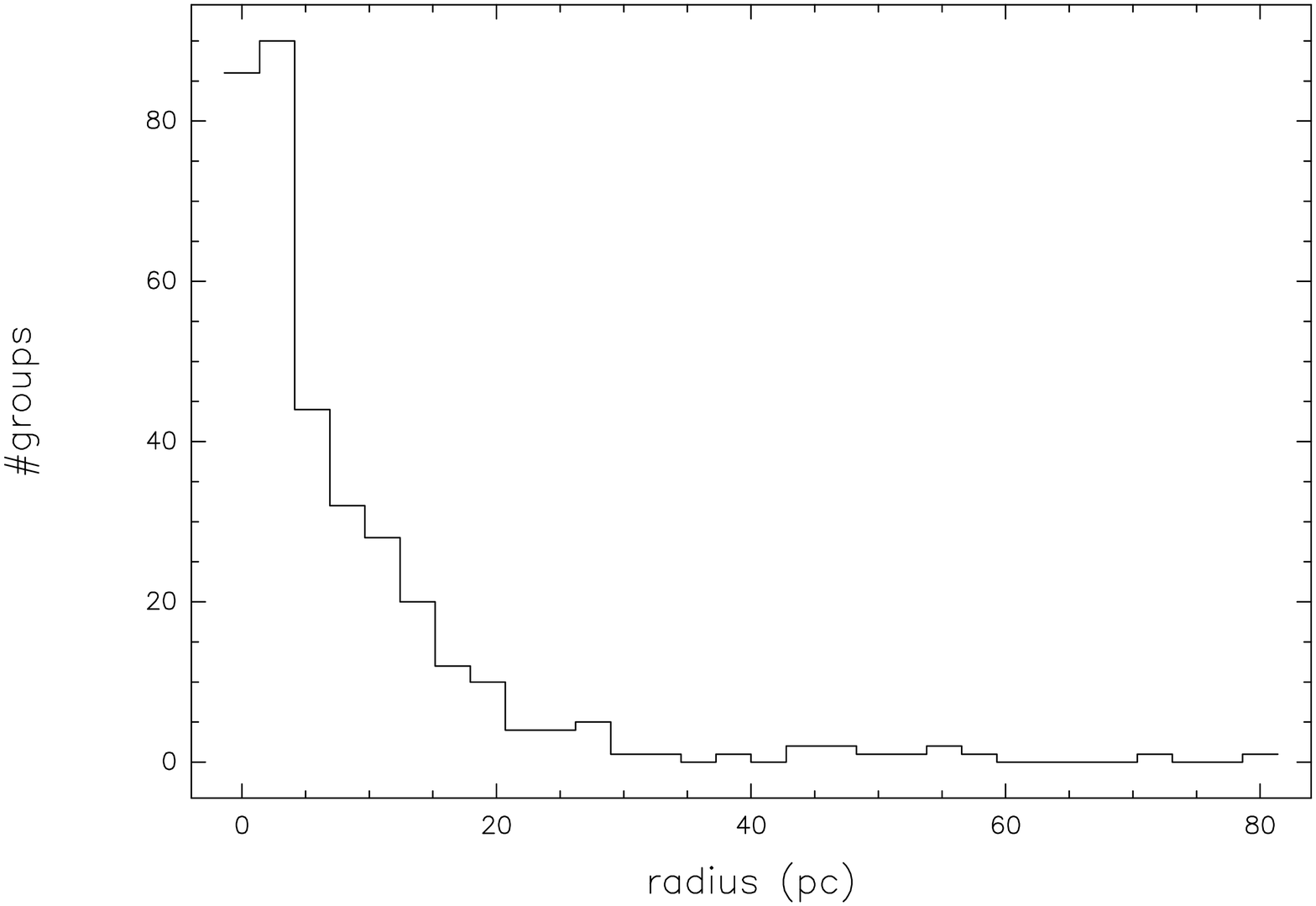}}\vspace*{0.5cm}

\caption[vdisp radius distribution]{Number distribution of ATLASGAL groups with the number of sources per group in \textbf{a)}, the velocity dispersion in \textbf{b)}, and radius in \textbf{c)}.}\label{vdisp radius}
\end{figure}

\begin{figure}
\centering
\includegraphics[angle=0,width=9.0cm]{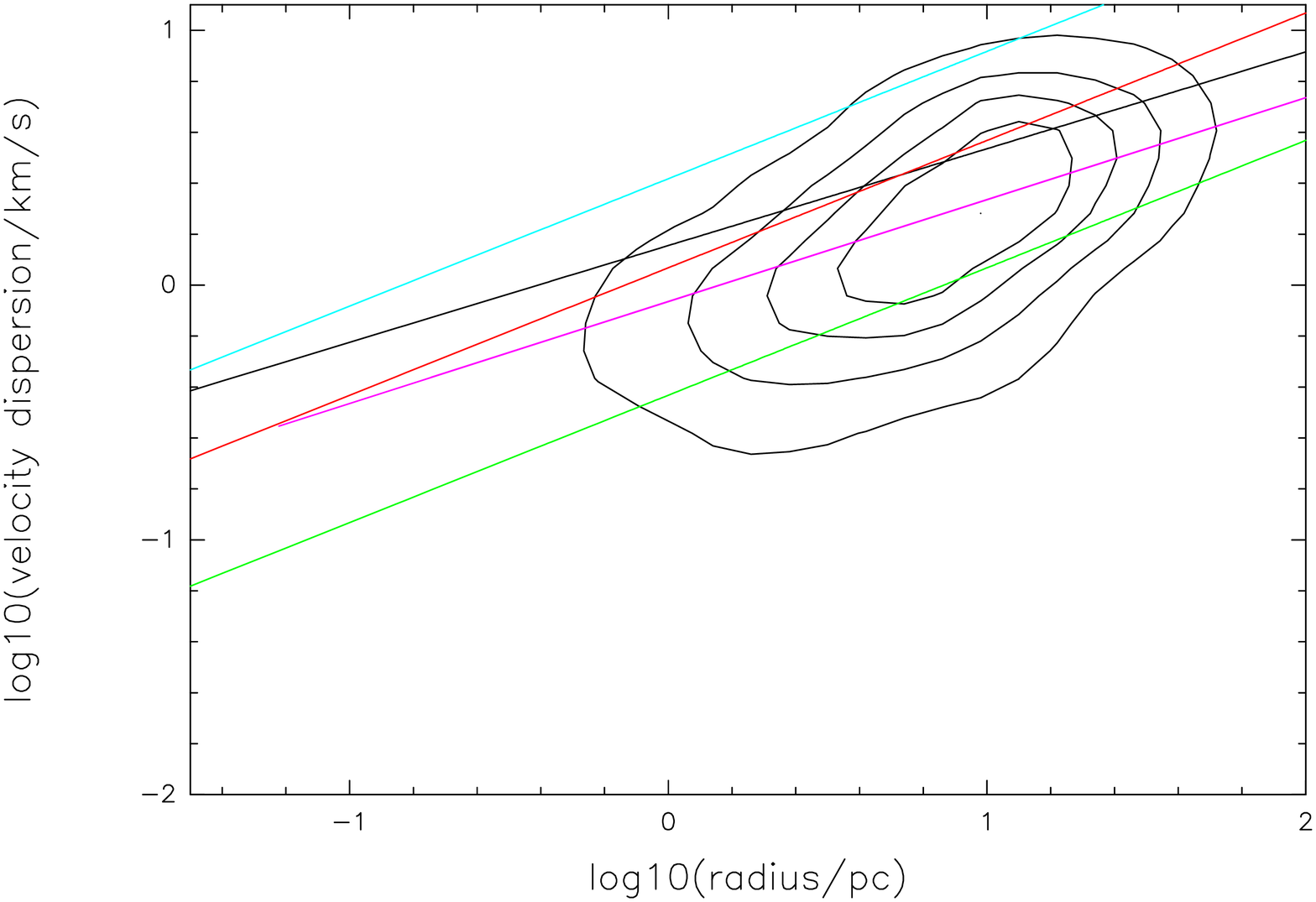}
\includegraphics[angle=-90,width=9.0cm]{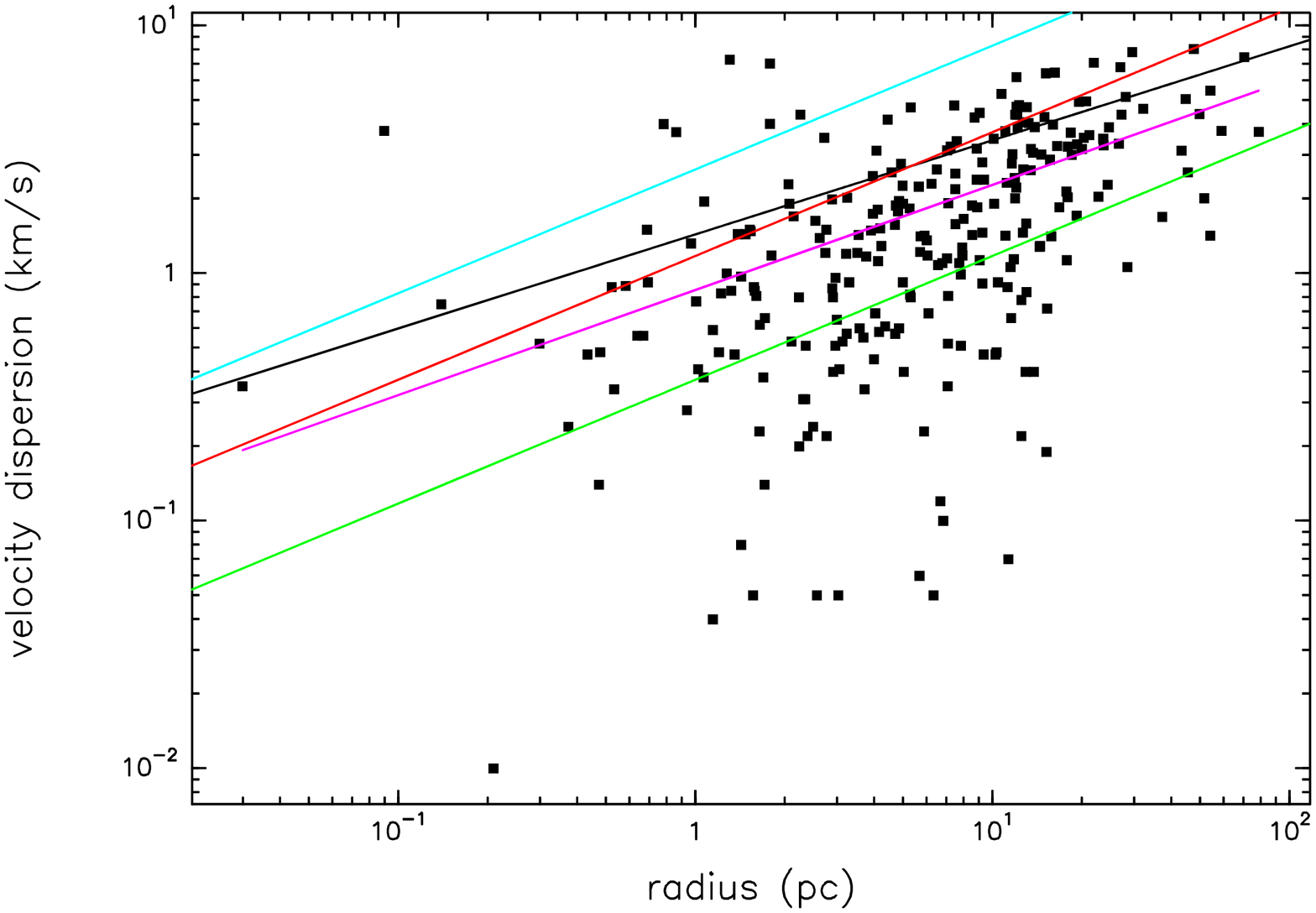}
\caption[size linewidth relation]{Correlation plot of the radius and velocity dispersion of all groups. The black line indicates the relation by \cite{1981MNRAS.194..809L}, the green, red, and blue lines show the formula by \cite{2011MNRAS.411...65B} for an H$_2$ column density of 10$^{21}$ cm$^{-2}$, 10$^{22}$ cm$^{-2}$, and $5\times 10^{22}$ cm$^{-2}$. The purple line shows a power-law fit to the data with an exponent of $0.4 \pm 0.02$. The contour plot is shown in the upper panel with the range of the logarithm of the radius and the velocity dispersion divided into bins of 0.3.}\label{size-linewidth}
\end{figure}

\section{Notes on individual sources}
\subsection{Sources with extreme properties}
\label{extreme sources}
\subsubsection*{Clumps with high masses}
The comparison of the kinematic distance with the gas mass in Fig. \ref{mass-size} reveals some extreme sources clustering at different distances that are known regions of massive star formation. Two clumps exhibiting a mass of $3.6 \times 10^4$ M$_{\odot}$ at a distance of 4.5 kpc are in the high mass star forming complex W33 \citep{1978A&A....64..341B}. Our KDA resolution yielding the near distance is consistent with trigonometric parallax measurements resulting in a distance of 2.53 kpc \citep{2013A&A...553A.117I}. The complex of sources at 5.5 kpc with high masses between $2 \times 10^4$ and $5 \times 10^4$ M$_{\odot}$ are located in W51 main. A similar distance of 5.4 kpc is derived from the trigonometric parallax \citep{2012IAUS..287..423S} (see Appendix \ref{gmc}). A cluster of ATLASGAL clumps at a distance around 8 kpc with gas masses higher than $10^4$ M$_{\odot}$ are in the W43 complex and W43 south (see Appendix \ref{gmc}). In addition, the source with a mass of $5.6 \times 10^4$ M$_{\odot}$ at 11 kpc is G10.47+0.03, an ultracompact HII region \citep{1989ApJS...69..831W} in the W31 complex (see Appendix \ref{gmc}).\\

\subsubsection*{Sources with large distances from the Galactic plane}
To investigate if individual complexes dominate extreme values for the distance from the Galactic mid-plane, an illustration of its radial distribution is shown for the ATLASGAL sample in the first quadrant in the top panel and for sources in the fourth quadrant in the lower panel of Fig. \ref{height samples}. Among the sources, which exhibit extremely high distances from the mid-plane indicated as red triangles, are the clumps at R$_{\mbox{\tiny Gal}} \approx 5.4$ kpc and exhibiting a height larger than 200 pc below the plane in the upper panel. These are located in complex 4 at $l \approx 6.02^{\circ}$ and $b \approx -1.23^{\circ}$ (see Table \ref{v11-dist}). Because none of these sources contains an embedded HII region, we cannot analyse the HI absorption and determine the far kinematic distance from missing HI self-absorption, which is more uncertain than using the two methods. The complex is also known as NGC 6530, which is a young open cluster, for which \cite{2001AstL...27..386L} derived a distance between 560 and 711 pc from Hipparcos trigonometric parallaxes, which would locate the region at a smaller height below the disk. We can also use only the HISA method for the clump at R$_{\mbox{\tiny Gal}} \approx 6.8$ kpc with a height of 154 pc below the plane, located in complex 262 at $l = 49.044^{\circ}$ and $b \approx -1.078^{\circ}$, for which we determine the far kinematic distance. This source is associated with IRAS 19230+1341 and $6.7-$GHz methanol maser emission \citep{2005A&A...432..737P}. Moreover, there is a complex of sources at  R$_{\mbox{\tiny Gal}} \approx 5$ kpc, which are located at a distance of 158 pc below the mid-plane. One of them at $l = 17.608^{\circ}$ and $b = -0.726^{\circ}$, also known as IRAS 18227$-$1358, is in complex 75, which consists only of this source. The other two clumps at $l \approx 18.30^{\circ}$ and $b \approx -0.72^{\circ}$ are in complex 80, which contains no embedded HII region. We thus locate them at the far distance from missing HI self-absorption. Because the displacement from the Galactic plane of star forming regions is confined to $\approx 120$ pc \citep{2011MNRAS.410.1237U}, the assignement of the far distance is unrealistic and the complexes are likely at the near distance. In addition, the HISA method is less reliable for the far distance than for the near distance. Moreover, the sources at a large height above the plane are not associated with 21 cm continuum sources either and we also use the HISA method to determine the far distance to them. Among those are the clump at R$_{\mbox{\tiny Gal}} \approx 4.6$ kpc with a distance of $\sim 185$ pc from the disk, located in complex 58 at $l \approx 15.029^{\circ}$ and $b \approx 0.851^{\circ}$, and one at R$_{\mbox{\tiny Gal}} \approx 5.5$ kpc and a height of 178 pc, which is in complex 33 at $l \approx 11.92^{\circ}$ and $b \approx 0.753^{\circ}$. Approximately 7\% of the sources in the first and fourth quadrant, which we assign to the far distance, have a displacement larger than 120 pc from the plane. We keep the far distance for these clumps to use a consistent method to resolve the KDA for all ATLASGAL sources. Statistically this small fraction does not influence the Galactic distribution.

Clumps in the fourth quadrant (see lower panel of Fig. \ref{height samples}) with an extremely high displacement from the plane of 210 pc and a galactocentric radius of 5.8 kpc are in complex 187 ($l = 330.044^{\circ}, b = 1.050^{\circ}$). These sources are embedded in the IRDC G330.030+1.019 \citep{2009A&A...505..405P}, which hints at the near distance, while we determined the far distance for this complex. However, our KDA resolution is uncertain because we could use only the HISA method, which is less reliable for the far distance solution than for the near distance. Additional complexes, which we locate at the far distance based on this method and which are therefore at a large height above the Galactic plane are complex 186 ($l = 329.718^{\circ}, b = 0.804^{\circ}$) with only one source at R$_{\mbox{\tiny Gal}} \approx 5.8$ kpc and a height of 160 pc. It is associated with the IRDC G329.717+0.785 \citep{2009A&A...505..405P}. Other examples are complex 280 ($l = 340.088^{\circ}, b = 0.928^{\circ}$) at R$_{\mbox{\tiny Gal}} \approx 4.2$ kpc and a distance from the plane of 180 pc, embedded in the IRDC G340.094+0.929 \citep{2009A&A...505..405P}, as well as complex 360 ($l = 350.816^{\circ}, b = 0.514^{\circ}$) at R$_{\mbox{\tiny Gal}} \approx 8.4$ kpc and a displacement from the plane of 150 pc, located within the IRDC G350.816+0.513 \citep{2009A&A...505..405P}.

\subsubsection*{Complexes with extremely large scale heights}
The ATLASGAL sample within a galactocentric radius between 4.5 and 6.5 kpc shows a flat scale height distribution in Fig. \ref{scale height galradius}, resulting mainly from a few complexes, which contain several sources. Approximately 7\% of the sources in the first quadrant in the galactocentric radius bins from 4 to 7 kpc are in complex 261 ($l = 49.2^{\circ}, b = -0.7^{\circ}$) (see Appendix \ref{gmc}) with a mean Galactic latitude of $-0.31^{\circ}$. It is located at a kinematic distance of 5.54 kpc and at a mean height of 30.25 pc below the Galactic plane. Two additional large complexes in the first quadrant are close to the Galactic plane and thus have small heights: Complex 172 ($l = 30.767^{\circ}, b = -0.050^{\circ}$) (see Appendix \ref{gmc}) contains $\sim 8$\% of the sources in the first quadrant with a mean latitude of 0.07$^{\circ}$, a distance of 8.4 kpc, and a mean height of 10.67 pc. Approximately 8\% of the clumps in the first quadrant are in complex 167 ($l = 29.96^{\circ}, b = -0.2^{\circ}$) (see Appendix \ref{gmc}) with a mean latitude of $-0.04^{\circ}$, a distance of 8 kpc, and a mean height of 5.84 pc below the mid-plane. In the fourth quadrant complex 242 ($l = 336.97, b = -0.010$) consists of $\sim 10$\% of the sources within a galactocentric radius between 4 and 7 kpc. It is located at a mean latitude of 0.01$^{\circ}$, nearby at a kinematic distance of 4.7 kpc, and thus has a small mean height of 1 pc below the Galactic plane. Complex 210 ($l = 333.19^{\circ}, b = -0.36^{\circ}$) (see Appendix \ref{gmc}) has the largest contribution, $\sim 12$\%, to the clumps in the fourth quadrant in the bins from 4 to 7 kpc. It exhibits a mean latitude of $-0.38^{\circ}$ and a distance of 3.52 kpc, which results in a mean height of 23.37 pc below the mid-plane.

\subsubsection*{Sources with high surface densities}
The effective radius is compared with the gas mass of the ATLASGAL sample in Fig. \ref{radius-mass}. Some sources are located close to the dotted blue line, which indicates a surface density of $\sim 1$ g cm$^{-2}$. Among those sources with high gas masses concentrated to small source radii is G35.2$-$0.74 with a mass of 423 M$_{\odot}$ and an effective radius of about 0.1 pc, which is associated with compact HII regions detected at 6 cm by \cite{1984MNRAS.210..173D}. Another clump at $l = 348.604^{\circ}, b = -0.912^{\circ}$, which has a mass of 114 M$_{\odot}$ and a radius of 0.08 pc, also exceeds the surface density of 1 g cm$^{-2}$. It is an UCHIIR, which is also observed in CS (2$-$1) by \cite{1996AAS..115...81B}. One source with a higher mass of $\sim 5 \times 10^4$ M$_{\odot}$ and a radius of $\sim 1.5$ pc is located in the W51 giant molecular cloud complex, associated with W51 IRS1 in W51 Main (see Appendix \ref{gmc}). It emits at 20 $\mu$m, 2 cm, and 6 cm, which shows that it is an extended HII region illuminated by early OB stars in the centre \citep{1982ApJ...255..527G}. Another clump with a slightly smaller mass of $\sim 3 \times 10^4$ M$_{\odot}$ and effective radius of 1.24 pc is in W51 IRS2, which is a compact infrared source near W51 IRS1 \citep{1982ApJ...255..527G}.

\subsection{Identification of complexes with known giant molecular clouds}
\label{gmc}
\subsubsection*{W43 Main}
Complex 172 has most ATLASGAL sources in the first quadrant and consists of 47 clumps . It is located in the well known W43 main star forming complex ($l = 30.767^{\circ}, b = -0.050^{\circ}$ \citep{2003ApJ...582..277M}), which contains a giant HII region. The association of giant molecular clouds with the W43 stellar cluster and the star formation efficiency of the molecular cloud complex of $\sim 25\%$ \citep{2003ApJ...582..277M}, which might result in a stellar density of $\sim$ 100 stars pc$^{-3}$, hint at a sudden ministarburst in W43. It can therefore be classified as a Galactic mini-starburst region and serve as a template to constrain cloud and gas properties in distant starburst galaxies \citep{2003ApJ...582..277M}.
\subsubsection*{W43-south}
We place 30 sources in the first quadrant in complex 167 ($l = 29.96^{\circ}, b = -0.2^{\circ}$ \citep{2011A&A...529A..41N}), which is W43-south. It is located at a slightly smaller Galactic longitude, $29.8^{\circ} < l < 30.5^{\circ}$, than the W43 main cloud (complex 172). Investigation of $^{13}$CO emission toward the two \citep{2011A&A...529A..41N} revealed that they might be connected by low-density gas and that they likely belong to the same giant molecular cloud. The distance to the W43 region has already been derived by several studies: \cite{1999ApJ...517..799P} obtained the near distance to the HII region G29.96$-$0.02, which is located in W43-south, by analysing extinction as well as formaldehyde absorption lines. Moreover, Motte et al. (in prep.) also derive the near distance to that source from large-scale HI absorption. \cite{2009ApJ...690..706A} determine the near distance to the W43 main cloud (complex 172) from HI self-absorption, while they locate G29.96$-$0.02 and another source in W43-south at the far distance from their investigation of HI self-absorption as well as HI absorption. These methods are also used by \cite{2011A&A...526A.151R}, who determine the near distance to several sources in the W43 main cloud and note that there is a disagreement in distances to G29.96$-$0.02 given in the literature. They use the near distance to that source derived from the maser parallax method. The W43 complex has also been observed by the BeSSeL Survey \citep{2009ApJ...693..397R}: From observations of three 12 GHz methanol masers and a 22 GHz water maser \cite{2014ApJ...781...89Z} derived a parallax distance of $5.5 \pm 0.37$ kpc. \cite{2003ApJ...582..756K} obtained the far distance to two sources in W43-south by analysing HI absorption. Using the same method \cite{2008A&A...486..191P} derived the far distance to some clumps located in the W43 main cloud. These distance assignments agree with our distance resolutions for the W43 main cloud (complex 172) and W43-south (complex 167), which we place at the far distance. Investigation of HI self-absorption (see Sect. \ref{HISA analysis}) toward the two complexes hints at the near distance, while HI absorption (see Sect. \ref{continuum absorption method}) toward four sources in the W43 main cloud and two clumps in W43-south reveals the far distance. Combining the two methods as described in Sect. \ref{KDA methods} gives the far distance of $8.4 \pm 0.82$ kpc to complex 172 and $7.9 \pm 0.64$ kpc to complex 167.
\subsubsection*{M17}
We locate 22 sources in complex 50, which is M17 (the Omega Nebula, the Swan Nebula) ($l = 15.10^{\circ}, b = -0.75^{\circ}$ \citep{2009MNRAS.399.2146W}). It is one of the giant molecular clouds with the largest sizes and masses in the Milky Way \citep{1976ApJS...32..603L}. A study of dust emission at near-infrared, far-infrared, and submm wavelengths \citep{1987A&A...181..378C} showed that M17 is associated with one of the most luminous compact HII regions in the Galaxy. \cite{1975ApJ...195..367L} observed 6-cm H$_2$CO aborption lines toward M17, which revealed a lane of dust obscuration west of the HII region. This part of the molecular cloud contains a small region of strong CO emission located south-west of the continuum peak at 6 cm (M17 SW) and is also detected in other molecular lines such as HCN and H$_2$CO at 140 GHz \citep{1974ApJ...189L..35L} probing high cloud densities of $\gtrsim 10^5$ cm$^{-3}$ \citep{1971ApJ...168L..59T}. Varying H$_2$CO linewidths \citep{1975ApJ...195..367L} and CO measurements in M17 SW, which give a lower mass limit of 6000 $M_{\odot}$ \citep{1974ApJ...189L..35L}, hint at a molecular cloud collapse. The 12 GHz CH$_3$OH maser line emitted in M17 SW was measured by \cite{2011ApJ...733...25X} to determine the trigonometric parallax. They obtain a distance of $2 \pm 0.13$ kpc and locate G15.03$-$0.68 in the Sagittarius spiral arm. We also derive a near distance of $2.4 \pm 0.61$ kpc from HI self-absorption and HI absorption toward two sources in the complex.
\subsubsection*{W51 Main}
Complex 261 with 22 ATLASGAL clumps is part of the W51 molecular cloud ($l = 49.2^{\circ}, b = -0.7^{\circ}$ \citep{2009BASI...37...45G}), known as W51 Main. The whole W51 complex is one of the most massive ($1.2 \times 10^6$ M$_{\odot}$) and largest ($\Delta l \times \Delta b = 83 \times 114$ pc) giant molecular clouds \citep{1998AJ....116.1856C}. In addition, W51 is one of the most luminous ($10^7$ L$_{\odot}$) star forming complexes and harbours many O stars \citep{1975LNP....42..443B}, which indicate that the region is currently forming an OB association. Moreover, \cite{2004MNRAS.353.1025K} discovered embedded star clusters from their near-infrared observations, which are associated with UCHIIRs. They obtain a high star formation efficiency of $\sim 10\%$ from an estimate of the W51 molecular cloud mass. \cite{2012IAUS..287..423S} observed the 22 GHz H$_2$O maser line in W51 Main to measure the trigonometric parallax, which results in a distance of $\sim 5.41$ kpc. The 21 cm continuum data toward four ATLASGAL sources show absorption lines at velocities between 65 and 73 km~s$^{-1}$, which are consistent with the higher velocity range of some of the absorption spectra in the W51 molecular cloud by \cite{1997ApJS..108..489K}. Previous studies placed the complex in the Sagittarius arm at the tangent point \citep{1997ApJS..108..489K,2012MNRAS.423..647E}, which agrees with our distance assignment at $5.5 \pm 1.67$ kpc.
\subsubsection*{W31}
We place 21 sources in complex 22, which is W31 ($l = 10.3262^{\circ}, b = -0.1432^{\circ}$ \citep{1970AuJPA..14....1G}). It is a massive star forming region, which was resolved into two components (G10.2$-$0.3 and G10.3$-$0.1) by \cite{1970AuJPA..14....1G} and \cite{1970AuJPA..14...77S} using observations at 5 GHz and 408 MHz. Complex 22 also includes two UCHIIRs \citep[G10.15$-$0.34 and G10.30$-$0.15,][]{1989ApJS...69..831W} located in the larger scale surrounding of the two subregions. \cite{1980A&AS...40..379D} compared velocities of the H110$\alpha$ hydrogen recombination lines of clumps in W31 with H$_2$CO absorption line velocities to resolve the KDA, which results in the near kinematic distance of 6 kpc. \cite{2001AJ....121.3149B} measure a spectrophotometric distance of 3.4 kpc, while \cite{1972A&A....19..354W} obtains H$_2$CO absorption lines at velocities, which are higher than the hydrogen recombination line velocity, and therefore assigns the complex to the far distance. \cite{1976A&A....49...57G} also measure absorption at velocities exceeding the velocity of the HII regions, which hints at the far distance of 18.7 kpc. In addition, analysis of H$_2$CO, OH, and HI absorption lines by \cite{1997ApJ...478..624C} indicates the far kinematic distance of 14.5 kpc. Examination of HI self-absorption toward the complex yields the near distance. In addition, HI absorption toward two clumps (G10.15$-$0.34 and G10.32$-$0.16) is at velocities, which are larger than the source velocity, but much smaller than the tangent point velocity and therefore hints at the near distance of $2 \pm 0.93$ kpc. However, this analysis cannot reject the far distance because absorption at velocities larger than $\sim 50$ km~s$^{-1}$ is unlikely given the lack of HI and CO gas at Galactic longitudes between 5$^{\circ}$ and 25$^{\circ}$ \citep[see e.g. Fig. 3 in][]{1987ApJ...322..706D,2004A&A...419..191C}. \cite{2014ApJ...781..108S} observed 22.2 GHz H$_2$O maser emission toward G10.62$-$00.38 to measure the trigonometric parallax within the BeSSeL Survey \citep{2009ApJ...693..397R}, which resulted indeed in a distance of 5 kpc and located W31 on the near edge of the hole in the gas distribution within a galactocentric radius of $\sim 3$ kpc.
\subsubsection*{N49}
Complex 156 consisting of 17 ATLASGAL clumps is associated with the HII region N49 ($l = 28.827^{\circ}, b = -0.229^{\circ}$ \citep{2006ApJ...649..759C}). It is one of the bubbles, which are detected by the investigation of GLIMPSE \citep{2003PASP..115..953B} images from \cite{2006ApJ...649..759C}. This region consists of a layer of dense neutral material, which accumulates as a shell around a radio HII region with an approximately spherical shape. 20 cm free-free emission probes the ionized gas of N49, the thermal continuum at 24 $\mu$m traces the hot dust within the HII region, which is surrounded by an outer layer of 8 $\mu$m emission \citep{2008ApJ...681.1341W}. A hole is revealed in the centre of the 24 $\mu$m and 20 cm emission, where the exciting star is located. This might be an O5V or O8III star, although the age estimate by \cite{2008ApJ...681.1341W} favours the O5V star with $\geq 10^5$ yr. Analysis of the 870 $\mu$m emission yields four condensations within the shell around the central HII region and a mass of 4200 M$_{\odot}$ for the whole shell \citep{2010A&A...523A...6D}. NH$_3$ velocities \citep{2012AA...544A.146W} are observed toward three condensations and the $^{13}$CO velocity \citep{2006ApJS..163..145J} is measured toward the fourth condensation indicating their assocation with the ionized gas. Two massive young stellar objects can be revealed at 24 $\mu$m, which are in an early evolutionary phase \citep{2008ApJ...681.1341W}, as well as an UCHIIR located in the surrounding of the most massive condensation \citep{2010A&A...523A...6D}. \cite{2009ApJ...690..706A} derived the near distance to the region from HI self-absorption and HI absorption. Because none of the observed ATLASGAL sources in N49 satisfied our criteria for radio continuum absorption, we could only look for HI self-absorption and we also determined the near distance of $5.7 \pm 0.41$ kpc.
\subsubsection*{M16}
We locate ten sources in complex 70, which is M16 (the Eagle Nebula) ($l = 16.94^{\circ}, b = 0.76^{\circ}$ \citep{1974AJ.....79..786Z}). The HII region M16 belongs to a larger star forming complex, which is mostly still embedded within the molecular cloud \citep{1976Ap&SS..41..105G}. Images taken by the Hubble Space Telescope reveal structures in the molecular cloud containing dense globules, which are photoevaporated more slowly than their lower density environment, forming ''molecular pillars''. The heating and ionization of M16 results from the young open star cluster NGC6611. The Eagle Nebula is associated with recent and ongoing high mass star formation. Its stellar population has been analysed by \cite{1993AJ....106.1906H}. They estimated an age of $(2 \pm 1) \times 10^6$ yr and locate the region at a distance of 2 kpc from spectroscopic parallaxes. To resolve the KDA we could only analyse HISA because complex 70 contains no 21 cm continuum source. We obtain a near distance of $2.2 \pm 0.56$ kpc, which agrees with the distance derived by \cite{1993AJ....106.1906H}.
\subsubsection*{G333}
Complex 210 consists of most ATLASGAL sources in the fourth quadrant containing 58 clumps ($l = 333.19^{\circ}, b = -0.36^{\circ}$ \citep{2006MNRAS.367.1609B}). It is known as the G333 giant molecular cloud complex or RCW106 region with several molecular clouds forming high mass stars as well as HII regions. We determine the near kinematic distance to complex 210 of $3.5 \pm 0.35$ kpc from HI self-absorption and HI absorption toward several sources. Our estimate is consistent with the kinematic distance of 3.6 kpc calculated by \cite{1979ApJ...232..761L} as well as the range from 3.3 to 3.8 kpc given by \cite{2012MNRAS.420.1656U} and with the spectrophotometric distance of 3.96 kpc obtained by \cite{2011MNRAS.411..705M}.
\subsubsection*{G305}
From our ATLASGAL sample in the fourth quadrant 41 sources are located in complex 29 ($l = 305.361^{\circ}, b = 0.056^{\circ}$ \citep{2004A&A...427..839C}), which is the rich high mass star forming complex G305. Some luminous HII regions are embedded within the molecular cloud complex, the analysis of MSX data by \cite{2004A&A...427..839C} yields strong radio sources with up to 31 O7V stars ionizing an HII region. \cite{2003A&A...397..133R} and \cite{2011MNRAS.411..705M} divided the region into two parts, G305.2+0.0 and G305.2+0.2, and derived distances to them. While \cite{2003A&A...397..133R} obtained a distance of 3.5 kpc for both, the spectrophotometric distance analysed by \cite{2011MNRAS.411..705M} is inconclusive for G305.2+0.0, but yields the same result as assigned by \cite{2003A&A...397..133R} for G305.2+0.2. Our HI self-absorption method also indicates the near distance to complex G305. However, the analysis of HI absorption toward one source of the region, G305.19+0.03 (see Fig. \ref{21cm continuum source}), results in the far distance. The composite of the two methods reveals the far distance of $6.6 \pm 1.24$ kpc to the complex.
\subsubsection*{NGC 6334}
Complex 356 with 28 ATLASGAL clumps is known as the NGC 6334 molecular cloud complex ($l = 351.317^{\circ}, b = 0.661^{\circ}$ \citep{1982ApJ...255..103R}) with embedded HII regions identified at 6 cm as compact to extended radio sources. \cite{1978A&A....69...51N} derived the distance to NGC 6334 from optical observations of young stars within this complex of 1.74$\pm$0.31 kpc and place it in the Sagittarius arm. They also give a kinematic distance of 0.7$\pm$1.9 kpc, which corresponds to the interarm region. The two values are within the uncertainty of the near kinematic distance of $0.8 \pm 1.03$ kpc, which we obtained. The large distance uncertainty originates from the unreliability of kinematic distances near a longitude of 0$^{\circ}$. Our result also agrees with the spectrophotometric distance of 1.82 kpc given by \cite{2011MNRAS.411..705M} and with the distance of 1.35 kpc derived recently from trigonometric parallax measurement by \cite{2014ApJ...783..130R}.
\subsubsection*{G338.4+0.12}
Complex 255 contains 24 ATLASGAL sources, which connects three small complexes. The complex is a giant HII region, where \cite{1987A&A...171..261C} measured hydrogen recombination lines near 5 GHz. \cite{1987A&A...171..261C} derived a vast range of recombination line velocities over 120 km~s$^{-1}$ indicating several features, which lie on the same line of sight to the observer at different distances. We obtain a narrower velocity range from $-30$ to $-42$ km~s$^{-1}$ of most sources and a few clumps exhibit $-46$ to $-56$ km~s$^{-1}$ in our smaller complex at a longitude of 338$^{\circ}$. \cite{1996A&AS..120...41G} also determined velocities and distances to two radio sources in the complex from observations of H$\alpha$ emission. They give kinematic distances of 2.2 kpc for 338.398+0.164 and 3.2 kpc for 338.450+0.061, which lie within the errors of our distance estimates. However, they should assign the far distance taking the OH absorption line into account. While \cite{2003A&A...397..133R} also places the complex at the far distance, our KDA resolution reveals the near distance of 3.06 kpc. The HI intensity map shows absorption, which hints at the near kinematic distance. In addition, the HI absorption method results in the near kinematic distance to one HII region and the far distance to four 21 cm continuum sources. The combination of the two techniques gives the near kinematic distance of $3.1 \pm 0.41$ kpc to the complex.
\subsubsection*{Gum 50}
We locate 23 sources in the fourth quadrant in complex 163, which is known as Gum 50 ($l = 328.573^{\circ}, b = -0.531^{\circ}$ \citep{1988iras....1.....B}). Observations of hydrogen recombination lines near 5 GHz by \cite{1987A&A...171..261C} reveal HII regions located in the complex such as IRAS 15539$-$5353 and G328.57$-$0.53. Moreover, a few sources in this high mass star forming region, e.g. IRAS 15525$-$5407, IRAS 15539$-$5353, and G328.34$-$0.53, are associated with IRDCs \citep{2009A&A...505..405P}, which indicates the near distance. This is in agreement with our HI self-absorption and HI absorption analysis resulting in the near kinematic distance of $3 \pm 0.39$ kpc.

\begin{table*}[htbp]
\caption[]{Associations of largest ATLASGAL complexes with known giant molecular clouds (see Sect. \ref{literature-complex}).}
\label{large-complex}
\centering
\begin{tabular}{l c c l c c}
\hline\hline
Sample &  Complex number & Number of clumps & Giant molecular cloud & Distance$_{\mbox{\tiny complex}}$ & $\mathrm{log_{10}}$ (Sum mass$_{\mbox{\tiny clumps}}$/M$_{\odot}$) \\
 & & & & (kpc) & \\ \hline 
in first quadrant & 172  & 47 & W43 Main & 8.4  & 5.8  \\ 
 & 167 & 30 & W43-south & 7.92 &  5.54 \\
 & 50 & 22 & M17 & 2.36 & 5.14 \\
 & 261 & 22 & W51 Main & 5.54 & 5.68 \\
 & 22 & 21 & W31 & 1.97 & 4.67 \\
 & 156 & 17 & N49 & 5.71 & 4.84 \\ 
 & 70 & 10 & M16 & 2.22 & 4.32 \\ \hline
in fourth quadrant & 210 & 58 & G333 & 3.52 & 5.49 \\
 & 29 & 41 & G305 & 6.58 & 5.52 \\
 & 356 & 28 & NGC6334 & 0.83 & 4.09 \\
 & 255 & 24 & G338.4+0.12 & 3.06 & 4.75 \\
 & 163 & 23 & Gum 50 & 2.96 & 4.61 \\ \hline 
\end{tabular}
\end{table*}

\begin{figure}[h]
\centering
\includegraphics[angle=-90,width=9.0cm]{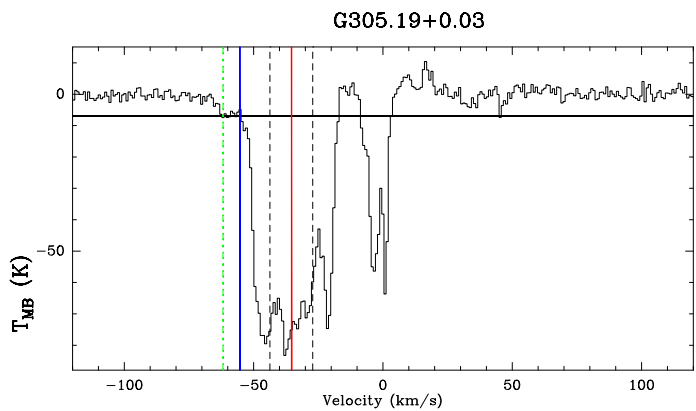}
\caption[21cm continuum absorption spectrum]{HI spectrum toward G305.19+0.03, an HII region in the high mass star forming complex G305 extracted from the SGPS. For explanation of the different lines see Fig. \ref{21cm continuum}. The HI absorption method yields the far distance.}\label{21cm continuum source}
\end{figure}

\section{\label{comp}Comparison of kinematic and trigonometric parallax distances}
Recently, more than a hundred distances to high mass star forming regions have been determined via measurements of trigonometric parallaxes of class II CH$_3$OH and H$_2$O masers with Very Long Baseline Interferometry (VLBI). The data were taken with the VLBI Exploration of Radio Astronomy (VERA) array\footnote{http://veraserver.mtk.nao.ac.jp} and with the NRAO Very Long Baseline Array in the course of Bar and Spiral Structure Legacy (BeSSeL) survey\footnote{http://bessel.vlbi-astrometry.org} \citep{2014ApJ...783..130R}. Parallaxes provide gold standard distance measurements since they result from  \textit{direct} measurements and basic trigonometry and do not require assumptions such as a rotation curve. 

Of the 103 sources for which \cite{2014ApJ...783..130R} present parallax distances, 48 are covered by ATLASGAL. For total of 27 of these we have determined kinematic distances. Our LSR velocities, determined from NH$_3$ spectra agree with maser velocities within the uncertainties assigned by \cite{2014ApJ...783..130R}, in 15 cases within 2 km~s$^{-1}$ and in all cases within 6 km~s$^{-1}$. In Fig. \ref{compfig} we compare the parallax distances with our kinematic distances. 

For a third of the 27 sources the kinematic and the parallax distance agree within the errors and all but two within a factor of two. For the latter two, G12.88$+$0.48 and G16.58$-$0.05, choosing the far kinematic distances, 12.9 and 11.7 kpc, respectively, obviously was a wrong decision, which for the former is reinforced by its high latitude. Both are found in complexes that do not contain radio continuum sources and thus have no HI absorption. They were placed at the far distance because they also did not exhibit HI self-absorption. 

Part of the scatter in Fig. \ref{compfig} may be caused by peculiar velocities, which of course are not accounted for in the rotation curve model used to determine the kinematic distances. A dramatic example for this is the case of W3OH for which the parallax distance is almost a factor of two smaller than the kinematic distance \citep{2006Sci...311...54X}.

\begin{figure}[h]
\centerline{\resizebox{\hsize}{!}{\includegraphics[angle=0]{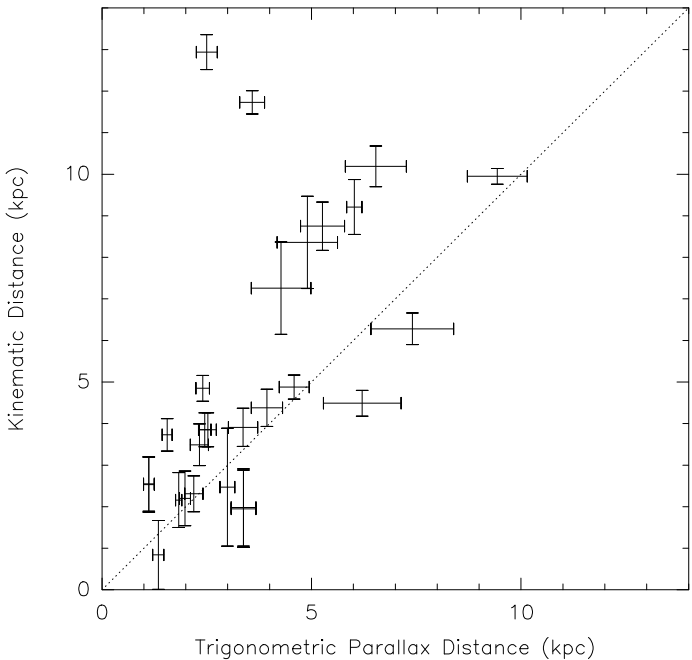}}}
\caption{Kinematic distances determined in our study vs. trigonometric parallax distances from the BeSSeL survey and VERA results. The dotted fiducial line marks equality.}  
\label{compfig}
\end{figure}

Some of the sources with measured parallax distance and resolved kinematic distance ambiguity are described in more detail.

\subsubsection*{G12.89+0.49}
G12.89+0.49 is in complex 44. This source is a high mass protostellar object \citep{2002ApJ...566..945B}, which is associated with IRAS 18089$-$1732 \citep{2002ApJ...566..931S}. It is well known to harbour strong H$_2$O \citep{1996ApJ...463..205A}, CH$_3$OH, \citep{1995MNRAS.272...96C} and OH masers \citep{1991MNRAS.250..611C}. Previous studies \citep{2002ApJ...566..931S,2002ApJ...566..945B} give the near and far kinematic distance, but did not solve the KDA. Using 12.2 GHz methanol masers \cite{2011ApJ...733...25X} derived a distance of 2.34 kpc from the trigonometric parallax. \cite{2013A&A...553A.117I} also measured a parallax distance from 22.2 GHz water maser observations, which is in agreement with the result of \cite{2011ApJ...733...25X}, and locate the high mass star forming region in the Scutum spiral arm. Because there is no 21 cm continuum source in complex 44, we could only analyse HISA, which reveals the far kinematic distance of $13 \pm 0.46$ kpc. However, the HISA method is less reliable for the far distance than for the near distance. 
\subsubsection*{G12.91$-$0.26 and G12.68$-$0.18}
These two sources are located in the W33 molecular cloud complex. It is a strong radio continuum source detected by \cite{1958BAN....14..215W} and consists of various evolutionary phases of high mass star formation from quiescent IRDCs to HII regions. Radio recombination line observations revealed two velocity components in different parts of W33, one at $\sim 36$ km~s$^{-1}$ and one at $\sim 58$ km~s$^{-1}$ \citep{1978A&A....64..341B}. We also obtain several velocities from NH$_3$ observations in this region and divide it into different complexes. We locate G12.91$-$0.26 in complex 38 with a mean radial velocity of 43 km~s$^{-1}$ and a velocity dispersion of 6.81 km~s$^{-1}$ and G12.68$-$0.18 in complex 41 with a mean velocity of 55.56 km~s$^{-1}$ and a velocity dispersion of 0.57 km~s$^{-1}$. \cite{2013A&A...553A.117I} used a water maser exhibiting velocities between 36 to 38 km~s$^{-1}$ to derive the parallax of their observation at $l = 12.909^{\circ}, b = -0.261^{\circ}$, which corresponds to a distance of 2.53 kpc. This is consistent with our investigation of HI self-absorption and HI absorption towards two sources in complex 38, which reveal the near distance of $4.2 \pm 0.39$ kpc. \cite{2013A&A...553A.117I} also fit the position of a water maser to obtain the parallax of G12.68$-$0.18, which results in a distance of 2.4 kpc. This is in agreement with our KDA resolution of complex 41, which gives the near kinematic distance of $4.9 \pm 0.3$ kpc using only the HISA method. 
\subsubsection*{G23.01$-$0.41}
G23.01$-$0.41 is located in complex 114. This high mass star forming region contains a 12 GHz methanol maser, which coincides with a H$_2$CO maser and NH$_3$ (3,3) peak emission. A fit to the positions of the two masers gives one parallax, which corresponds to a distance of 4.59 kpc \citep{2009ApJ...693..424B}. Different distances to this source are given by previous studies so far: \cite{1997A&A...325..282C}, \cite{1999A&AS..137...43F} and \cite{1983AuJPh..36..417C} used the far kinematic distance, although the distance ambiguity was not resolved. In contrast, \cite{1998A&AS..132..211H} locate the region at the near kinematic distance. This is consistent with our KDA resolution resulting in $4.9 \pm 0.29$ kpc from HI self-absorption.
\subsubsection*{G23.46$-$0.18}
G23.46$-$0.18 belongs to complex 117. This is a star forming region, where 6.7 GHz methanol maser emission observed by \cite{1998MNRAS.301..640W} hints at massive young stellar objects. Previous studies revealed different velocity components of the molecular cloud harbouring G23.44$-$0.18: \cite{2012PASJ...64...74O} obtained three components at 101, 103, and 104 km~s$^{-1}$ from their H$^{13}$CO$^+$ spectra, while \cite{1989ApJS...71..469L} derived the radio recombination line velocity of 103 km~s$^{-1}$ for an HII region, which is likely associated with the molecular cloud. \cite{2009ApJS..182..131R} give two $^{13}$CO velocities at 101 km~s$^{-1}$ and $\sim 104$ km~s$^{-1}$, for which \cite{2009ApJ...699.1153R} determined kinematic distances of 6.43 kpc to the component at 101 km~s$^{-1}$ and of 6.65 kpc to the 104 km~s$^{-1}$ component from HISA and HI absorption. \cite{2009ApJ...693..424B} fitted the positions of four 12 GHz methanol masers to measure the parallax of their observation at $l = 23.4398^{\circ}, b = -0.1822^{\circ}$. This yields a distance of 5.88 kpc and places the region near the end of the galactic bar in the Norma arm. We measured a velocity of 98.4 km~s$^{-1}$ toward G23.46$-$0.18, which is close to the velocity of the methanol maser at 97.6 km~s$^{-1}$ \citep{2009ApJ...693..424B}. We derive the near kinematic distance of $5.9 \pm 0.3$ kpc using the HISA method, which is in agreement with the distance from the trigonometric parallax and also similar to the kinematic distance of 6.43 kpc of \cite{2009ApJ...699.1153R}.
\subsubsection*{G23.96$-$0.11}
G23.96$-$0.11 is in complex 121, which is associated with G23.657$-$0.127 observed with the BeSSeL survey. G23.657$-$0.127 is a massive protostar, where \cite{2005A&A...442L..61B} detected 6.7 GHz methanol maser emission between 77 and $\sim 88$ km~s$^{-1}$. Analysis of an EVN image showed that the methanol masers are arranged in an approximately circular ring \citep{2005A&A...442L..61B}. \cite{2008A&A...490..787B} observed 12.2 GHz methanol masers with the VLBA towards G23.657$-$0.127, which have all 6.7 GHz counterparts and are also distributed spherically symmetric. Fitting 19 positions of 12.2 GHz methanol masers, which have approximately the same velocities as the 6.7 GHz methanol masers, reveals a distance of 3.19 kpc from the trigonometric parallax \citep{2008A&A...490..787B}. We obtain a similar velocity range of complex 121 with a mean velocity of 78 km~s$^{-1}$ and a velocity dispersion of 3.7 km~s$^{-1}$. The HISA method and HI absorption towards one source in complex 121 results in the near kinematic distance of $4.9 \pm 0.31$ kpc, which is similar to the distance derived by \cite{2008A&A...490..787B}.
\subsubsection*{G35.19$-$0.74}
G35.19$-$0.74 is located in complex 214 and associated with G35.2$-$0.7, which is observed by the BeSSeL survey. The complex is a massive star forming region, where \cite{2003MNRAS.339.1011G} observed CO, SiO, and radio emission indicating several outflows from G35.2$-$0.7. In addition, VLA observations at 3.5 and 6 cm showed fragmentation of the radio continuum into multiple YSOs embedded within dense clumps along the outflow \citep{2003MNRAS.339.1011G}. \cite{1987ApJ...319..730S} derived the near distance of 3.3 kpc because the far distance would locate the source at a larger height than 150 pc from the plane. The positions of two 12 GHz masers with radial velocities of 27.9 and 27.5 km~s$^{-1}$ were fitted by \cite{2009ApJ...693..419Z} to obtain the trigonometric parallax, which leads to a distance of 2.19 kpc. The mean velocity of complex 214 is 34.1 km~s$^{-1}$ with a velocity dispersion of 2.02 km~s$^{-1}$, which agrees with the velocities of the masers. Because complex 214 contains no 21 cm continuum source, we use only the HISA method, which reveals the near kinematic distance of $2.3 \pm 0.43$ kpc. Our KDA resolution is therefore consistent with the distances given by \cite{1987ApJ...319..730S} and \cite{2009ApJ...693..419Z}.

\end{appendix}

\newpage

\bibliography(atlasgal-distances)
\bibliographystyle{aa}

\end{document}